\pdfoutput=1

\documentclass{article}
\usepackage{arxiv}

\usepackage[T1]{fontenc}

\usepackage{cite}  

\usepackage{xcolor}
\usepackage{graphicx}
\usepackage{hyperref}
\usepackage{amsfonts}
\usepackage{amsmath}

\usepackage[normalem]{ulem}
\usepackage{amsmath, amssymb}
\usepackage{stmaryrd}
\usepackage[labelfont=bf, textfont=it]{caption}
\usepackage[labelfont=it, textfont=it]{subcaption}
\usepackage{multirow}
\usepackage{siunitx}
\usepackage{tikz}
\usetikzlibrary{positioning,spy,arrows,backgrounds}
\tikzstyle{every picture}+=[remember picture]
\tikzstyle{na} = [baseline=0ex]
\usepackage[outline]{contour}
\contourlength{0.7pt}

\graphicspath{{graphics/}}

\hypersetup{
    pdftoolbar=true,        
    pdfmenubar=true,        
    pdffitwindow=false,     
    pdfstartview={FitH},    
    pdftitle={Harmonics Virtual Lights},    
    pdfauthor={Pierre Mézières},     
    pdfsubject={Fast projection of luminance field on spherical harmonics},   
    colorlinks=true,       
    linkcolor=red,          
    citecolor=blue,        
    filecolor=magenta,      
    urlcolor=cyan           
}

\newcommand{\etal}{\emph{et al.}}

\newcommand{\figp}[1]{(Fig.~#1)}
\newcommand{\Fig}[1]{Fig.~#1}
\newcommand{\fig}[1]{Fig.~#1}
\newcommand{\equp}[1]{(Eq.~#1)}

\newcommand{\equ}[1]{Eq.~#1}
\newcommand{\sectp}[1]{(Sec.~#1)}

\newcommand{\sect}[1]{Sec.~#1}
\newcommand{\ie}{\emph{i.e.} }
\newcommand{\eg}{\emph{e.g.} }

\newcommand{\diff}[1]{\mathop{\mathrm{d}#1}}
\newcommand{\point}{x}

\newcommand{\sphere}{\Omega}
\newcommand{\function}{S} %
\newcommand{\Function}{\mathbf{S}}
\newcommand{\radiance}{L}
\newcommand{\Radiance}{\mathbf{L}}
\newcommand{\RadianceZH}{\mathbf{\tilde{L}}}
\newcommand{\reflectance}{F}
\newcommand{\Reflectance}{\mathbf{F}}

\newcommand{\Visibility}{\mathbf{V}}
\newcommand{\brdf}{f_s}
\newcommand{\sh}{Y}
\newcommand{\order}{l}
\newcommand{\maxorder}{{N+1}}
\newcommand{\degree}{m}
\newcommand{\legendre}{P}
\newcommand{\factor}{K}
\newcommand{\surfaceProj}{Q}
\newcommand{\angleProj}{a}
\newcommand{\dir}{\omega}
\newcommand{\dirlight}{\omega_{light}}
\newcommand{\vplNumber}{M}
\newcommand{\norm}[1]{\left\lVert#1\right\rVert}
\newcommand{\abs}[1]{\left\lvert#1\right\rvert}

\newcommand{\tab}{\mathrm{Tab}}

\newlength{\imgvspace}
\newenvironment{tightequation}{%
    \begin{equation}%
}{%
    \end{equation}%
}
\newenvironment{tightequation*}{%
    \begin{equation*}%
}{%
    \end{equation*}%
}
\makeatletter

\makeatletter
\newenvironment{tightalign*}{%
    \start@align\@ne\st@rredtrue\m@ne%
}{%
    \endalign%
}

\newcommand{\tightcaption}[1]{%
    \caption{#1}%
}

\newcommand{\tightsubcaption}[1]{%
    \caption{#1}%
}

\begin{document}

\title{
Harmonics Virtual Lights : fast projection of luminance field on spherical harmonics for efficient rendering.
}
\date{}

\newcommand{\IRIT}{IRIT, Universit\'{e} de Toulouse, CNRS, UT3, Toulouse, France}

\author{ 
    \href{https://orcid.org/0000-0003-4981-4649}{\includegraphics[scale=0.06]{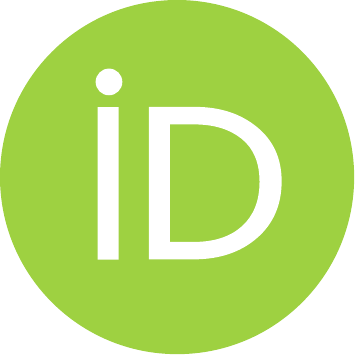}\hspace{1mm}}Pierre Mézières\\
\IRIT \\
\texttt{pierre.mezieres@irit.fr} \\
\And
\href{https://orcid.org/0000-0002-6246-7146}{\includegraphics[scale=0.06]{orcid.pdf}\hspace{1mm}} François Desrichard \\
\IRIT \\
\texttt{françois.desrichard@irit.fr}
\And
\href{https://orcid.org/0000-0003-0506-6036}{\includegraphics[scale=0.06]{orcid.pdf}\hspace{1mm}} David Vanderhaeghe \\
\IRIT \\
\texttt{david.vanderhaeghe@irit.fr}
\And
\href{https://orcid.org/0000-0001-5606-9654}{\includegraphics[scale=0.06]{orcid.pdf}\hspace{1mm}} Mathias Paulin \\
\IRIT \\
\texttt{mathias.paulin@irit.fr} \\
}

\renewcommand{\undertitle}{}
\renewcommand{\shorttitle}{Harmonics Virtual Lights}
\renewcommand{\headeright}{}
\maketitle

\title{Harmonics Virtual Lights}

\begin{abstract}
In this paper, we introduce Harmonics Virtual Lights (HVL), to model indirect light sources for interactive global illumination of dynamic 3D scenes. Virtual Point Lights (VPL) are an efficient approach to define indirect light sources and to evaluate the resulting indirect lighting. 
Nonetheless, VPL suffer from disturbing artifacts, especially with high frequency materials.
Virtual Spherical Lights (VSL) avoid these artifacts by considering spheres instead of points but estimates the lighting integral using Monte Carlo which results to noise in the final image.
We define HVL as an extension of VSL in a Spherical Harmonics (SH) framework, defining a closed form of the lighting integral evaluation.
We propose an efficient SH projection of spherical lights contribution faster than existing methods. Computing the outgoing luminance requires $\mathcal{O}(n)$ operations when using materials with circular symmetric lobes, and  $\mathcal{O}(n^2)$ operations for the general case, where $n$ is the number of SH bands.
HVL can be used with either parametric or measured BRDF without extra cost and offers control over rendering time and image quality, by either decreasing or increasing the band limit used for SH projection.
Our approach is particularly well designed to render medium-frequency one-bounce global illumination with arbitrary BRDF in interactive time.
 
\end{abstract}



\section{Introduction}
Fast and high-quality rendering of 3D environments has numerous applications: 
real-time rendering allows seamless interactions in video games and virtual reality applications, while interactive rendering is used by content creators to assess the look of a scene in seconds, and progressively refine it before moving on to offline rendering.
These applications require very detailed 3D geometry and displays realistic materials including high specularity and advanced effects such as anisotropy, represented by a Bidirectional Reflectance Distribution Function (BRDF).

\begin{figure*}
    \includegraphics[width=\linewidth]{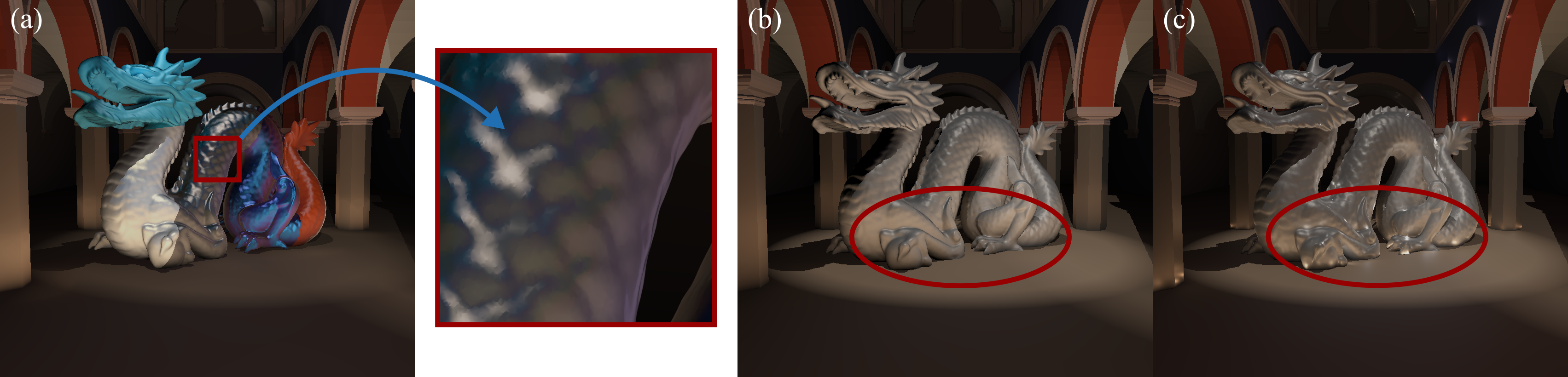}%
    \tightcaption{(a)~Harmonics Virtual Lights (HVLs) can render global illumination with arbitrary BRDF at no extra cost, while for example VPLs have an additional cost to use numerical material. (b)~HVLs do not produce the classic artifacts of (c)~the Virtual Point Lights methods.}
    \label{fig:Teaser}%
\end{figure*}

Efficient rendering of such scenes often relies on lighting algorithms based on virtual light sources, which compute the final picture in two steps.
In the first step, light propagation is performed from the scene light sources, \ie{} the primary sources, and yields Virtual Point Lights (VPLs) distributed on surface of the scene, \ie the secondary sources that act as secondary light sources.
Afterwards, the estimation of global lighting at each point of the scene is reduced to the simpler problem of direct lighting from the secondary sources, which can be handled using rasterization. 

Many-lights methods allow to manage a big number of secondary light sources in real-time thus resulting in an efficient rendering algorithm that simulates realistic light transport with inter-reflections.
The key point of these family of algorithms concern the way virtual light sources are distributed into the scene and how their properties are represented so that the cost of direct lighting remains very low.
Our proposal, agnostic to the 
BRDF representation, consists in an efficient projection of the luminance field on spherical harmonics such that VPLs are replaced by what we refer as Harmonic Virtual Lights.

VPLs provide unbiased results at interactive framerates, but the resulting lighting presents several limitations.
Several millions of VPLs must be distributed inside the scene to capture detailed inter-reflections; at each pixel of the image, efficient summation of secondary contributions becomes a computational challenge that must be tackled for these approaches to scale.
Due to the singularities of VPLs, they exhibit several artefacts during lighting evaluation, a glossy surface will show the reflection of a VPL if the density of secondary sources is not sufficient and artefacts appear in the vicinity of VPL position. 
The unbounded contributions results in artefacts around secondary sources
because luminance follows an inverse square relation with the distance between the two points.
As a simple workaround, limiting the glossiness of materials in the scene makes the unbounded term less noticeable in most cases, at the expense of the specular component.
Virtual Spherical Lights (VSLs) consider spherical secondary light sources to avoid VPLs singularities, and thus limits the spikes contribution.
However, VSLs evaluation relies on Monte Carlo integration and is hardly appropriate for interactive to real-time rendering.

We propose a solution compatible with interactive scenarios that addresses these issues while increasing the computational efficiency of virtual lights.
We propose to craft a finer representation of secondary emitters from spherical harmonics (SH) which we call \emph{Harmonics Virtual Lights} (HVLs).
Spherical harmonics represent spherical functions in the frequency domain, allowing a band limited approximation by taking only the first harmonics coefficients.

\noindent Our main contributions are:
\begin{itemize}
    \item An efficient projection of spherical lights contributions on SHs. 
    \item The harmonics virtual light (HVL).
    \item Two heuristics to efficiently estimate the HVL radius.
\end{itemize}

HVLs follow the same pipeline as previous many-lights approaches: secondary lights are sampled for each primary light sources onto objects surfaces, these secondary light sources are then evaluated and the results combined to obtain the final image \sectp{\ref{sec:overview}}.
HVL evaluation is based on Spherical Harmonics (SHs) to solve the rendering equation \sectp{\ref{sec:SHframework}} and in particular we use Zonal Harmonics (ZHs) sub-basis of SHs \sectp{\ref{sec:projRadAndMat}} to efficiently project materials and spherical lights on the SH basis. 
This projection allows the use of the spherical harmonics lighting framework with hundreds to thousands of lights, using a full spherical harmonics convolution ($\mathcal{O}(n^2)$).
To compute indirect lighting, we distribute HVLs in the 3D scene using reflective shadow maps and calculate their contribution \sectp{\ref{sec:HVL}} using SHs.
We propose two heuristics to manage the HVL radius as it has a crucial role in the final image quality \sectp{\ref{sec:HVLradius}}.
Finally, we analyse the results produced by HVLs and compare them with state of the art method VPL/VSL/VSGL \sectp{\ref{sec:results}}. 
While comparing with VSGL, we derive from existing work to reduce the complexity of convolution on zonal harmonics (becoming $\mathcal{O}(n)$) \sectp{\ref{sec:VSGL}}.
Our proposed HVLs model handle arbitrary materials and suffer less artefacts than VPLs while being less expensive than VSLs.

\section{Related Work}
\label{sec:RelatedWork}
\paragraph*{Spherical harmonics lighting.}

In the field of computer graphics, many signals are naturally defined on a spherical domain. 
Orthogonal bases make it possible to represent these functions efficiently and suit the rendering equation~\cite{Kajiya1986} particularly well.
In this paper, we focus on the spherical harmonics basis which has a rich history. 
Ramamoorthi and Hanrahan~\cite{Ramamoorthi2001a} propose to use SHs to efficiently compute and store illuminance environment maps; Sloan~\etal~\cite{Sloan2002} extend this principle to store complex luminous transfer functions, and Kautz~\etal~\cite{Kautz2002} further develop the method to handle arbitrary BRDFs.
The real-time performances of their approach come at the cost of static scene, with only dynamic view point.
To remain fully dynamic, they compute the SH projection on points in the scene and interpolate the result.
Thus, they can perform exact calculations simply by applying a rotation on the environment map.
This means that the need of an efficient way to rotate the SHs coefficients on the fly is required.
Even with effective derivations to apply the rotation as proposed by Nowrouzezahrai~\etal~\cite{Nowrouzezahrai2012}, heavy computation is necessary on high bands. In our approach, we avoid this costly step by integrating the rotation into the SH projection~\cite{Sloan2005}.

Real-time projection of arbitrary functions is one of the main goals of spherical harmonics lighting methods.
Wang and Ramamoorthi~\cite{Wang2018} extend the principle of rotation proposed by Nowrouzezahrai~\etal~\cite{Nowrouzezahrai2012} to efficiently project polygonal domains on spherical harmonics.
They reduce the integration of a polygonal light to the integration of its contours; in turn, contour integration fits the integration of zonal harmonics, allowing real-time performances. Wu~\etal~\cite{Lifan2020} extends this by computing the SH coefficients and their gradients, in the case of polygonal sources, on a 3D grid, thus allowing Hermite interpolation of the SH coefficients at each shaded point instead of computing them explicitly.
We propose a simpler mathematical framework where integration on zonal harmonics is straightforward \sectp{\ref{sec:projRadAndMat}} when dealing with spherical lights.

\paragraph*{Virtual lights.}

The seminal work of Keller~\cite{Keller1997} introduces the concept of point-based global illumination. 
The main idea is to approximate global illumination from its discretization on the surface of the scene using virtual lights.
One of the common ways to manage virtual lights is to consider them as virtual point lights.
To place the virtual lights in real-time for dynamic scenes, the most efficient state of the art technique in rasterization is the reflective shadow maps (RSM) of Dachsbacher~\etal~\cite{Dachsbacher2005}.
The principle is straightforward, and reflective shadow maps allow efficient placement of VPLs at the first bounce from the light.
Today, new graphics cards are also able to dispatch virtual lights with ray tracing, as used by Lin and Yuksel~\cite{Lin2019}.

When considering virtual point lights, there are two main kinds of artefacts. 
First, if there are not enough virtual lights, flickering will appear in dynamic scenes.
But spike artefacts are a more troublesome problem: to compute the illumination of a point, the squared inverse of the distance to the virtual point light is used. When the point is close to the VPL, this term will be very high and cause strong light spikes in the image.
A similar problem occurs in the presence of glossy materials: when the view aligns with the reflection of the virtual point lights and their density is not sufficient, the reflection of lights on glossy materials clearly appears.
The naive way to overcome this problem is to clamp the light contribution of virtual point lights, losing glossy materials appearance.
To address this problem, Ha\v{s}an~\etal~\cite{Hasan2009} propose to replace virtual point lights with virtual spherical lights.
Instead of considering the null integration domain of a single point, they consider a disk centered on that point.
The contribution of light is averaged over the disk, drastically diminishing artefacts.
Alternatively, Tokuyoshi proposes virtual spherical gaussian lights (VSGL)~\cite{Yusuke2015} and approximates the contribution of a set of VPLs. 
The lack of VPLs is compensated by approximating a set of VPL with spherical gaussian. 
While the use of spherical gaussians had been the subject of earlier work \cite{Xu2014a}, the use of spherical harmonics coupled to VPL had never been realized. 
In contrast of VSGL, we consider lights individually; our idea is to make up for the lack of VPLs by replacing point light sources with spherical light sources and covering the scene to avoid artefacts. 
At the same time, each HVL account for more information than simple VPL, as the main idea of Rich-VPL \cite{Simon2015}.

For complementary information about point-based global illumination and many-light rendering in general, we refer the reader to the two most recent states of the art on the topic of global illumination~\cite{Ritschel2012,Dachsbacher2014}.

\section{HVL Lighting Pipeline Overview}
\label{sec:overview}

\begin{figure}[]
\begin{minipage}[c]{0.47\linewidth}
    \centering%
        \includegraphics[width=0.76\linewidth]{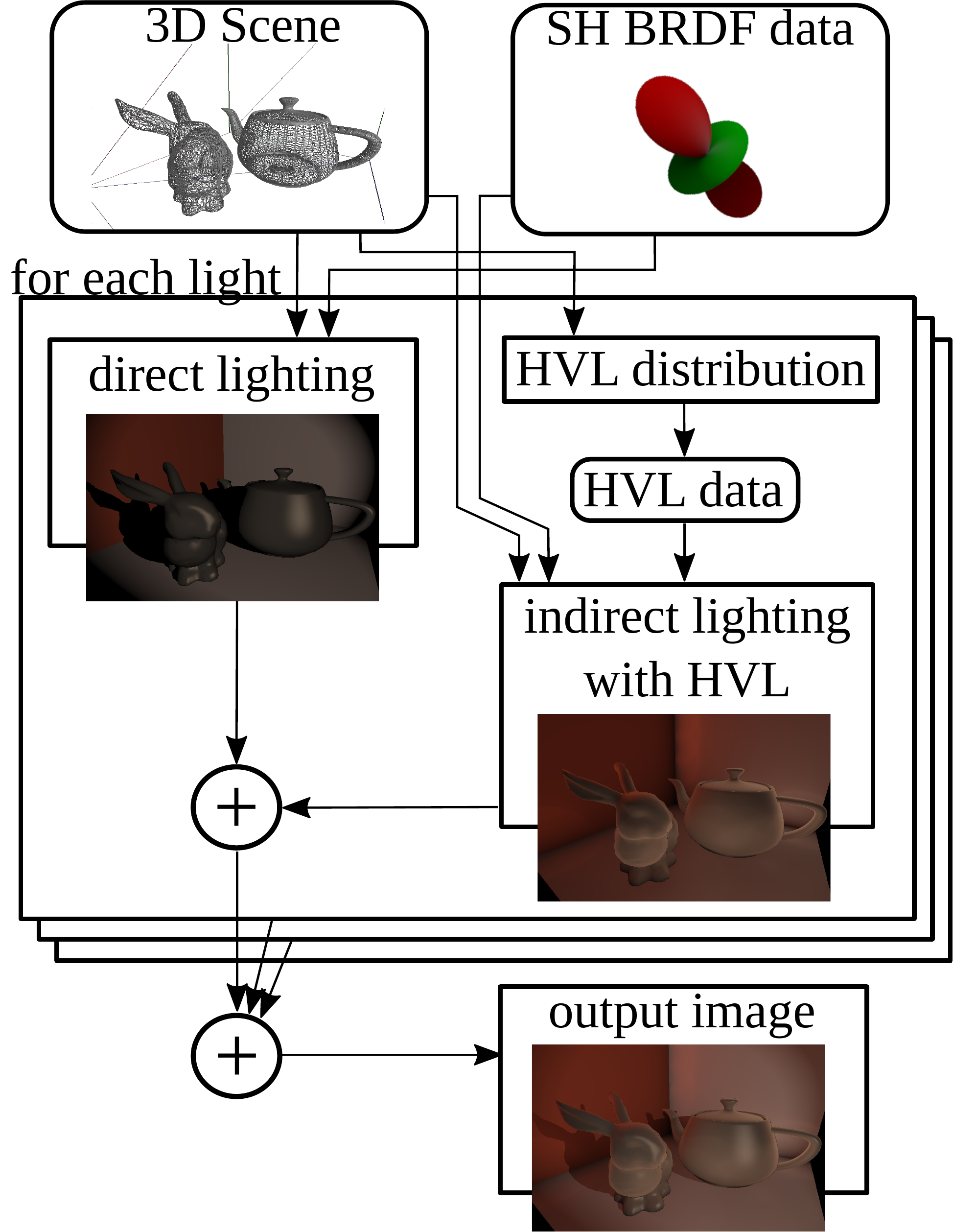}%
        \tightcaption{\label{fig:pipeline}%
        Overview of the HVLs rendering pipeline. The inputs are a 3D scene with dynamic objects and lights, along with precomputed BRDF spherical harmonics coefficients. For each primary light of the scene, direct lighting is computed using SHs framework. Indirect lighting starts with the distribution of HVLs from primary lights, followed by the gathering of HVLs contributions at each pixel. This pipeline allows to use parametric like measured BRDF at the same cost.}
\end{minipage}
\hfill
\begin{minipage}[c]{0.47\linewidth}
    \centering%
        \includegraphics[width=0.999\linewidth]{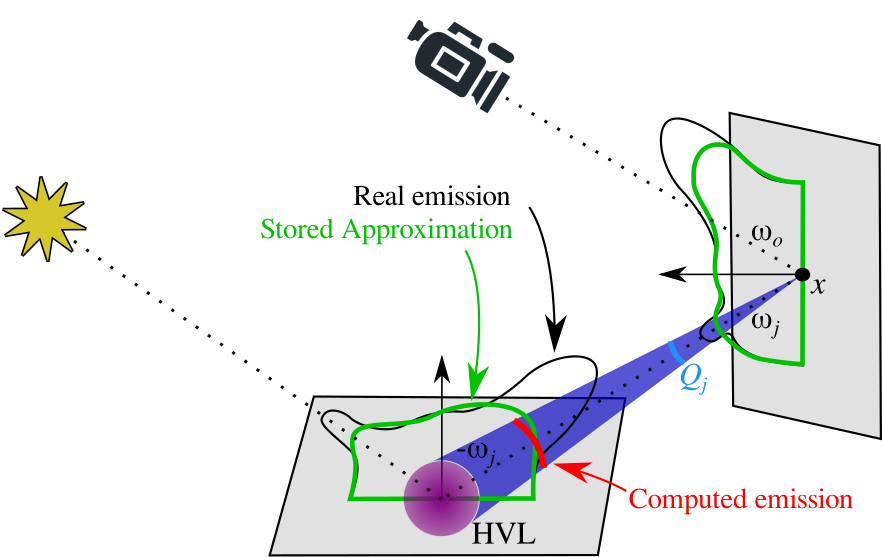}%
        \tightcaption{\label{fig:lightPathHVL}%
        Indirect lighting evaluation for one fragment and one HVL. The HVL, in purple, uses a low frequency approximation of the light emission corresponding to the first bounce of light emitted by the primary light.
        It's contribution on the fragment at $x$ is evaluated by projecting the HVL luminance onto SH and then convoluting with the BRDF at fragment's location. The HVL emission is computed as constant over its subtended solid angle.}

\end{minipage}
\end{figure}

This section focus on indirect lighting, indeed, in our approach, direct lighting can be computed with any method that supports parametric or measured material. 
The HVLs rendering pipeline, illustrated by \fig{\ref{fig:pipeline}} is built over an unified representation of materials : the SH projection of the BRDF.
This representation is pre-computed only once for each material.
From there, our two passes indirect lighting pipeline is similar to previous many-lights approaches. 
At each frame, the first pass distributes HVLs according to primary light sources; it stores the SH coefficients of the reflected luminance field of each HVL in the HVLs cache.
The second pass gathers the contributions of the HVLs by computing the convolution between light emitted by HVLs and the material at fragment location. 
The light emitted by HVL is the result of the reconstruction of the HVL reflected luminance field in the direction of the fragment.
Thus, HVLs are anisotropic spherical lights whose emission intensity depends on the direction of the receiving fragment ($\omega_{i}$).
\Fig{\ref{fig:lightPathHVL}} shows a typical one-bounce light path evaluated by our technique.
The final image is then computed as the sum of direct and indirect contributions from each primary light source. 

We formulate the indirect lighting $\radiance_o(\point, \dir_o)$, leaving the fragment $x$ in the direction $\omega_o$ as
\begin{equation}
\label{eq:sumIndirect}
\radiance_o(\point, \dir_o) = \sum_{s \in S} \Big( \sum_{j \in H_s} L_{j,o}(\point, \omega_o, \omega_j) \Big)
\end{equation}
where $S$ is the set of primary sources, $H_s$ the set of HVLs generated from $s$, $\dir_j$ the direction from $x$ to the $j$-th HVL and $L_{j,o}(\point, \dir_o, \dir_j)$ its luminance contribution. According to the rendering equation~\cite{Kajiya1986}, this contribution is expressed as
\begin{equation}
    L_{j,o}(\point, \dir_o, \dir_j) = \int_{\Omega} L_j(\point, \dir_i)\,\reflectance(\point, \dir_i, \dir_o) \diff{\omega_i}
    \label{eq:exactIndirect}
\end{equation}
where $\reflectance=\cos(\theta_i)\brdf(\point, \omega_i, \omega_o)$ is the reflectance, $\brdf$ being the BRDF, and $L_j(\point,\omega_i)$ is the incoming luminance at $\point$ emitted by the HVL $j$.
As this luminance is only perceived in the solid angle subtended by the HVL, the integral domain is reduced to this solid angle $Q_j$.
We finally approximate this equation by considering that the HVL emission is constant over the solid angle.
\begin{equation} 
    L_{j,o}(\point, \dir_o, \dir_j) \approx L_j(\point, \dir_j) \int_{Q_j} \reflectance(\point, \dir_i, \dir_o) \diff{\dir_i}
    \label{eq:approxIndirect}
\end{equation}
This approximation allows us to propose a computationally efficient projection of spherical lights contributions on SHs \sectp{\ref{sec:projRadAndMat}} that we use to evaluate the lighting for each HVL \sectp{\ref{sec:HVLcontribution}}.

Practical spherical harmonics evaluation being band limited, we rely on the number of bands as a control parameter of the trade-off between accuracy and speed.
To reach interactive performance, we typically use 0-5 bands to evaluate HVLs emission and 0-10 bands for the SH convolution on fragments. 
This allows to represent low frequency luminance field for HVLs and medium frequency BRDF for the fragment.
Our technique scales well toward high band limit and is able to render thousands of HVLs on 0-20 bands in a few seconds.

\section{Lighting with Spherical Harmonics}
\label{sec:SHframework}

The computational challenge in \equ{\ref{eq:exactIndirect}} and \ref{eq:approxIndirect} is the evaluation of the integral term, which is a motivation for using the SHs framework.
Considering \equ{\ref{eq:exactIndirect}}, when computing $\radiance_{j,o}(\point, \dir_o)$ and fixing both $\point$ and $\dir_o$, we evaluate the integral of a product of two spherical functions of $\dir_i$. Using spherical harmonics, this integral reduces to a dot product between vectors of coefficients dependent on $\point$ and $\dir_o$. For conciseness, we consider the dependence on $\point$ and $\dir_o$ implicit and omit it in the following remaining of the paper.

\subsection{Spherical Harmonics Review}
\label{sec:SHdecomp}

\begin{figure}[]%
\begin{minipage}[c]{0.47\linewidth}
    \centering%
        \includegraphics[width=0.90\linewidth]{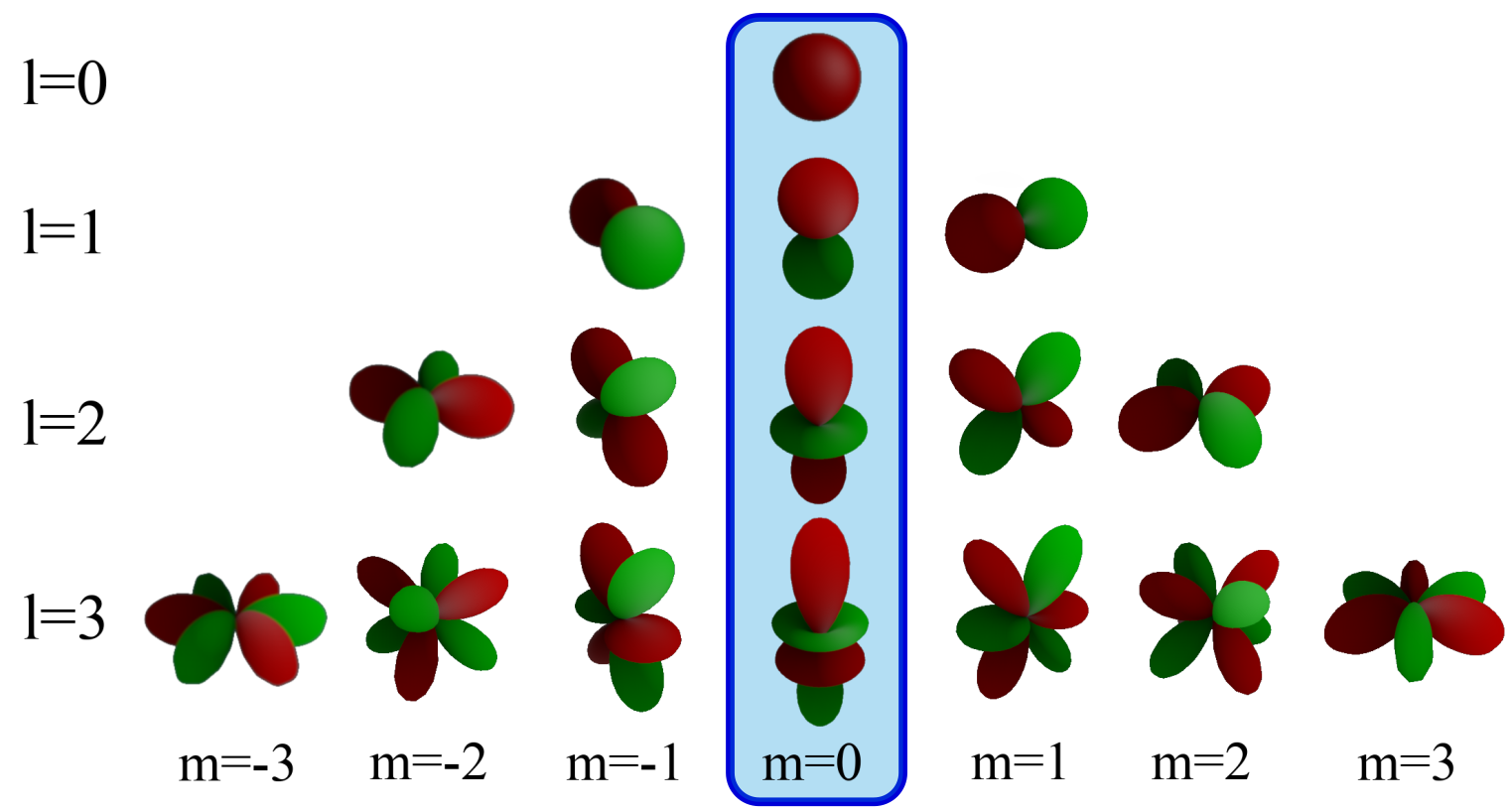}%
        \tightcaption{Representation of the first spherical harmonics base functions. Lobes are colored according to their sign, with positive values in red and negative in green. The sub-basis of zonal harmonics for which $m = 0$ is outlined in the middle.}
        \label{fig:HS}%
\end{minipage}
\hfill
\begin{minipage}[c]{0.47\linewidth}
    \centering%
        \includegraphics[width=0.90\linewidth]{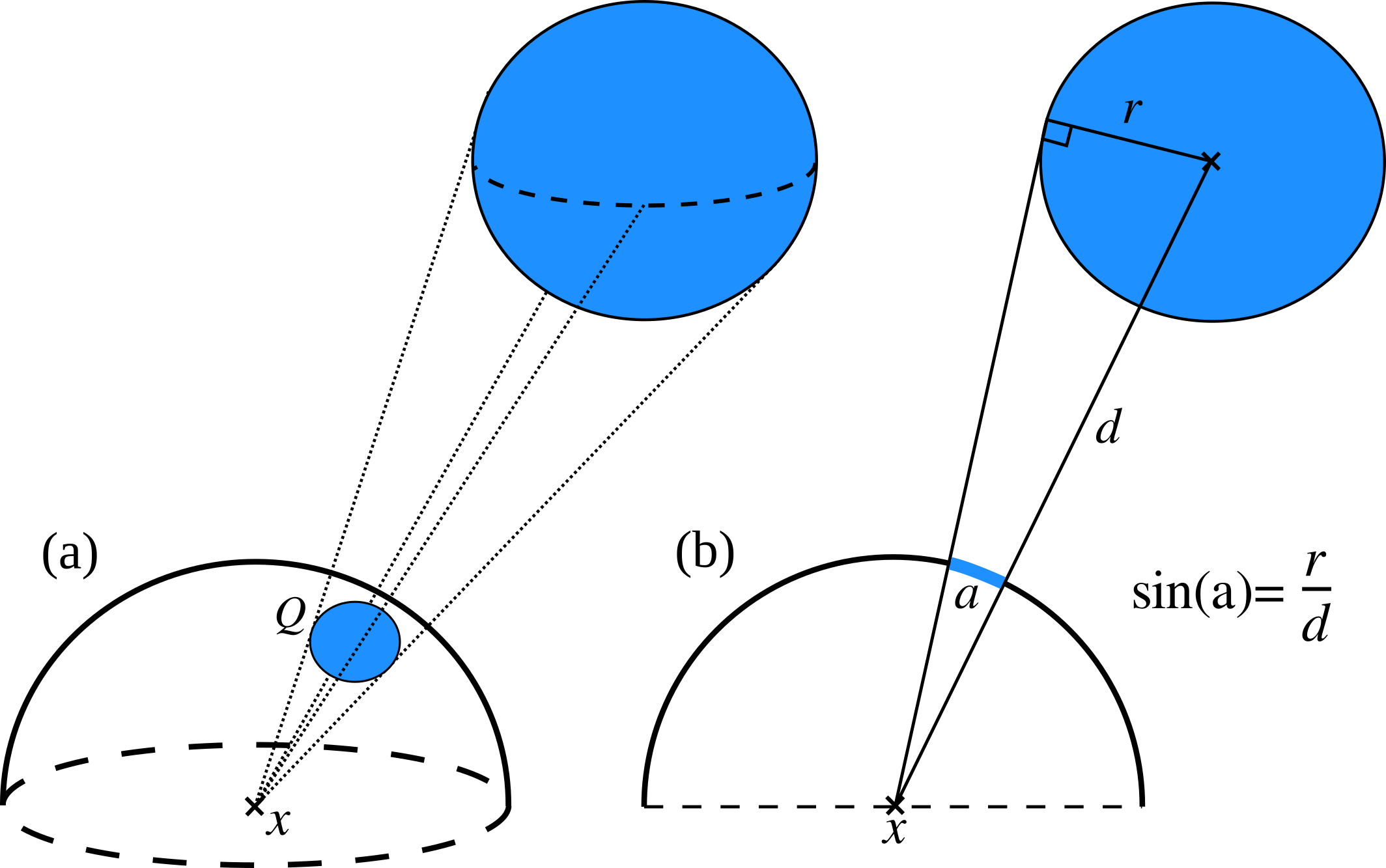}%
        \caption{Geometric setup of an HVL luminance contribution. (a)~The HVL in blue projects to a spherical cap $Q$ on the unit sphere centered at the shading point $\point$. (b)~2D cut showing the HVL radius $r$, the distance $d$ with the HVL, and the half-angle $a$ of the cap.}
        \label{fig:sphericalLight}%
\end{minipage}
\end{figure}

Let's rewrite \equ{\ref{eq:exactIndirect}} in term of spherical harmonics.
We note $\Radiance$ and $\Reflectance$ the projection of $\radiance$ and $\reflectance$ on the real spherical harmonics basis. 
The dot product $\Radiance \cdot \Reflectance$ is defined by:
\begin{tightequation}
\label{eq:LFdotproduct}
    \int_\sphere
        \radiance(\dir_i)\,\reflectance(\dir_i)
    \diff{\dir_i}\,=\,
    \Radiance \cdot \Reflectance
\end{tightequation}%
To obtain the coefficients $\Radiance_l^m$ (resp. $\Reflectance_l^m$) of the  vector $\Radiance$ (resp. $\Reflectance$), we project the original functions on
the SHs basis \{$Y_l^m(\omega)$\} which is indexed by order $\order$ and degree $\degree$.
As shown in \fig{\ref{fig:HS}}, successive orders $\order \geq 0$ correspond to increasing frequency bands, in which degrees span $-\order \leq \degree \leq \order$.
Projection to and reconstruction from SHs of the spherical function $\function(\dir)$ (\ie $\Radiance$ or $\Reflectance$) obey the following equations:
\begin{gather}
\label{eq:projectionSH}
    \Function_\order^\degree =
    \int_\sphere
        \function(\dir)\,\sh_\order^\degree(\dir)
    \diff{\dir}
\\
\label{eq:reconstructionSH}
    \function(\dir) =
    \sum_{\order = 0}^{+\infty}\,
        \sum_{\degree = -\order}^\order
            \Function_\order^\degree\,\sh_\order^\degree(\dir)
    \approx
    \sum_{\order = 0}^N\,
        \sum_{\degree = -\order}^\order
            \Function_\order^\degree\,\sh_\order^\degree(\dir)
\end{gather}

In practice, SHs coefficients are band limited to $l=N$  and the reconstruction becomes an approximation with fixed maximum frequency.
In this case, the integral of \equ{\ref{eq:exactIndirect}} reduces to a sum of $(\maxorder)^2$ terms, allowing for an efficient evaluation of the equation.

Using spherical coordinates, real SH $\sh_\order^\degree$ are expressed as
\begin{tightequation}
    \sh_\order^\degree(\theta, \phi) =
    \left\{
        \begin{array}{ll}
            \sqrt{2}\factor_\order^{|\degree|}\,
            \legendre_\order^{|\degree|}(\cos\theta)\,
            \sin(|\degree|\phi)
            & \degree < 0 \\
            \factor_\order^0\,
            \legendre_\order^0(\cos\theta)
            & \degree = 0 \\
            \sqrt{2}\factor_\order^\degree\,
            \legendre_\order^\degree(\cos\theta)\,
            \cos(\degree\phi)
            & \degree > 0
        \end{array}
    \right.
\end{tightequation}%
where $\legendre_\order^\degree$ denotes the associated Legendre polynomial, and $\factor_\order^\degree$ is a normalizing factor:
\begin{tightequation}
    \factor_\order^\degree =
    \sqrt{
        \frac{2\order + 1}{4\pi}
        \frac{(\order - \degree)!}{(\order + \degree)!}
    }
\end{tightequation}
\subsection{Computing Luminance and Material Coefficients}
\label{sec:projRadAndMat}

To compute the vector $\Radiance$ at a shaded point $\point$, we need to project the incoming luminance field $\radiance(\dir_i)$ on the SH basis. 
As stated \equ{\ref{eq:approxIndirect}} we consider the luminance emitted constant over the solid angle subtended by the light source.

\paragraph*{Simple light primitive SH projection}
In the case of simple primitives, such as point, spot, or directional lights, the luminance contribution of each source corresponds to a Dirac distribution.
Thus, the computation of the projection comes down to evaluate the SH in the Dirac direction and scaling by the light source's intensity.

\paragraph*{Efficient spherical light SH projection}
When dealing with spherical light sources, and our framework aims at handling thousands of them, the projection is less direct and more expensive.
We define the projection using Zonal Harmonics only as they provide efficient projection and reconstruction framework.

For the hemisphere around a shaded point, the solid angle $Q$ subtended by the light source corresponds to a spherical cap \figp{\ref{fig:sphericalLight}}.
Considering the incident luminance as constant, and equal to $1$, over $Q$ and $0$ elsewhere, the projection on SH of the incident luminance function simplifies to :
\begin{equation}
    \Radiance_\order^\degree = \int_\surfaceProj \sh_\order^\degree(\dir) \diff{\dir}
    \label{eq:initialEquation}
\end{equation}

\begin{figure}[]%
    \centering%
        \includegraphics[width=0.41\linewidth]{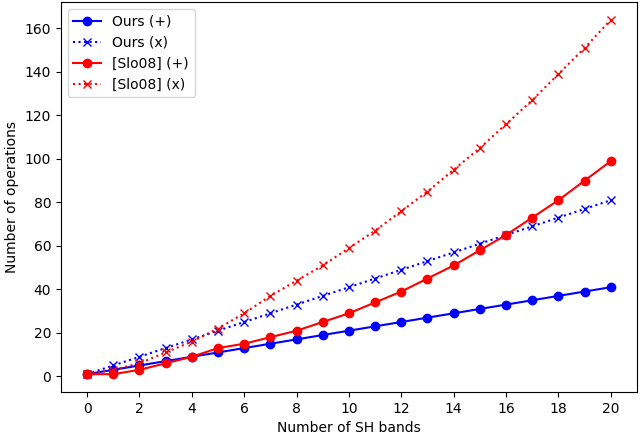}%
        \caption{Number of arithmetic operations needed to compute the vector of ZH coefficients up to a certain band. Our method \equp{\ref{eq:zonalHarmonicsLight}} is compared with the analytic Sloan's method~\cite{Sloan2008}. Number of operations calculated assumes that the recurrence loop is unrolled to avoid operations corresponding to constants for each ZH bands. (\eg$\sqrt{\pi/(2l+1)}$ counts as one multiplication for each bands.)}
        \label{fig:arithmeticOperations}%
\end{figure}

We define $\dirlight$ as the direction from the shaded point to the center of the spherical light.
Hence, $Q$ induces a radially symmetric luminance field around $\dirlight$.
We align the z-axis of the shading frame with $\dirlight$. Hence the computation of Zonal Harmonics boils down to the computation of $\RadianceZH_\order$ coefficients, since the other coefficients are null.
\begin{tightequation}
    \RadianceZH_\order = \factor_\order \int_\surfaceProj \legendre_\order(\dirlight \cdot \dir) \diff{\dir}
\end{tightequation}%
where $\factor_\order = \factor_\order^0$ and $\legendre_\order = \legendre_\order^0$.
Using spherical coordinates, this equation becomes
\begin{tightequation}
    \RadianceZH_\order = \factor_\order \int_{\theta=0}^{\angleProj} \int_{\phi=0}^{2\pi} \legendre_\order(\cos(\theta)) \sin(\theta) \diff{\phi} \diff{\theta}
\end{tightequation}%
where $a$ is the half-angle subtended by the spherical light source \figp{\ref{fig:sphericalLight}}.
Integrating on $\phi$, we obtain%
\begin{tightequation}
    \RadianceZH_\order = 2\pi \factor_\order \int_{\theta=0}^{\angleProj} \legendre_\order(\cos(\theta)) \sin(\theta) \diff{\theta}
    \label{eq:ZHcoeffsTheta}
\end{tightequation}%
As detailed in Appendix \ref{sec:legendreIntegral}, the value of this integral is
\begin{tightequation}
    \int_{\theta=0}^a \legendre_\order(\cos(\theta)) \sin(\theta) \diff{\theta} = \frac{-\legendre_{\order+1}(\alpha)+\legendre_{\order-1}(\alpha)}{2\order+1}
    \label{eq:integralLegendre}
\end{tightequation}%
where $\alpha=\cos(a)$. By re-injecting this into \equ{\ref{eq:ZHcoeffsTheta}}, the computation of the zonal harmonics coefficients for a spherical light in the coordinates frame aligned with the emission direction only requires evaluating Legendre polynomials and by simplifying, we obtain
\begin{equation}
    \label{eq:zonalHarmonicsLight}
\RadianceZH_\order = \left\{\begin{aligned}
    \sqrt{\frac{\pi}{2l+1}}(\legendre_{\order-1}(\alpha)-\legendre_{\order+1}(\alpha)) &\quad\mbox{ if } l \neq 0\\
    \sqrt{\pi}(1-\alpha) &\quad\mbox{ otherwise}
    \end{aligned}\right.
\end{equation}%
The SHs coefficients are then obtained by rotating the ZHs coefficients as proposed by Sloan~\etal~\cite{Sloan2005} :
\begin{tightequation}
\label{eq:SHcoeffSL}
    \Radiance_\order^\degree = \sqrt{\frac{4\pi}{2l+1}} \sh_\order^\degree(\dirlight) \RadianceZH_\order
\end{tightequation}%
As stated in \fig{\ref{fig:arithmeticOperations}}, beyond 5 SH bands, computing the ZH coefficients $\RadianceZH_\order$ using the closed form derived by Sloan~\cite{Sloan2008} exhibits more arithmetic operations than our general recurrence \equp{\ref{eq:zonalHarmonicsLight}}.
To evaluate the $\sh_\order^\degree$ basis function, we use the work of Sloan~\cite{Sloan2013SH}.

\paragraph*{BRDF coefficients}

To compute BRDF coefficients, we use the same principle as proposed by Kautz~\etal~\cite{Kautz2002}. 
A BRDF is defined by two directions $\omega_i$ and $\omega_o$, yielding a 4-dimensional function.
When we evaluate $\reflectance$, the direction $\dir_o$ is fixed; the BRDF becomes 2-dimensional and is projected on SHs.

In a pre-processing step, we tabulate SHs coefficients according to discrete samples of the direction $\dir_o = (\theta_o, \phi_o)$.
\begin{tightequation}
\label{eq:precomputeBRDF}
    \tab[\theta_o, \phi_o] = \Reflectance_{\dir_o}
\end{tightequation}%
where $\Reflectance_{\dir_o}$ is the SHs coefficient vector of $\Reflectance$ computed for the output direction $\dir_o$.
We uniformly sample $\theta$ and $\phi$ with one degree steps, and build a database of materials projected on spherical harmonics from the MERL database \cite{Matusik2003} and the database produced by Dupuy and Jakob\cite{Dupuy2018}.
With such uniform sampling, a material has $90 \times 360$ samples of $\Reflectance_{\dir_o}$ and accessing the coefficients requires a simple array lookup. The one degree step allows to capture subtle material variations. Other sampling and reconstruction strategies might be used but are out of scope of this paper.
Appendix~\ref{sec:materialsList} lists the materials we used for the pictures in this paper.

As an optimization compared to the approach of Kautz~\etal~\cite{Kautz2002}, we store isotropic material in a one-dimensional array. 
Isotropic materials depend on the $\Delta\phi=\abs{\phi_i-\phi_o}$ rather than the actual values of $\phi_i$ and $\phi_o$.
Hence by using the isotropic parametrization of $\reflectance$
\begin{align}
\reflectance_{iso}(\theta_o, \theta_i, \Delta\phi)
  &= \reflectance(\theta_o, 0, \theta_i, \Delta\phi)
\end{align}
we drop the explicit dependence on $\phi_o$ and only discretize $\theta_o$. This way, the storage of isotropic material drops down to $90$ samples.
\begin{align}
 \tab_{iso}[\theta_o] &= \Reflectance_{\dir_o} \mbox{ with }\dir_o=(\theta_o, 0)
\end{align}
%
%
$\Reflectance$ is precomputed in a certain space, to have $\Radiance$ in the same space, in order to apply the convolution, $\Radiance$ is simply rotated by applying the corresponding rotation on $\dirlight$ \equp{\ref{eq:SHcoeffSL}}. This avoids the use of calculatory methods for rotating SH, such as the $zxzxz$ rotation~\cite{Kautz2002}.
When projecting BRDF on SH basis using the above method, the weakening factor of the rendering equation $\cos(\theta_i)$ is accounted in the projected function.
As we need the BRDF without this factor for the HVL evaluation \sectp{\ref{sec:HVLcontribution}}, in the following of the paper, we note $\Reflectance$ to indicate the BRDF projection with the factor (i.e $\cos(\theta_i)\brdf(\point, \omega_i, \omega_o)$) and $\Reflectance^\prime$ without.

SH are limited in their handling of high frequencies and the band-limit is prone to ringing artefacts.
To mitigate this problem, we use a window to filter the SH coefficients of the BRDF~\cite{Sloan2008} during pre-processing step, before storing them.

\section{Harmonics Virtual Light (HVL)}
\label{sec:HVL}

We define a HVL as a spherical light whose emitted luminance is stored as SH coefficients so that its contribution is evaluated efficiently using the projection defined in \sect{\ref{sec:projRadAndMat}}.
As we use simple primitive for primary light sources, incident luminance field at HVL location is then a dirac and the storage of the reflected luminance at HVL reduces to a reference on the projected BRDF. This will be reconstructed and scaled to define the emission of HVL according to the outgoing direction.

\subsection{HVL Luminance Contribution}
\label{sec:HVLcontribution}

\begin{figure}[]%
    \centering%
        \includegraphics[width=0.4\linewidth]{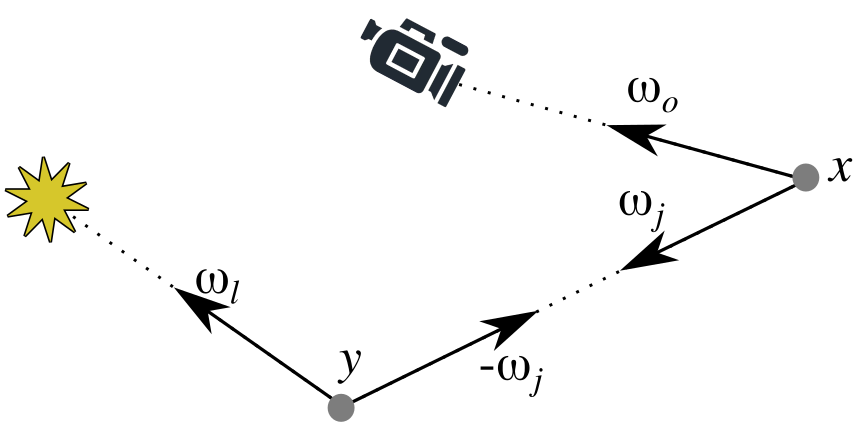}%
        \caption{\label{fig:notationLightPath} Notations for a light path, $y$ is an HVL and $x$ a shaded fragment.}%
\end{figure}
The luminance contribution of each HVL corresponds to the computation of \equ{\ref{eq:approxIndirect}}. Using our representation, this equation is expressed as
\begin{equation}
\radiance_{j,o}(x,\omega_o,\omega_j) \approx \radiance_j(x, \omega_j) \Radiance \cdot \Reflectance_{\omega_o,x}
\label{eq:contribHVL}
\end{equation}%
where $\Radiance$ corresponds to the coefficients vector of \equ{\ref{eq:SHcoeffSL}}. The incoming luminance at $\point$ emitted by the $j$-th HVL is defined as
\begin{equation}
    \radiance_j(x, \omega_j) = \Phi_j ( \Reflectance^{\prime}_{\omega_l,y} \cdot \sh(-\omega_j)) G_j(-\omega_j)
    \label{eq:hvlRadiance}
\end{equation}
where $y$ and $\omega_l$ are respectively defined as the position of the HVL and its direction to primary source \figp{\ref{fig:notationLightPath}}. $\Phi_j$ is the power of the HVL, corresponding to the power of the primary source divided by the number of HVL. The dot product corresponds to the reconstruction of the emission function towards the shaded fragment \equp{\ref{eq:reconstructionSH}}. Note that if the HVL is considered as isotropic, only the band 0 is needed for the reconstruction, resulting in no SH evaluation at all. The geometric factor $G_j(-\omega_j)$ correct the energy as we use spherical sources.
\begin{equation}
    G_j(-\omega_j) = \frac{1}{\pi r_j^2} \max(0, \langle n_j | -\omega_j \rangle) H(n_x,\omega_j)
    \label{eq:geometricFactor}
\end{equation}%
where $n_j$ is the normal of the $j$-th HVL, $r_j$ its radius, $\pi r_j^2$ is an approximation of the surface area, the scalar product corrects the energy according to the surface orientation.
The factor H represents the proportion of the HVL that lies in the hemisphere oriented by $n_x$, the normal at point $x$, and is defined by 
\begin{equation}
    H(n_x,\omega_j) = S\Big(\frac{\Bar{a}-\min(\max(\cos^{-1}(n_x\cdot\omega_j),\underline{a}),\Bar{a})}{2a}\Big) 
\end{equation}
where $a$ is the half-angle subtended by the spherical light source, $\Bar{a}=a+\frac{\pi}{2}$, $\underline{a}=a-\frac{\pi}{2}$ and $S(x) = 3x^2-2x^3$ is the Smoothstep function. 
To obtain this equation we adapt the general equation that computes the approximation of the intersection of two spherical caps \cite{Oat2007} to our case, where one spherical cap is the hemisphere.
In our real-time experimentation, we use this approximation of the intersection, for an offline implementation, H could be computed with the exact equation given in \cite{Oat2007}.

When evaluating the HVL emission \equp{\ref{eq:hvlRadiance}}, we use the BRDF projection without the $\cos(\theta_i)$ that is computed with the geometric factor \equp{\ref{eq:geometricFactor}}. This ensures that, when using low-frequency reconstruction (\ie a small number of SH bands), the cosine is not over-smoothed and then, the energy is preserved.

The visibility term, although not present in the equations for the sake of clarity, is approximated by the visibility of the center of the HVL, as the VSL approach does.

\begin{figure}[b]%
    \centering%
    \hspace{3cm}
    \begin{subfigure}[t]{0.24\linewidth}%
        \includegraphics[width=\linewidth]{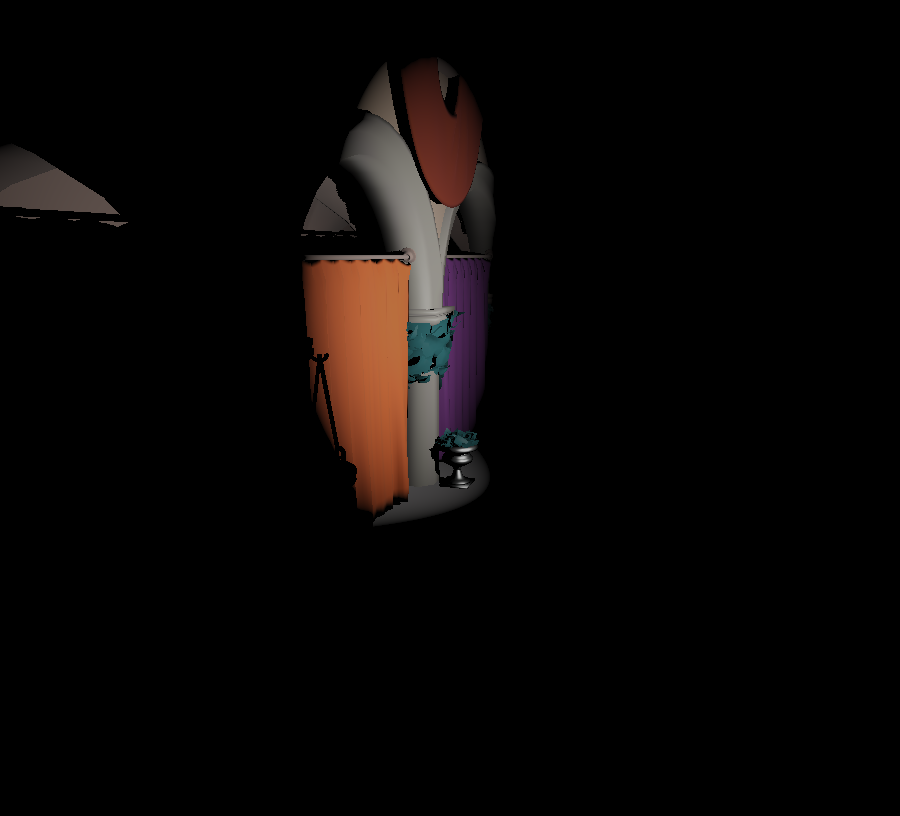}%
        \tightsubcaption{Direct lighting only}%
        \label{fig:directExample}%
    \end{subfigure}%
    \begin{subfigure}[t]{0.24\linewidth}%
        \includegraphics[width=\linewidth]{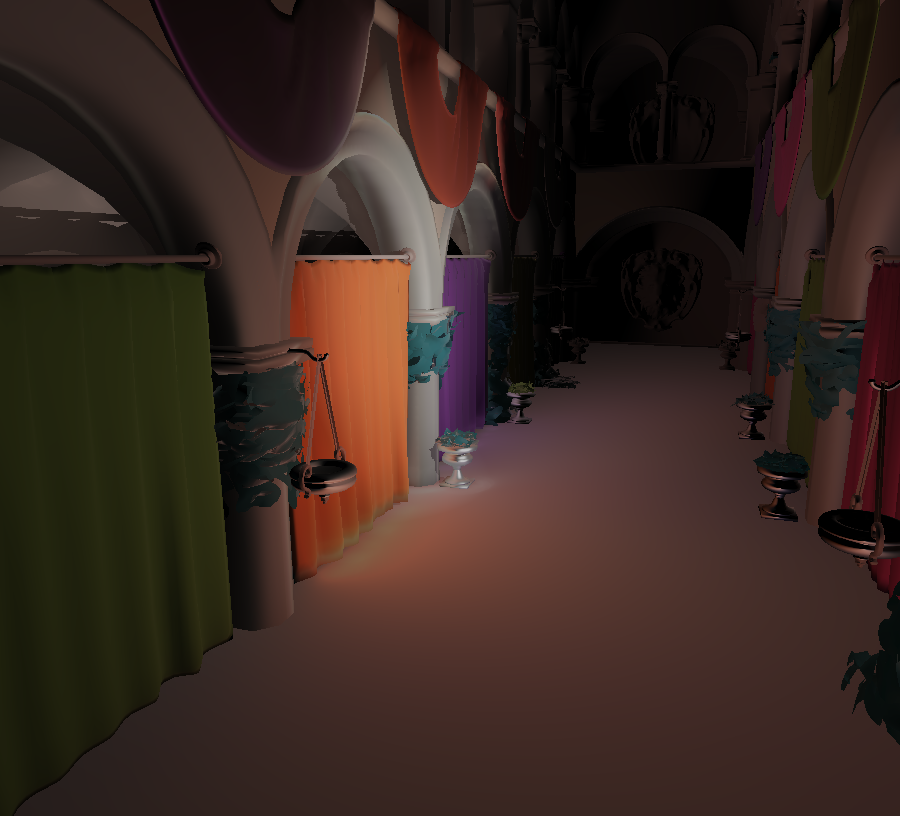}%
        \tightsubcaption{Global illumination}%
        \label{fig:directIndirectExample}%
    \end{subfigure}%
    \hspace{3cm}
    \tightcaption{Global illumination using our HVLs on the Sponza scene. (a)~Shows direct lighting only, rendered from realistic materials projected on spherical harmonics. (b)~Global illumination is obtained by distributing HVLs from the primary light source to the surface of the scene, and gathering their contribution.}%
    \label{fig:GIexamples}%
\end{figure}

\subsection{Distribution and Caching}
\label{sec:distribution}

For one HVL, we need to store its emission function, its position and the normal on the underlying surface.
Our implementation uses the principle of Reflective Shadow Maps (RSM) to generate HVLs \cite{Dachsbacher2005}. 
Our RSM store, in 2D GPU textures, world space coordinates and normal, as seen from the light source as well as the luminous flux and the reflectance function at each RSM pixel. 
The pre-computed material SH coefficients are tabulated in a collection of 2-dimensional textures, where each texture corresponds to a material for easy indexing.
To fully define the emission function of a HVL from its reflectance, we replace the diffuse and specular components of the standard RSM by a single material index, reducing the memory requirements for our RSM.

\paragraph*{Memory requirements}

The memory requirements of RSM being negligible on current hardware, we focus the discussion on the storage cost for materials.
For an isotropic material, $90$ directions are sampled; using 10 harmonic bands and 32-bit float RGB coefficients, one material represents 108kB.
Using 200 different materials for instance, the required memory is 21MB. On the other hand, in the presence of anisotropic materials, the storage for one material is  360 times larger and raise to 388MB for one material.

\subsection{Radius Heuristics}
\label{sec:HVLradius}

The radius of an HVL is an important parameter that has an impact on the final image quality.
Intuitively, if HVLs do not sufficiently cover the scene, the artefacts related to VPLs will appear. 
But if their radius is too large, details will be lost because the HVLs contributions will overlap, and the final rendering will miss high frequencies of the shading.
To prevent this, we have to adapt HVLs radius to the underlying geometry so that there is the right amount of overlapping.

\begin{figure}[]%
    \centering%
    \hspace{3.3cm}
    \begin{subfigure}[t]{0.1\linewidth}%
        \includegraphics[width=\linewidth]{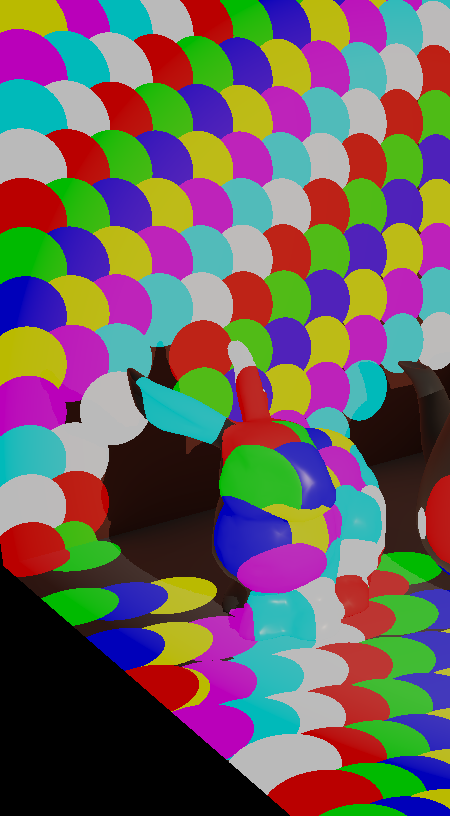}%
    \end{subfigure}%
    \hfill%
    \begin{subfigure}[t]{0.1\linewidth}%
        \includegraphics[width=\linewidth]{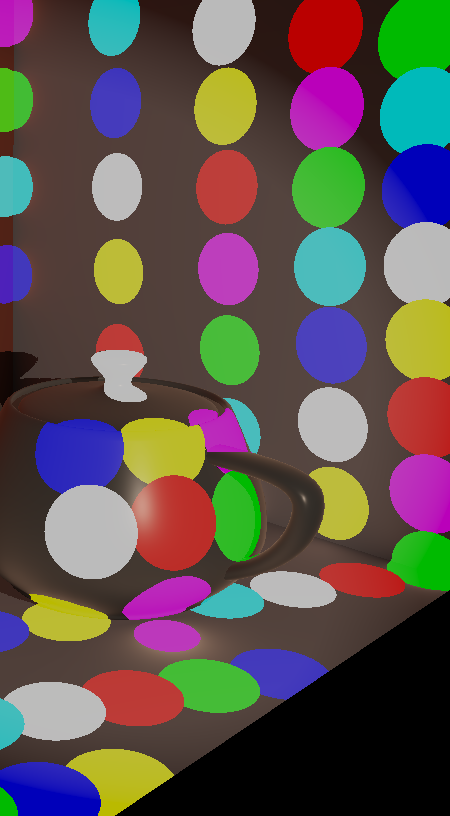}%
    \end{subfigure}%
    \hfill%
    \begin{subfigure}[t]{0.2\linewidth}%
        \includegraphics[width=\linewidth]{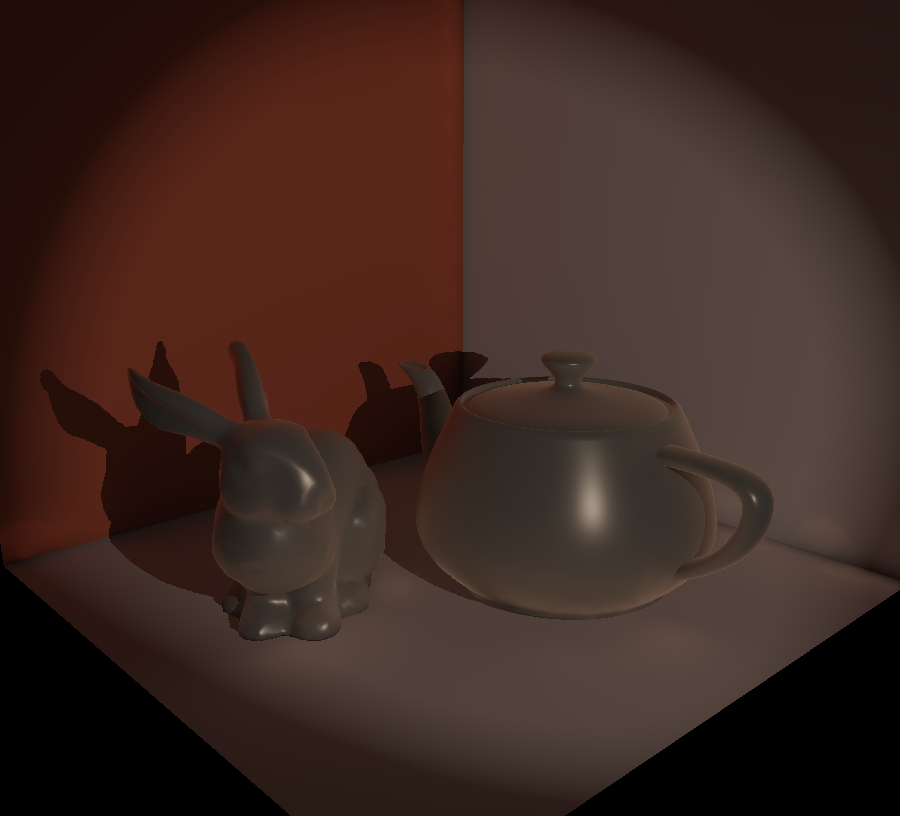}%
    \end{subfigure}%
    \hfill%
    \begin{subfigure}[t]{0.2\linewidth}%
        \includegraphics[width=\linewidth]{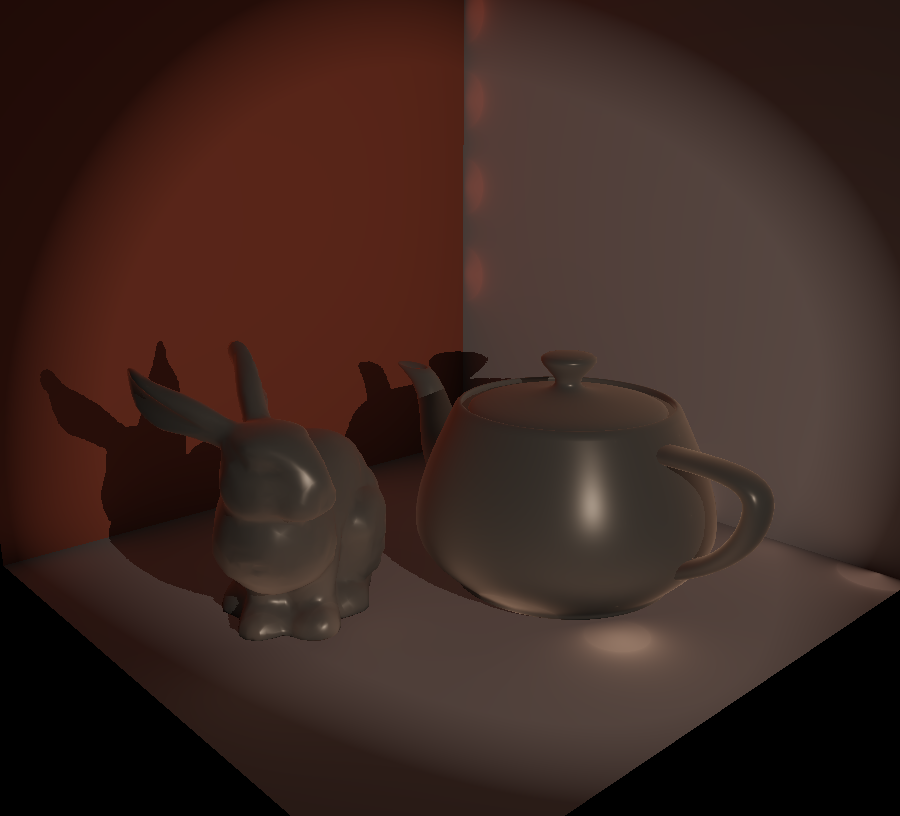}%
    \end{subfigure}
    \hspace*{3.3cm}
    \\%
    \hspace{3.3cm}
    \vspace{0.02cm}
    \begin{subfigure}[t]{0.1\linewidth}%
        \includegraphics[width=\linewidth]{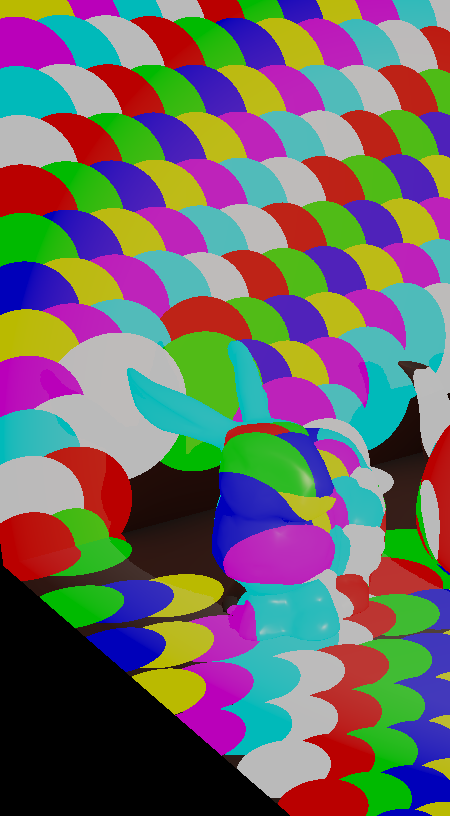}%
    \end{subfigure}%
    \hfill%
    \begin{subfigure}[t]{0.1\linewidth}%
        \includegraphics[width=\linewidth]{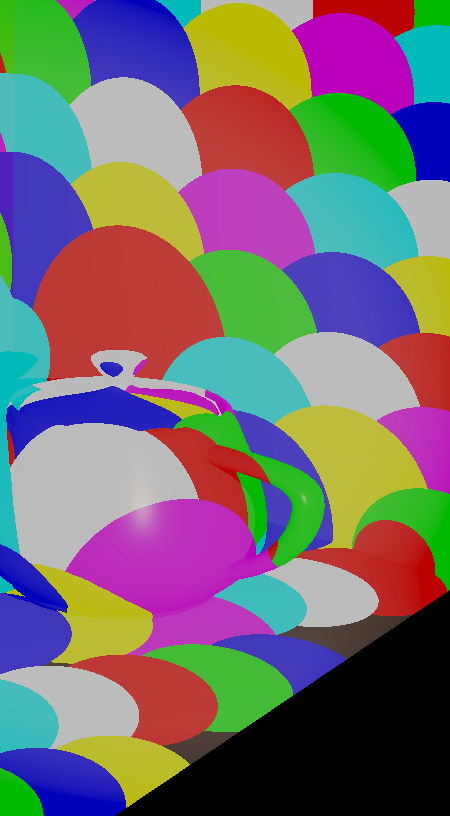}%
    \end{subfigure}%
    \hfill%
    \begin{subfigure}[t]{0.2\linewidth}%
        \includegraphics[width=\linewidth]{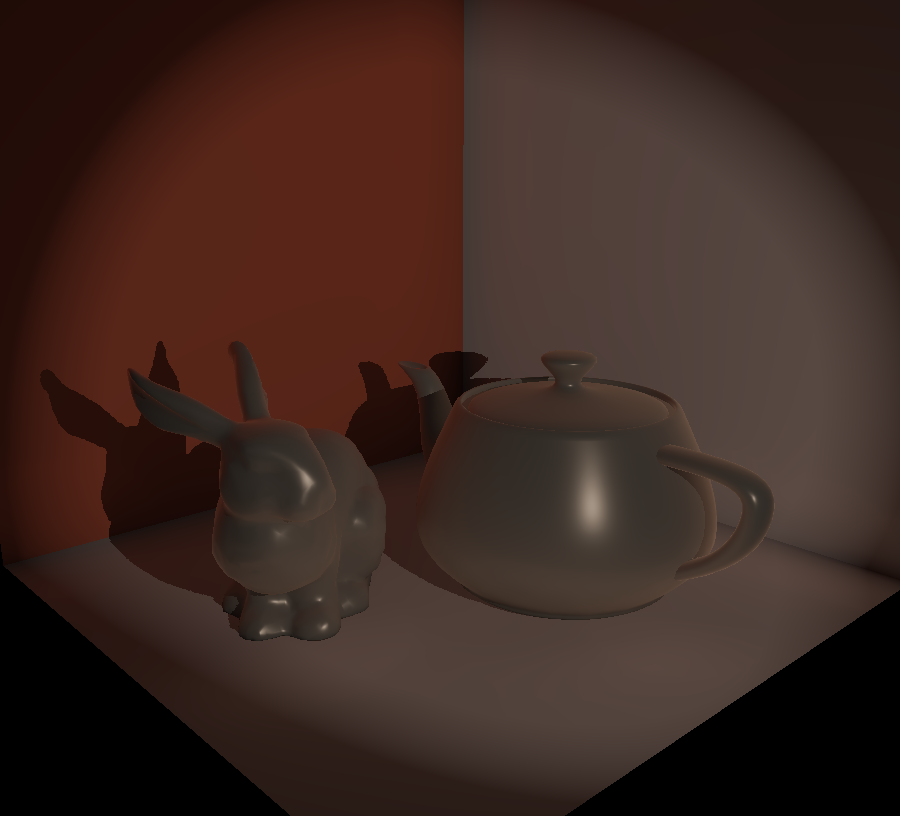}%
    \end{subfigure}%
    \hfill%
    \begin{subfigure}[t]{0.2\linewidth}%
        \includegraphics[width=\linewidth]{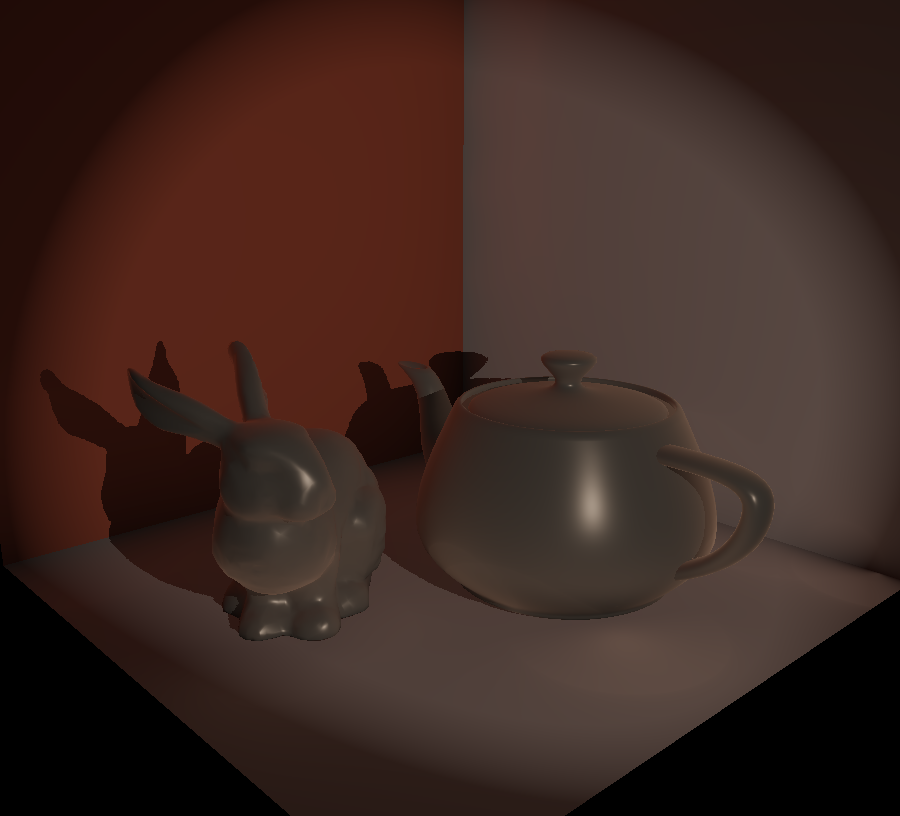}%
    \end{subfigure}
    \hspace*{3.3cm}
    \\%
    \hspace{3.3cm}
    \vspace{0.02cm}
    \centering%
    \begin{subfigure}[t]{0.1\linewidth}%
        \includegraphics[width=\linewidth]{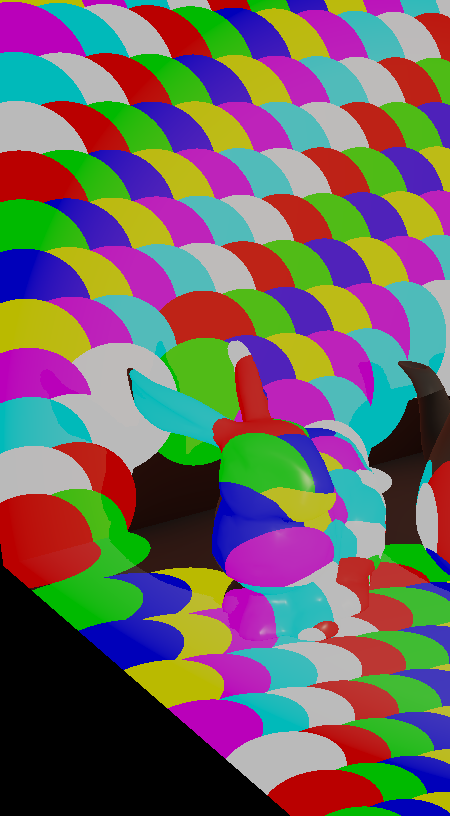}%
    \end{subfigure}%
    \hfill%
    \begin{subfigure}[t]{0.1\linewidth}%
        \includegraphics[width=\linewidth]{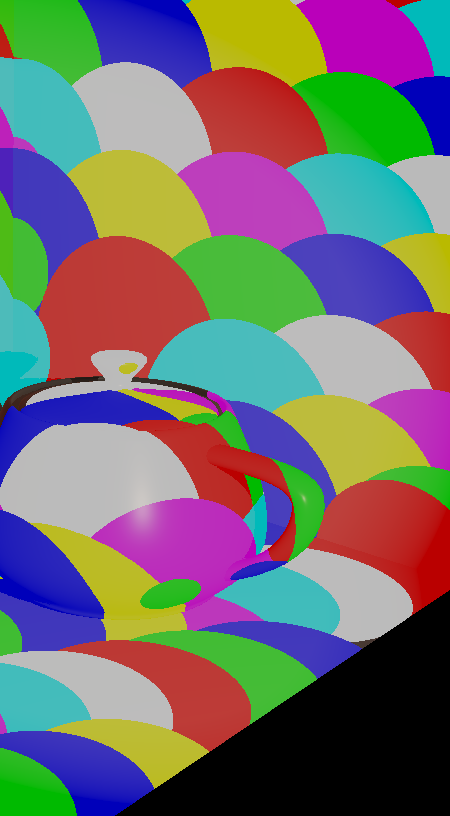}%
    \end{subfigure}%
    \hfill%
    \begin{subfigure}[t]{0.2\linewidth}%
        \includegraphics[width=\linewidth]{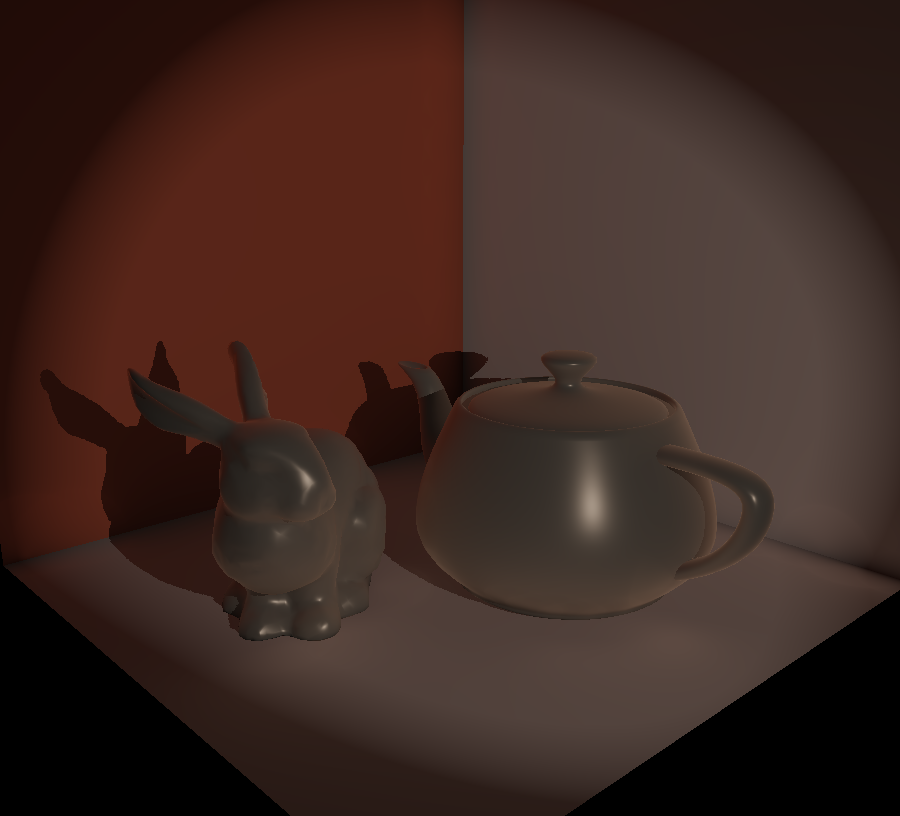}%
    \end{subfigure}%
    \hfill%
    \begin{subfigure}[t]{0.2\linewidth}%
        \includegraphics[width=\linewidth]{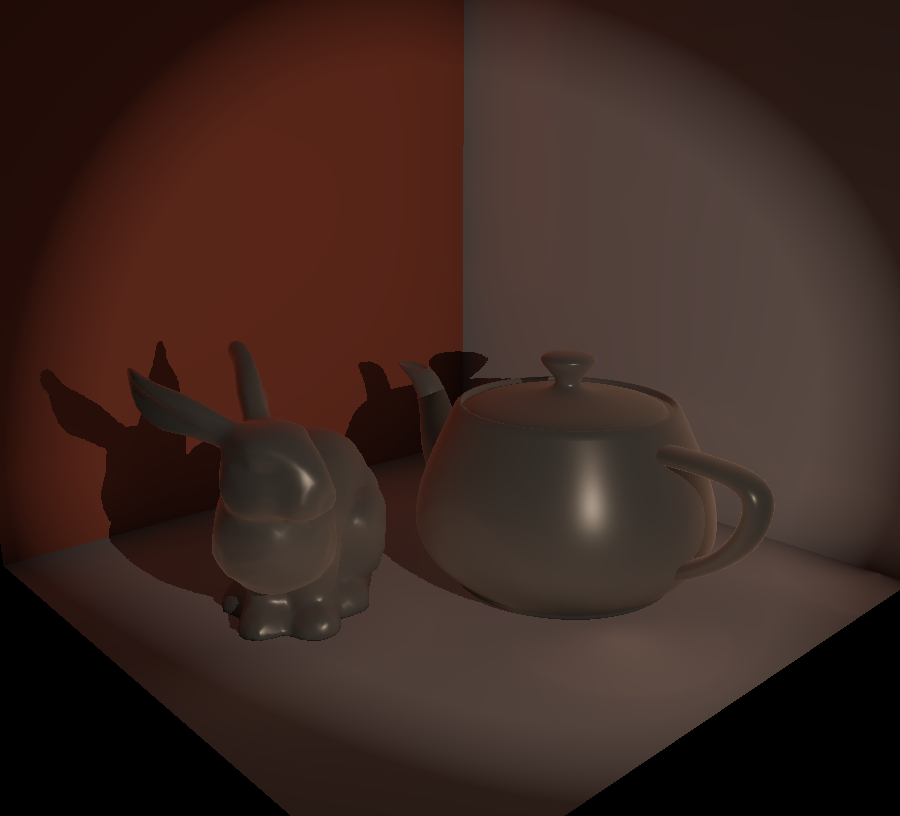}%
    \end{subfigure}
    \hspace*{3.3cm}
    \\%
    \tightcaption{
    HVLs coverage. Each colored circle corresponds to one HVL. From left to right: HVLs coverage with 400 and 100 HVLs; resulting image with 400 HVLs and then 100 HVLs. From top to bottom: fixed radius of 0.25, heuristic 1, heuristic 2 respectively.}%
    
    \label{fig:radiusHVL}%
\end{figure}

In a similar way than VSL, that are also spherical sources and exhibit the same requirements than HVL, we propose to choose the HVL radius proportionally to the local density of secondary light sources \cite{Hasan2009}. 
Instead of searching for the 10 closest neighbors, as proposed for VSL, we expose two dynamic heuristics relying on the simplicity of sampling the RSM to obtain the HVL locations.
Our first heuristic estimates the radius of an HVL from an approximation of the HVLs local density :
\begin{tightequation}
    r_1 = \frac{\sum_{n} w_n \norm{p-p_n} }{\sum_{n} w_n} k,\quad w_n = \frac{1}{\abs{d-d_n}+\epsilon}
\end{tightequation}%
where $r_1$ is the radius of the HVL, $d$ its depth in light space, $p$ denotes the position in world space coordinates and $\sum_n$ sums over the 8-connected neighbors of the HVL.
As proposed for VSL, the resulting radius is multiplied by a user-specified constant $k$ to ensure better coverage on complex scenes.
This heuristic averages the distance of a HVL to its neighbors so that it overlap them and thus avoid spike artefacts. 
We weight the contribution of neighbors by the depth difference between HVLs in light space to avoid too large overlapping because of the scene geometries, such as the hole formed by the rabbit's ears in \fig{\ref{fig:radiusHVL}}. 
$\epsilon$ is added to avoid a division by $0$, should two HVLs depths be the same.

This first heuristic is based on the neighborhood of the HVL, making it expensive to compute. We propose a second heuristic, more suitable for real-time scenarios, independant of the HVL neighborhood by considering the FOV angle $\lambda$ used to generate a square RSM instead. First, we approximate the angle $\gamma$ formed by two diagonally adjacent HVLs by considering that the two HVLs are at the same depth. 
\begin{tightequation}
    \gamma = \sqrt{2}\frac{\lambda}{\sqrt{M}}
\end{tightequation}%
where $\vplNumber$ is the total number of HVLs and $\sqrt{2}$ is the length of the diagonal. 
Starting from this, we approximate the distance between those HVLs, and hence the radius $r_2$ of an HVL, by considering a right triangle between the primary source and the two HVLs, where the right angle is at the HVL whose radius is sought.
\begin{tightequation}
    r_2= d \tan(\gamma) k 
    \label{eq:Heuristic2}
\end{tightequation}%
where $d$ is the depth of the HVL and $k$ is the same user-specified constant as for $r_1$. To avoid any trigonometric function evaluation, as $\gamma$ is close from $0$, we approximate the $\tan$ function by its Taylor series $\tan(x)\approx x+{x^3}/{3}$. The heuristic then becomes
\begin{tightequation}
    r_2 \approx  d \Big(\gamma + \frac{\gamma^3}{3} \Big) k
    \label{eq:approxHeuristic2}
\end{tightequation}%
This approximation of \equ{\ref{eq:Heuristic2}} is robust. Indeed, in order to have an error of less than $0.01$ for the $\tan$ evaluation from its Taylor series, calculus hints that we must have $\gamma \lesssim 0.58$ and thus ${\lambda}/{\sqrt{M}} \lesssim 0.2$. 
So, for a spot light of half-angle $45^{\circ}$, $\sqrt{M}\gtrsim 3.92$, means that we need at least $16$ HVLs for precise evaluation of the $\tan$ function. For a spot light of half-angle $90^{\circ}$, we need at least $64$ HVLs.
These requirements are well below the number of HVLs required in practice (\ie at least a few hundred).
As shown \figp{\ref{fig:radiusHVL}}, these heuristics allow to have HVLs that overlap well, and thus covers the directly illuminated scene geometry, without having one HVL that swallow another one, which would reduce the impact of the number of HVLs.
However, the balance remains fragile and artefacts may still appear, especially when the depth distribution of HVLs is heterogeneous because holes may appear in the coverage. It is the role of the user specified parameter $k$ to prevent the appearance of artefacts.

\section{Results}
\label{sec:results}

We implemented our proposal using OpenGL / GLSL 4.5 on an Intel Xeon 2.10 GHz processor and a RTX 2080 graphics card. We did not take into account the HVLs visibility in our renderings.

\subsection{Experimental results}
\label{subsec:quality}

\begin{figure}[t]%
    \centering%
    \hspace{3.3cm}
    \begin{subfigure}[t]{0.2\linewidth}%
        \includegraphics[width=\linewidth]{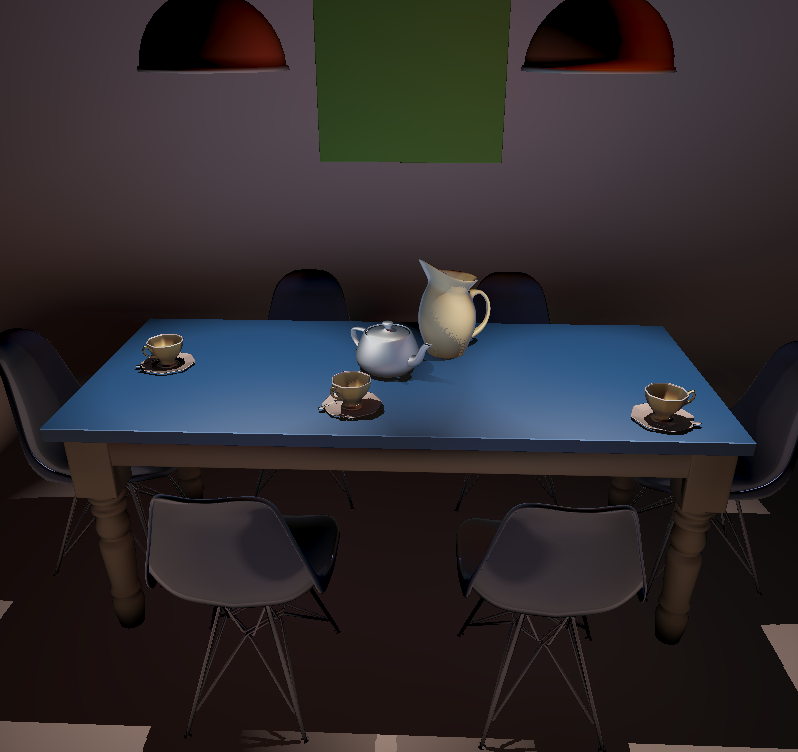}%
    \end{subfigure}%
    \hfill%
    \begin{subfigure}[t]{0.2\linewidth}%
        \includegraphics[width=\linewidth]{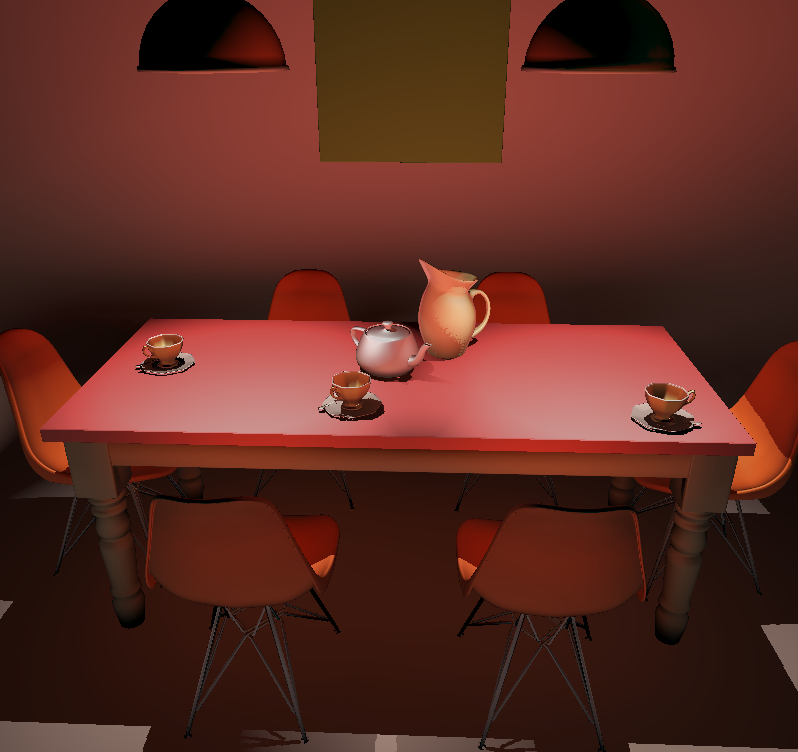}%
    \end{subfigure}%
    \hfill%
    \begin{subfigure}[t]{0.2\linewidth}%
        \includegraphics[width=\linewidth]{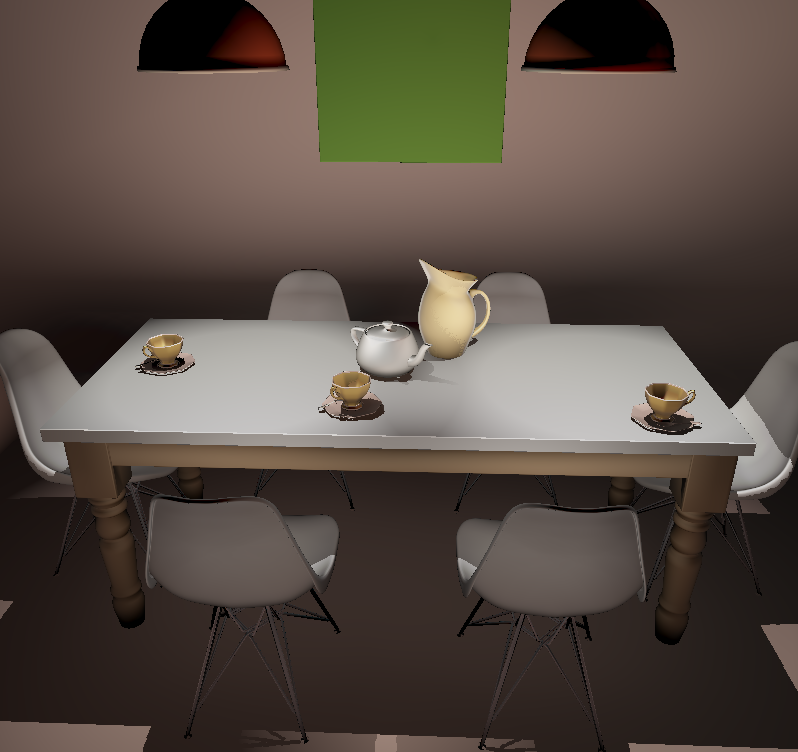}%
    \end{subfigure}%
    \hspace{3.3cm}
    \tightcaption{\label{fig:differentMaterial}%
Example of different realistic materials on the chairs and table. In our framework, materials may change dynamically without affecting render times.}%
\end{figure}

\Fig{\ref{fig:differentMaterial}} illustrates that HVLs allow any material to be used, and the effect they have on the render.
Our method, through the use of SH, applies a low-pass filter on the lighting, thus providing a smooth and artefact-free results \figp{\ref{fig:comparisonHVLVPLVSL}}.

As each HVL is accounted independently of the others, increasing their number has a linear impact on the computational cost. Increasing the SH band limit has a quadratic cost as the number of SH coefficient is equal to the square of the bands number. 
Indeed, we can observe these trends in Table~\ref{table:IndirectLight}.
Thus, the computation time for HVLs is predictable, but is greatly impacted by the number of integrated SH bands. Furthermore, if low frequency BRDF are used, increasing the band limit will have no impact on the final picture quality. Using an adaptive band limit might keep the computational cost minimal but will also suffer from code divergence in the shader evaluation of the SH projection.

\subsection{Comparison with VPL, VSL and path-traced reference}

\begin{figure*}[]%
    \vspace*{-0.25cm}
    \centering%
    \hspace{0.5cm}
    \begin{subfigure}[b]{0.465\linewidth}%
        \begin{tikzpicture}
          \begin{scope}[spy using outlines=
              {rectangle, magnification=3.9, width=2.3cm, height=1.8cm, connect spies, rounded corners}]
                \node[anchor=south west,inner sep=0] (img) {\includegraphics[width=\textwidth]{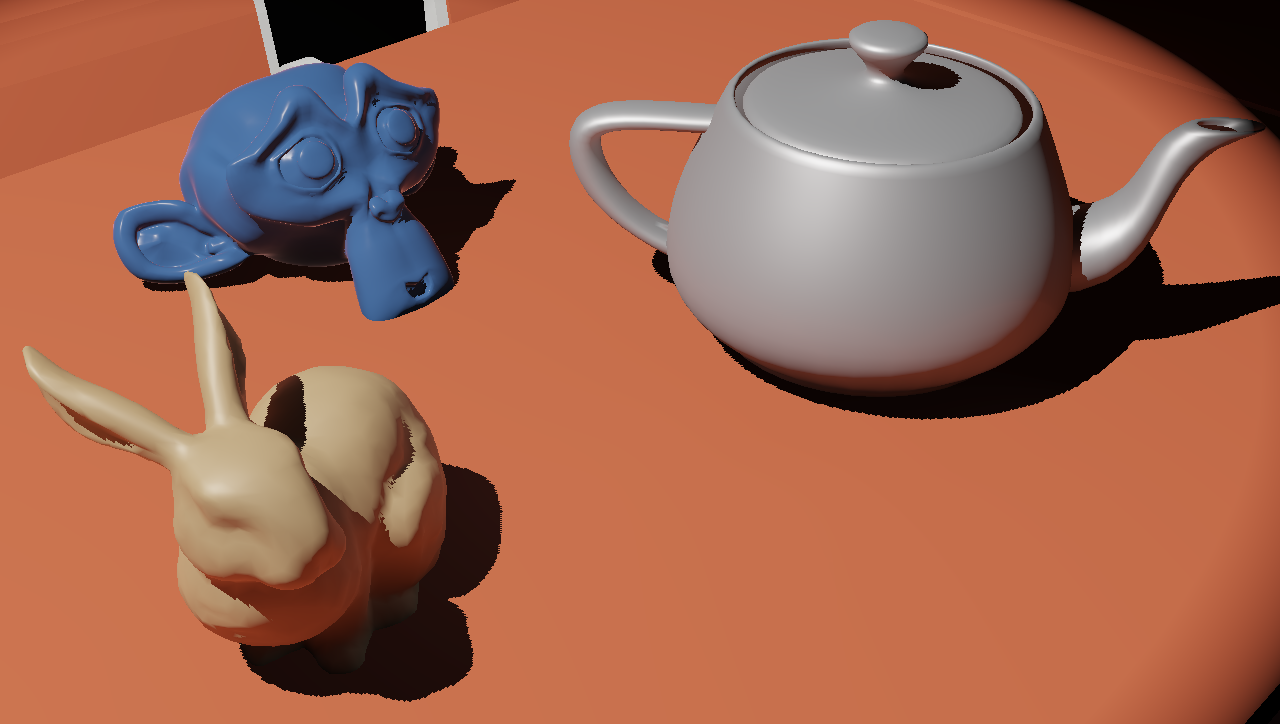}};
                \node [below right] at (img.152) [scale=1.2] {\hspace{0.01cm}\textcolor{white}{(a) HVL}};
                \node [below left] at (img.30) [scale=1.2] {\hspace{0.01cm}\textcolor{white}{\contour{black}{\shortstack{rmse 0.0398, psnr 28.010, ssim 0.9805}}}};
                \node [above right] at (img.208) [scale=1.2] {\textcolor{white}{61ms}};
                \spy [red] on (2,3.35) in node (a) [right] at (3.7,1.05);
                \spy [blue] on (6.45,2.45) in node [right] at (6.2,1.05);
                \spy [green ,magnification=3.1, width=1.1cm, height=1.1
                cm] on (2.38,0.5) in node [right] at (0.05,1.2);
                \coordinate (DG) at (img.west);
                \coordinate (DD) at (img.east);
          \end{scope}
          \node[circle,draw=orange,ultra thick,inner sep=0,minimum size=5pt,
      scale=5,xshift=-0.12cm,yshift=-0.063cm] (circle) at (a.center) {};
        \end{tikzpicture}
        \phantomcaption{}
        \label{subfig:HVLVPLVSLa}
    \end{subfigure}
    \hspace*{0.5cm}
    \\%
    \vspace{-0.37cm}
    \hspace*{0.5cm}
    \begin{subfigure}[b]{0.465\linewidth}%
        \begin{tikzpicture}
          \begin{scope}[spy using outlines=
              {rectangle, magnification=3.9, width=2.3cm, height=1.8cm, connect spies, rounded corners}]
                \node[anchor=south west,inner sep=0] (img) {\includegraphics[width=\textwidth]{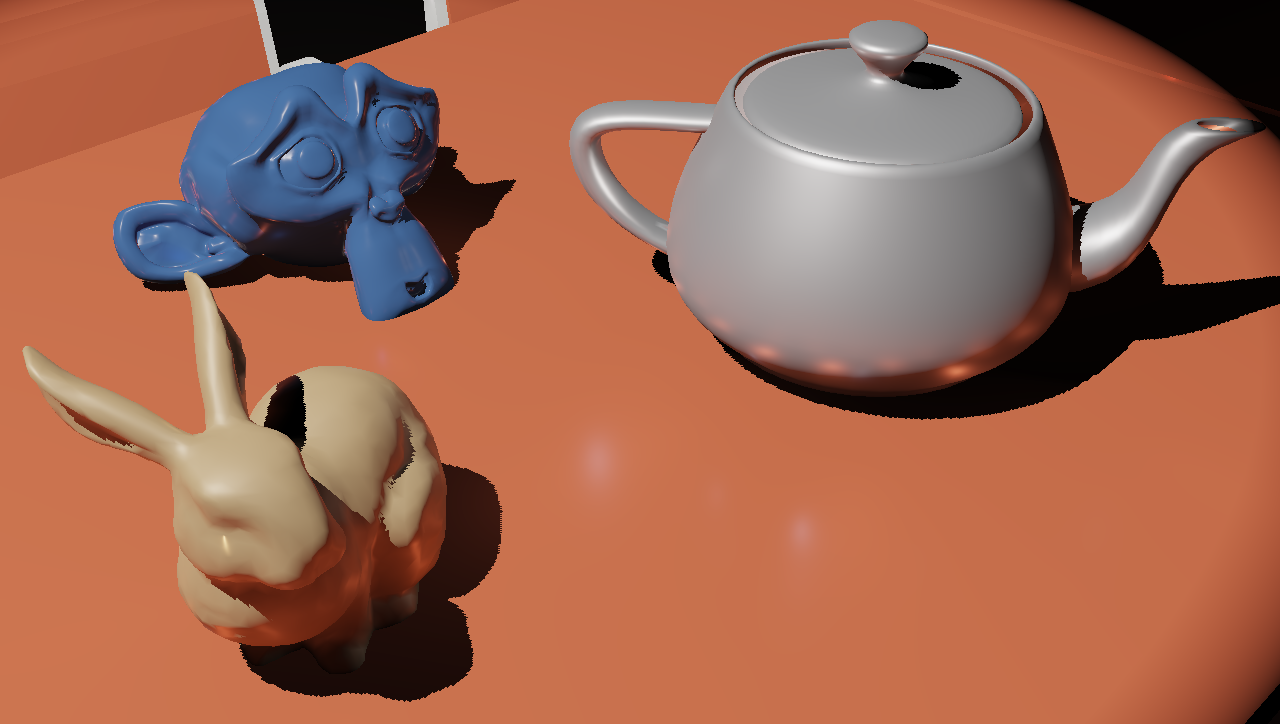}};
                \node [below right] at (img.152) [scale=1.2] {\hspace{0.01cm}\textcolor{white}{(b) VPL}};
                \node [below left] at (img.30) [scale=1.2] {\hspace{0.01cm}\textcolor{white}{\contour{black}{\shortstack{rmse 0.0549, psnr 25.202, ssim 0.9772}}}};
                \node [above right] at (img.208) [scale=1.2] {\textcolor{white}{13ms}};
                \spy [red] on (2,3.35) in node (a) [right] at (3.7,1.05);
                \spy [blue] on (6.45,2.45) in node [right] at (6.2,1.05);
                \spy [green ,magnification=3.1, width=1.1cm, height=1.1
                cm] on (2.38,0.5) in node [right] at (0.05,1.2);
                \coordinate (FG) at (img.135);
          \end{scope}
          \node[circle,draw=orange,ultra thick,inner sep=0,minimum size=5pt,
      scale=5,xshift=-0.12cm,yshift=-0.063cm] (circle) at (a.center) {};
        \end{tikzpicture}
        \phantomcaption{}
        \label{subfig:HVLVPLVSLb}
    \end{subfigure}%
    \hfill%
    \begin{subfigure}[b]{0.465\linewidth}%
        \begin{tikzpicture}
          \begin{scope}[spy using outlines=
              {rectangle, magnification=3.9, width=2.3cm, height=1.8cm, connect spies, rounded corners}]
                \node[anchor=south west,inner sep=0] (img) {\includegraphics[width=\textwidth]{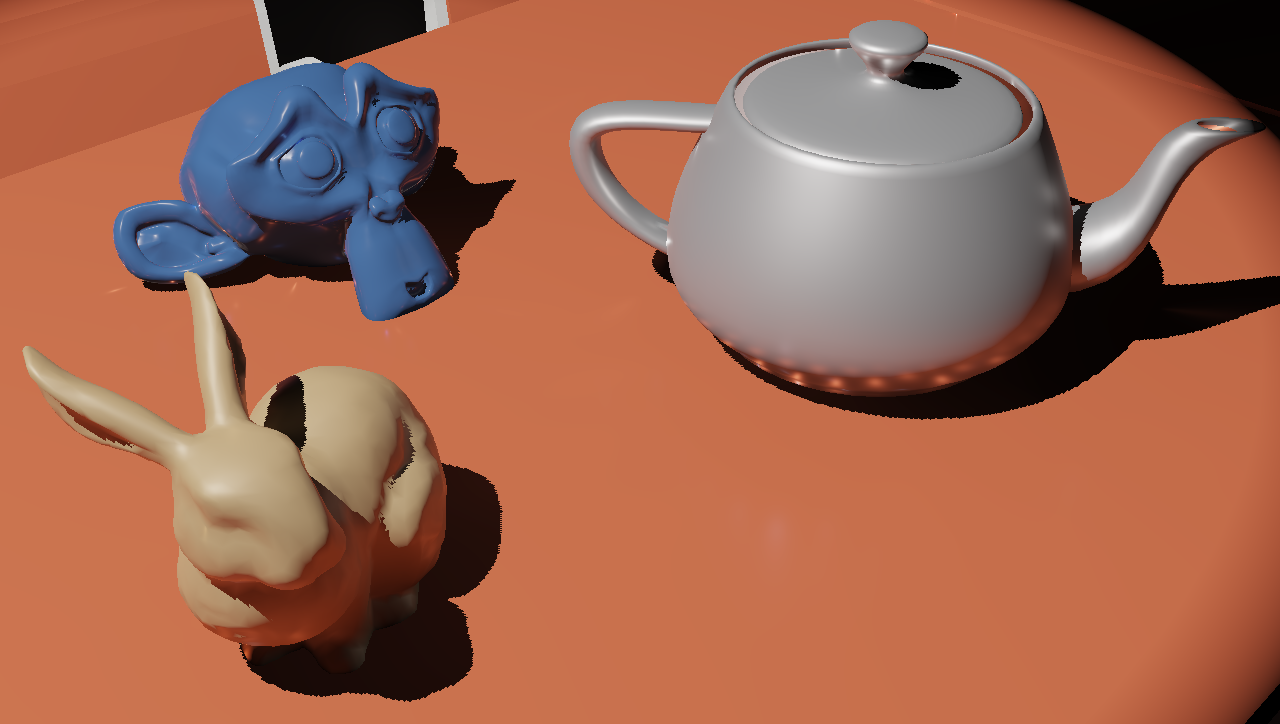}};
                \node [below right] at (img.152) [scale=1.2] {\hspace{0.01cm}\textcolor{white}{(c) VPL}};
                \node [below left] at (img.30) [scale=1.2] {\hspace{0.01cm}\textcolor{white}{\contour{black}{\shortstack{rmse 0.0880, psnr 21.107, ssim 0.9811}}}};
                \node [above right] at (img.208) [scale=1.2] {\textcolor{white}{61ms}};
                \spy [red] on (2,3.35) in node (a) [right] at (3.7,1.05);
                \spy [blue] on (6.45,2.45) in node [right] at (6.2,1.05);
                \spy [green ,magnification=3.1, width=1.1cm, height=1.1
                cm] on (2.38,0.5) in node [right] at (0.05,1.2);
                \coordinate (FD) at (img.45);
          \end{scope}
          \node[circle,draw=orange,ultra thick,inner sep=0,minimum size=5pt,
      scale=5,xshift=-0.12cm,yshift=-0.063cm] (circle) at (a.center) {};
        \end{tikzpicture}
        \phantomcaption{}
        \label{subfig:HVLVPLVSLc}
    \end{subfigure} 
    \hspace*{0.5cm}
    \\%
    \vspace{-0.37cm}
    \hspace*{0.5cm}
    \begin{subfigure}[b]{0.465\linewidth}%
        \begin{tikzpicture}
          \begin{scope}[spy using outlines=
              {rectangle, magnification=3.9, width=2.3cm, height=1.8cm, connect spies, rounded corners}]
                \node[anchor=south west,inner sep=0] (img) {\includegraphics[width=\textwidth]{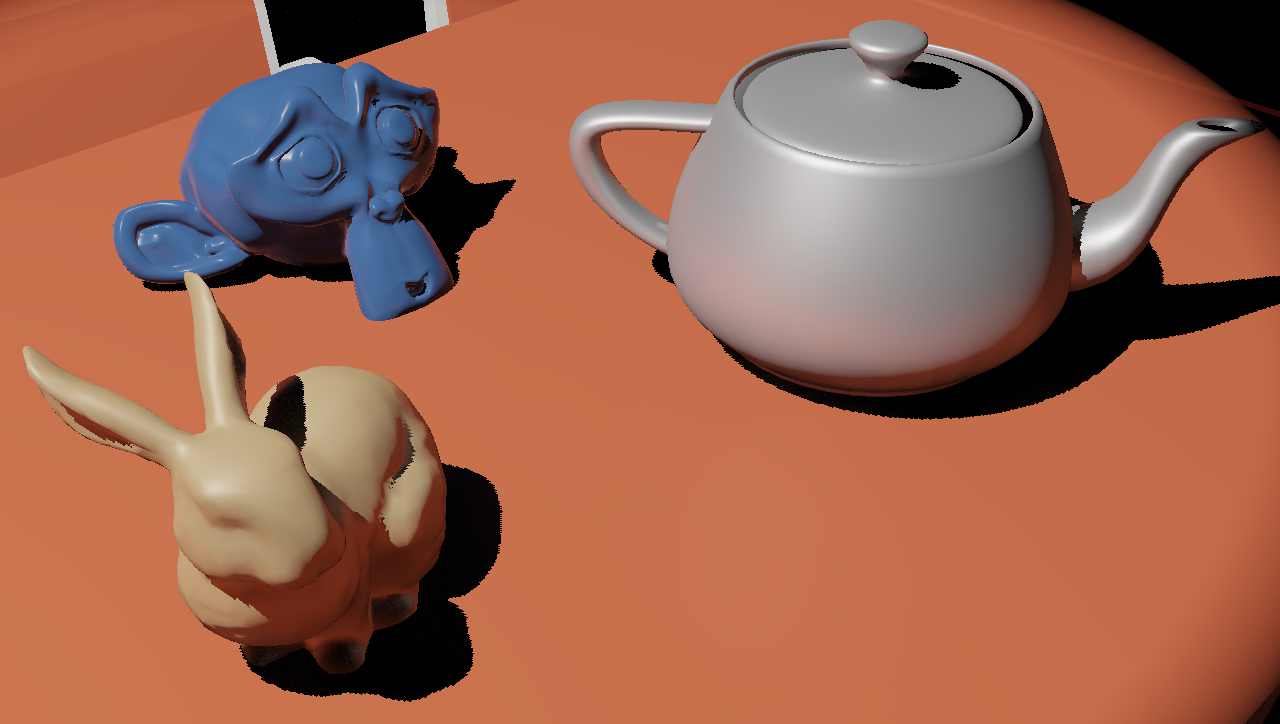}};
                \node [below right] at (img.152) [scale=1.2] {\hspace{0.01cm}\textcolor{white}{(d) VSL}};
                \node [below left] at (img.30) [scale=1.2] {\hspace{0.01cm}\textcolor{white}{\contour{black}{\shortstack{rmse 0.0434, psnr 27.243, ssim 0.9745}}}};
                \node [above right] at (img.208) [scale=1.2] {\textcolor{white}{254ms}};
                \spy [red] on (2,3.35) in node (a) [right] at (3.7,1.05);
                \spy [blue] on (6.45,2.45) in node [right] at (6.2,1.05);
                \spy [green ,magnification=3.1, width=1.1cm, height=1.1
                cm] on (2.38,0.5) in node [right] at (0.05,1.2);
          \end{scope}
          \node[circle,draw=orange,ultra thick,inner sep=0,minimum size=5pt,
      scale=5,xshift=-0.12cm,yshift=-0.063cm] (circle) at (a.center) {};
        \end{tikzpicture}
        \phantomcaption{}
        \label{subfig:HVLVPLVSLd}
    \end{subfigure}%
    \hfill%
    \begin{subfigure}[b]{0.465\linewidth}%
        \begin{tikzpicture}
          \begin{scope}[spy using outlines=
              {rectangle, magnification=3.9, width=2.3cm, height=1.8cm, connect spies, rounded corners}]
                \node[anchor=south west,inner sep=0] (img) {\includegraphics[width=\textwidth]{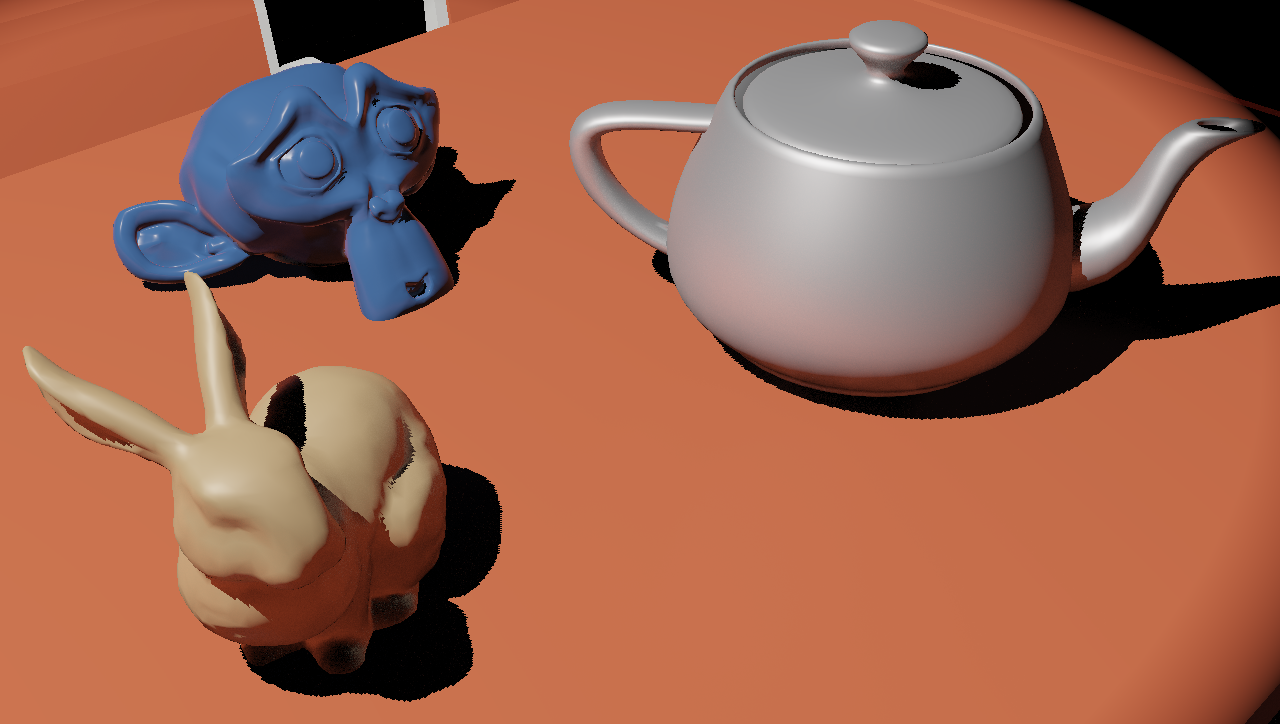}};
                \node [below right] at (img.152) [scale=1.2] {\hspace{0.01cm}\textcolor{white}{(e) VSL }};
                \node [below left] at (img.30) [scale=1.2] {\hspace{0.01cm}\textcolor{white}{\contour{black}{\shortstack{rmse 0.0433, psnr 27.260, ssim 0.9720}}}};
                \node [above right] at (img.208) [scale=1.2] {\textcolor{white}{63ms}};
                \spy [red] on (2,3.35) in node (a) [right] at (3.7,1.05);
                \spy [blue] on (6.45,2.45) in node [right] at (6.2,1.05);
                \spy [green ,magnification=3.1, width=1.1cm, height=1.1
                cm] on (2.38,0.5) in node [right] at (0.05,1.2);
          \end{scope}
          \node[circle,draw=orange,ultra thick,inner sep=0,minimum size=5pt,
      scale=5,xshift=-0.12cm,yshift=-0.063cm] (circle) at (a.center) {};
        \end{tikzpicture}
        \phantomcaption{}
        \label{subfig:HVLVPLVSLe}
    \end{subfigure} 
    \hspace*{0.5cm}
    \\%
    \vspace{-0.37cm}
    \hspace*{0.5cm}
    \centering%
    \begin{subfigure}[b]{0.465\linewidth}%
        \begin{tikzpicture}
          \begin{scope}[spy using outlines=
              {rectangle, magnification=3.9, width=2.3cm, height=1.8cm, connect spies, rounded corners}]
                \node[anchor=south west,inner sep=0] (img) {\includegraphics[width=\textwidth]{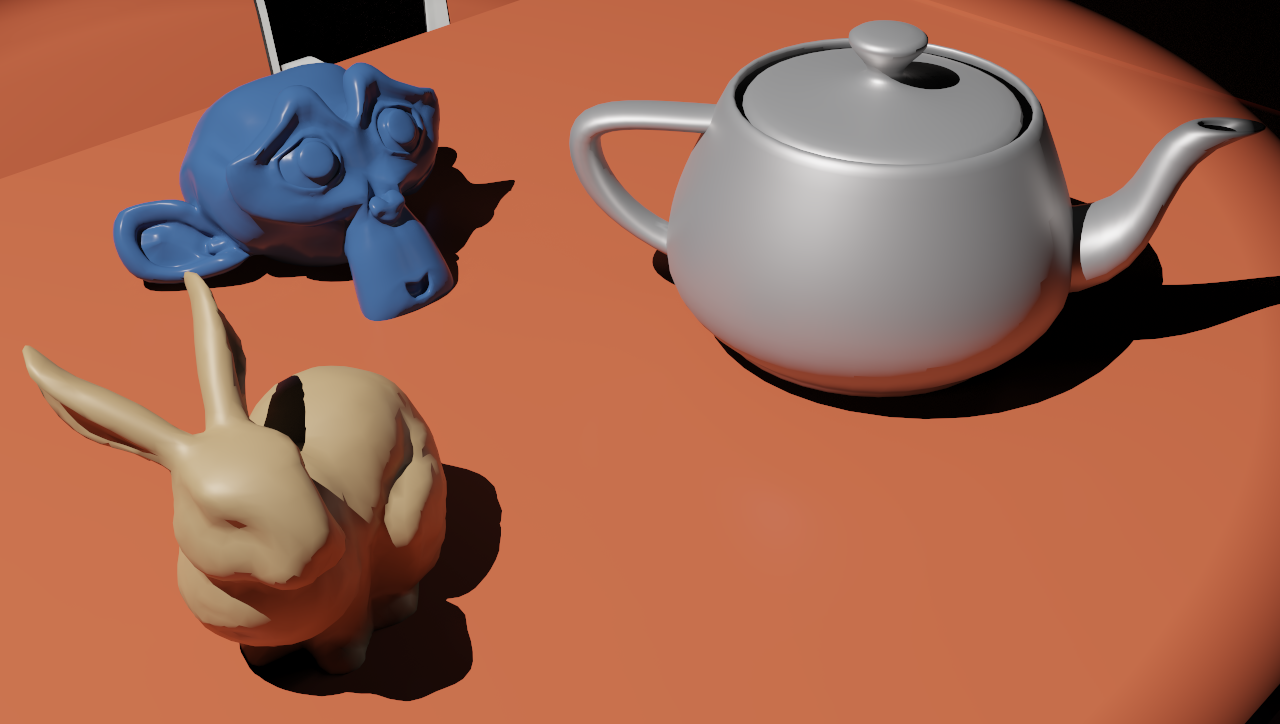}};
                \node [below right] at (img.151) [scale=1.2] {\hspace{0.01cm}\textcolor{white}{(f) Path-tracing}};
                \node [below left] at (img.30) [scale=1.2] {\hspace{0.01cm}\textcolor{white}{\contour{black}{\shortstack{Reference}}}};
                \spy [red] on (2,3.35) in node (a) [right] at (3.7,1.05);
                \spy [blue] on (6.45,2.45) in node [right] at (6.2,1.05);
                \spy [green ,magnification=3.1, width=1.1cm, height=1.1
                cm] on (2.38,0.5) in node [right] at (0.05,1.2);
          \end{scope}
          \node[circle,draw=orange,ultra thick,inner sep=0,minimum size=5pt,
      scale=5,xshift=-0.12cm,yshift=-0.063cm] (circle) at (a.center) {};
        \end{tikzpicture}
        \phantomcaption{}
        \label{subfig:HVLVPLVSLf}
    \end{subfigure} 
    \begin{tikzpicture}[overlay]
        \path[->,ultra thick] (DG) edge [bend right=45] node [sloped,above=0.1em] {\Large same number} (FG);
        \path[->,ultra thick] (DD) edge [bend left=45] node [sloped,above=0.1em] {\Large same time} (FD);
    \end{tikzpicture}
    \hspace*{0.5cm}
    \\%
    \caption{Comparison between HVL, VPL, VSL and Path-tracing.
    (a) uses 5 SH bands for convolution on fragment, 3 bands for the emission function and 400 HVLs, (b) and (c) are resp. render with 400 and 1928 VPLs. (d) and (e) are render resp. with 400 and 256 VSLs; $25$ and $9$ importance samples of the fragment's BRDF. HVLs and VSLs use heuristic $r_2$. (f) is a path-traced reference. Images are rendered at 1280x1024. The orange circle highlights a light leak artefact that can appears on virtual sources based method. They result from a combination of factors : the visibility of virtual sources, the nature of the sources and the low frequency approximation due to SH.}%
    \label{fig:comparisonHVLVPLVSL}%
\end{figure*}

\begin{figure*}[]%
    \centering%
    \hspace{0.5cm}
    \vspace*{-0.15cm}
    \begin{subfigure}[b]{0.465\linewidth}%
        \begin{tikzpicture}
          \begin{scope}[spy using outlines=
              {rectangle, magnification=3.9, width=2.3cm, height=1.8cm, connect spies, rounded corners}]
                \node[anchor=south west,inner sep=0] (img) {\includegraphics[width=\textwidth]{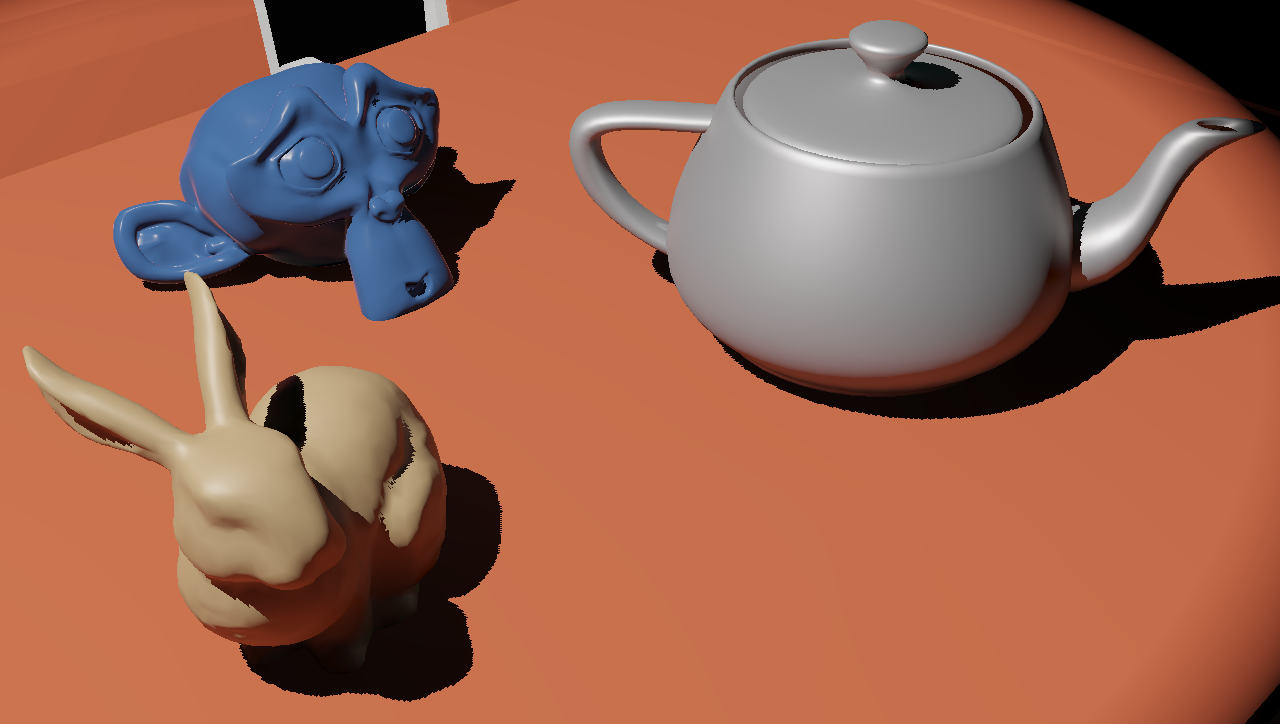}};
                \node [below right] at (img.152) [scale=1.2] {\hspace{0.01cm}\textcolor{white}{(a) HVL}};
                \node [below left] at (img.30) [scale=1.2] {\hspace{0.01cm}\textcolor{white}{\contour{black}{\shortstack{rmse 0.0397, psnr 28.032, ssim 0.9812}}}};
                \node [above right] at (img.208) [scale=1.2] {\textcolor{white}{$\approx$120s}};
                \spy [red] on (2,3.35) in node (a) [right] at (3.7,1.05);
                \spy [blue] on (6.45,2.45) in node [right] at (6.2,1.05);
                \spy [green ,magnification=3.1, width=1.1cm, height=1.1
                cm] on (2.38,0.5) in node [right] at (0.05,1.2);
          \end{scope}
          \node[circle,draw=orange,ultra thick,inner sep=0,minimum size=5pt,
      scale=5,xshift=-0.12cm,yshift=-0.063cm] (circle) at (a.center) {};
        \end{tikzpicture}
        \phantomcaption{}
        \label{subfig:HVLVPLVSL2a}
    \end{subfigure}%
    \hfill%
    \begin{subfigure}[b]{0.465\linewidth}%
        \begin{tikzpicture}
          \begin{scope}[spy using outlines=
              {rectangle, magnification=3.9, width=2.3cm, height=1.8cm, connect spies, rounded corners}]
                \node[anchor=south west,inner sep=0] (img) {\includegraphics[width=\textwidth]{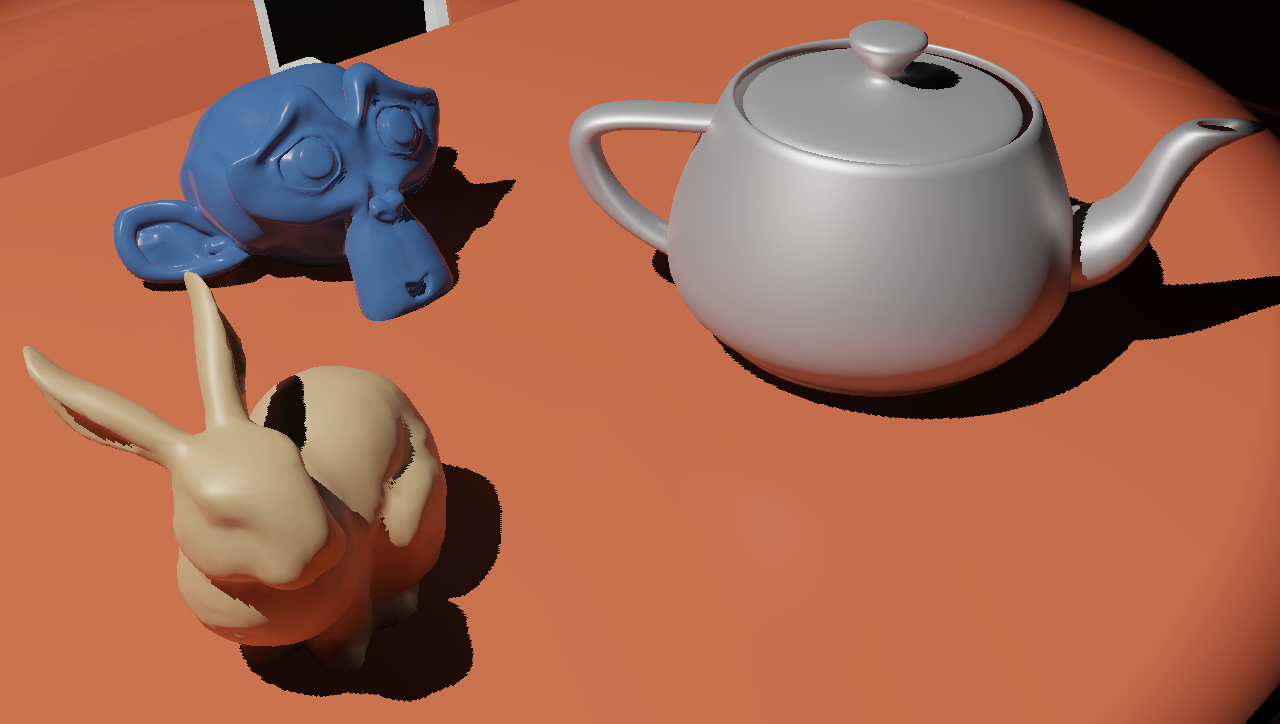}};
                \node [below right] at (img.152) [scale=1.2] {\hspace{0.01cm}\textcolor{white}{(b) VPL}};
                \node [below left] at (img.30) [scale=1.2] {\hspace{0.01cm}\textcolor{white}{\contour{black}{\shortstack{rmse 0.0368, psnr 28.679, ssim 0.9866}}}};
                \node [above right] at (img.208) [scale=1.2] {\textcolor{white}{$\approx$120s}};
                \spy [red] on (2,3.35) in node (a) [right] at (3.7,1.05);
                \spy [blue] on (6.45,2.45) in node [right] at (6.2,1.05);
                \spy [green ,magnification=3.1, width=1.1cm, height=1.1
                cm] on (2.38,0.5) in node [right] at (0.05,1.2);
          \end{scope}
          \node[circle,draw=orange,ultra thick,inner sep=0,minimum size=5pt,
      scale=5,xshift=-0.12cm,yshift=-0.063cm] (circle) at (a.center) {};
        \end{tikzpicture}
        \phantomcaption{}
        \label{subfig:HVLVPLVSL2b}
    \end{subfigure}
    \hspace*{0.5cm}
    \\%
    \caption{Comparison of HVL and VPL at same time using same resolution and same heuristic for HVL as the figure above. (a) is rendered with 350.000 HVLs and uses 9 SH bands for convolution and 3 SH bands for the emission of HVLs. (b) is rendered with 3 millions of VPLs. }%
    \label{fig:comparisonHVLVPLVSL2}%
\end{figure*}

To fairly compare HVL, VPL and VSL, we use only microfacet based BRDF (\fig{\ref{fig:comparisonHVLVPLVSL}} and~\ref{fig:comparisonHVLVPLVSL2}) and for clarity, we compile timings in Table~\ref{table:HVLVPLVSL}. Indeed, if VPL/VSL computation time might be greatly impacted by the use of measured, SH encoded BRDF, our HVL method is agnostic to the used BRDF model and performs similarly on parametric or measured BRDF.
In our comparison, we use the GGX microfacet distribution and evaluate the Fresnel reflectance using the exact term. 
We used material from MERL database that were fitted on GGX on the work of Ribardière \etal \cite{ribardiere2017}.
The three methods are also compared against path-traced references, obtained using pbrt-v3 \cite{pbrt-v3}, with three metrics : RMSE, PSNR and SSIM.

\begin{table}[bt]%
    \setlength{\tabcolsep}{8.7pt}%
    \centering%
    \begin{tabular}{|c |c | c c c c|}%
        \hline%
        \multirow{2}{*}{Scene} & HVL & \multicolumn{4}{c|}{Convolution band limit}\\\cline{3-6}
         & number & 3 & 5 & 7 & 9 \\\hline\hline%
        
        \multirow{3}{*}{\shortstack{Teaser\\\Fig{\ref{fig:Teaser}}(a)}} & 64 & 6.6 & 7.8 & 10.3 & 18.6 \\\cline{2-6}
         & 196 & 19.8 & 23.5 & 30.4 & 56.1 \\\cline{2-6}
        & 400 & 41.0 & 48.6 & 62.7 & 115.3 \\\hline

    \end{tabular}%
    \tightcaption{Timings (ms) for indirect lighting gathering from various HVLs numbers and SH bands for convolution on fragments. The emission function for each HVL is computed with 3 bands.}%
    \label{table:IndirectLight}%
\end{table}

\Fig{\ref{subfig:HVLVPLVSLa}}, \ref{subfig:HVLVPLVSLb} and \ref{subfig:HVLVPLVSLd} are computed using the same budget of virtual lights. 
To avoid a visual comparison bias, when using the same number of virtual lights, they are placed at the exact same locations for all three methods.
Our method avoids the classical VPLs artefacts but is more computationally expensive. Using $25$ samples per virtual light, VSL is even more expensive while still exhibiting noise. 
Aiming at interactive rendering, VSLs must use low sample count and results on noisy pictures (\fig{\ref{subfig:HVLVPLVSLe}}). To remove this noise, particularly visible on the bunny's feet, the number of samples must be at least tripled, with a similar increase on the computation time \figp{\ref{subfig:HVLVPLVSLd}}. Nonetheless, VSLs captures well high frequencies in the lighting and the result is close to the result produced by VPL in 120s using far less time \figp{\ref{subfig:HVLVPLVSL2b}}.

When given equal time to generate the picture (\fig{\ref{subfig:HVLVPLVSLa}}, \ref{subfig:HVLVPLVSLc} and \ref{subfig:HVLVPLVSLe}), VPLs still exhibit artefacts despite using more virtual lights whereas HVLs produce a smoother but artefact-free result. VSLs produce a noisy result as only a low number of samples might be used. Thus, HVLs are an interesting alternative to produce a fast low-frequency global lighting and gradually raise the final image in frequency. \Fig{\ref{subfig:HVLVPLVSL2a}} and \ref{subfig:HVLVPLVSL2b} are computed given a time budget of 120s. Even if less noticeable, VPLs still exhibit artefacts despite the big number of virtual lights while HVLs give a smooth result.

Light leaks artefacts may appear in some situations due to a combination of several factors (orange circle in \fig{\ref{fig:comparisonHVLVPLVSL}} and \ref{fig:comparisonHVLVPLVSL2}).
Inaccurate visibility estimation is one of these factors when using virtual sources. The artefact highlighted here is mainly due to non visible virtual sources behind Suzanne. For HVL and VSGL, another factor is due to the spherical nature of the sources. Indeed, when a shaded point is inside the sphere, the whole spherical source contributes to its shading thus overestimating the indirect lighting. Using more sources, thus reducing their radius, alleviates this problem. Finally, the frequency band limit resulting from the SH order for HVL produces a smoothing of these artefacts.

As shown by these experiments, HVLs always give a smooth results whatever the number of virtual light or the time budget given. This robustness of our method is its biggest advantage for interactive lighting design.

\begin{table}[b]%
    \setlength{\tabcolsep}{1.6pt}%
    \centering%
    \begin{tabular}{|c r|c c c|c c| c |c|c|}%
        \hline%
        \multicolumn{2}{|c|}{\multirow{3}{*}{Figures}} & \multicolumn{3}{c|}{Parameters} &  \multicolumn{2}{c|}{Timings} & \multicolumn{3}{c|}{Error metrics}\\\cline{3-10}
        & & \multirow{2}{*}{lights} & VSL & SH & \multirow{2}{*}{RSM*} & indirect  & \multirow{2}{*}{rmse} & \multirow{2}{*}{psnr} & \multirow{2}{*}{ssim} \\%
        & &  & samp. & bands &  & light  & & &\\\hline\hline%

        \multirow{3}{*}{VPL} & \ref{subfig:HVLVPLVSLb} & 400 & - & - & 0.59 & 13 & 0.054 & 25.20 & 0.977\\\cline{3-10}
        & \ref{subfig:HVLVPLVSLc} & 1928 & - & - & 0.59 & 61 & 0.088 &21.10& 0.981 \\\cline{3-10}
        & \ref{subfig:HVLVPLVSL2b} & 3M & - & - & \textbf{2.44} & 120k & 0.036 &28.67& 0.986\\\hline
         
        \multirow{2}{*}{VSL} &  \ref{subfig:HVLVPLVSLd} & 400 & $25$ & - & 0.59 &  254 & 0.043 &27.24& 0.974 \\\cline{3-10}
        & \ref{subfig:HVLVPLVSLe} & 256 & $9$ & - & 0.59 & 63 & 0.043 &27.26& 0.972\\\hline
         
        \multirow{2}{*}{HVL} & \ref{subfig:HVLVPLVSLa} & 400 & - & 5 & 0.38 & 61 & 0.039 &28.01& 0.980\\\cline{3-10}
        & \ref{subfig:HVLVPLVSL2a} & 350k & - & 9 & \textbf{1.6} & 120k & 0.039 &28.03& 0.981\\\hline
        
        \multicolumn{10}{c}{\small* RSM resolution is $1024^2$ pixels (\textbf{bold=$\mathbf{2048^2}$})}
         
    \end{tabular}%
    \tightcaption{Timings (ms) and error metrics (path-tracing used as reference) detail for \fig{\ref{fig:comparisonHVLVPLVSL}} and \ref{fig:comparisonHVLVPLVSL2}. RSM for HVLs are cheaper because less data are manipulated (see \sect{\ref{sec:distribution}} for details).}%
    \label{table:HVLVPLVSL}%
\end{table}

\subsection{Comparison with VSGL}
\label{sec:VSGL}

\begin{figure*}[]%
    \centering%
    \hspace{0.5cm}
    \begin{subfigure}[b]{0.465\linewidth}%
        \begin{tikzpicture}
  \begin{scope}[spy using outlines=
      {rectangle, magnification=3, width=3.3cm, height=2.5cm, connect spies, rounded corners}]
        \node[anchor=south west,inner sep=0] (img) {\includegraphics[width=\textwidth]{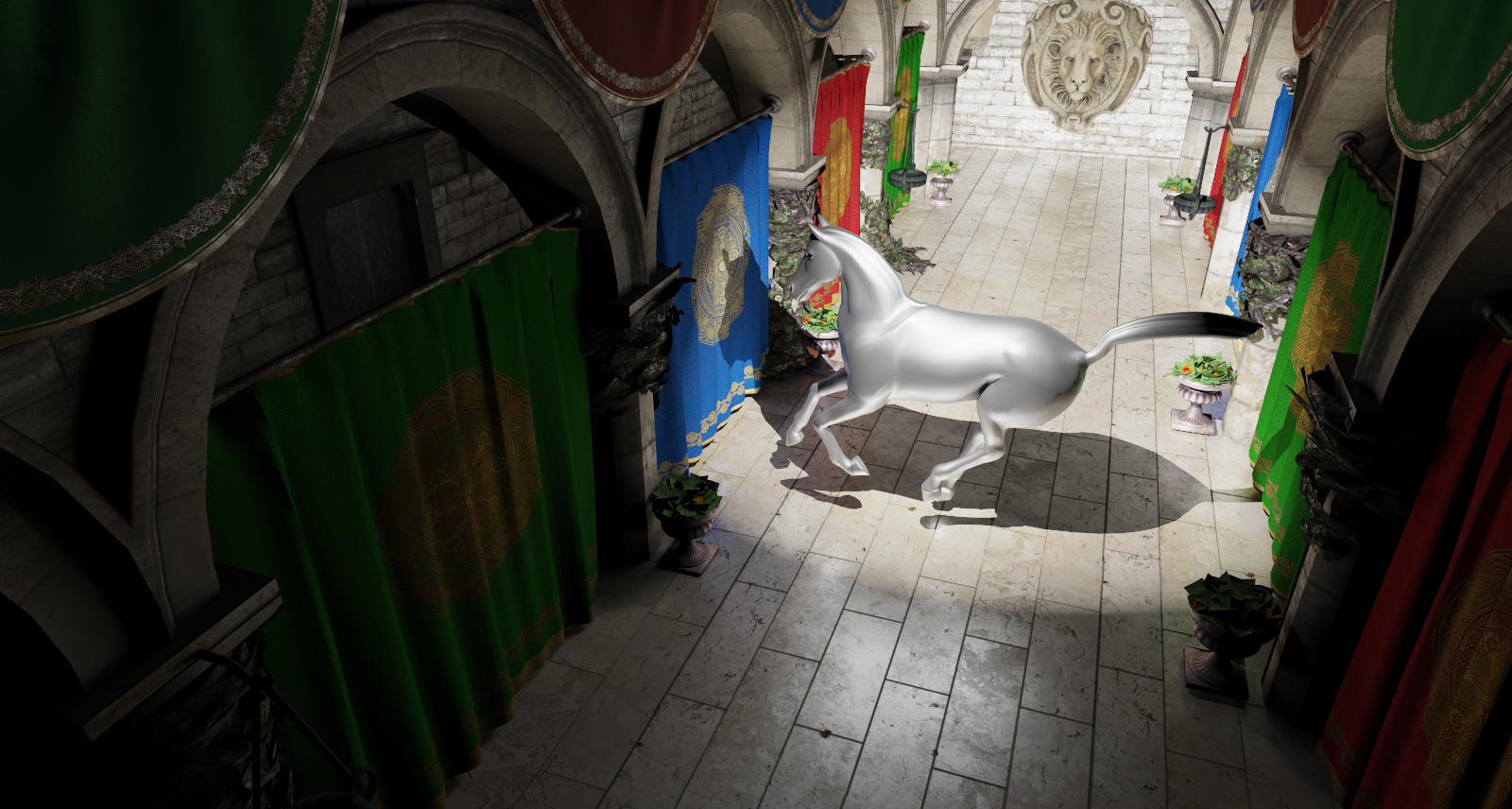}};
        \node [below right] at (img.152) [scale=1.5] {\hspace{0.2cm}\textcolor{white}{Path-tracing}};
        \spy [red] on (6.18,3.8) in node [left] at (3.4,2.5);
        \node [above left] at (img.-28) [scale=1.2] {\hspace{0.01cm}\textcolor{white}{reference}};
  \end{scope}
        \end{tikzpicture}
    \end{subfigure}%
    \hfill%
    \begin{subfigure}[b]{0.465\linewidth}%
        \begin{tikzpicture}
  \begin{scope}[spy using outlines=
      {rectangle, magnification=3, width=3.3cm, height=2.5cm, connect spies, rounded corners}]
        \node[anchor=south west,inner sep=0] (img) {\includegraphics[width=\textwidth]{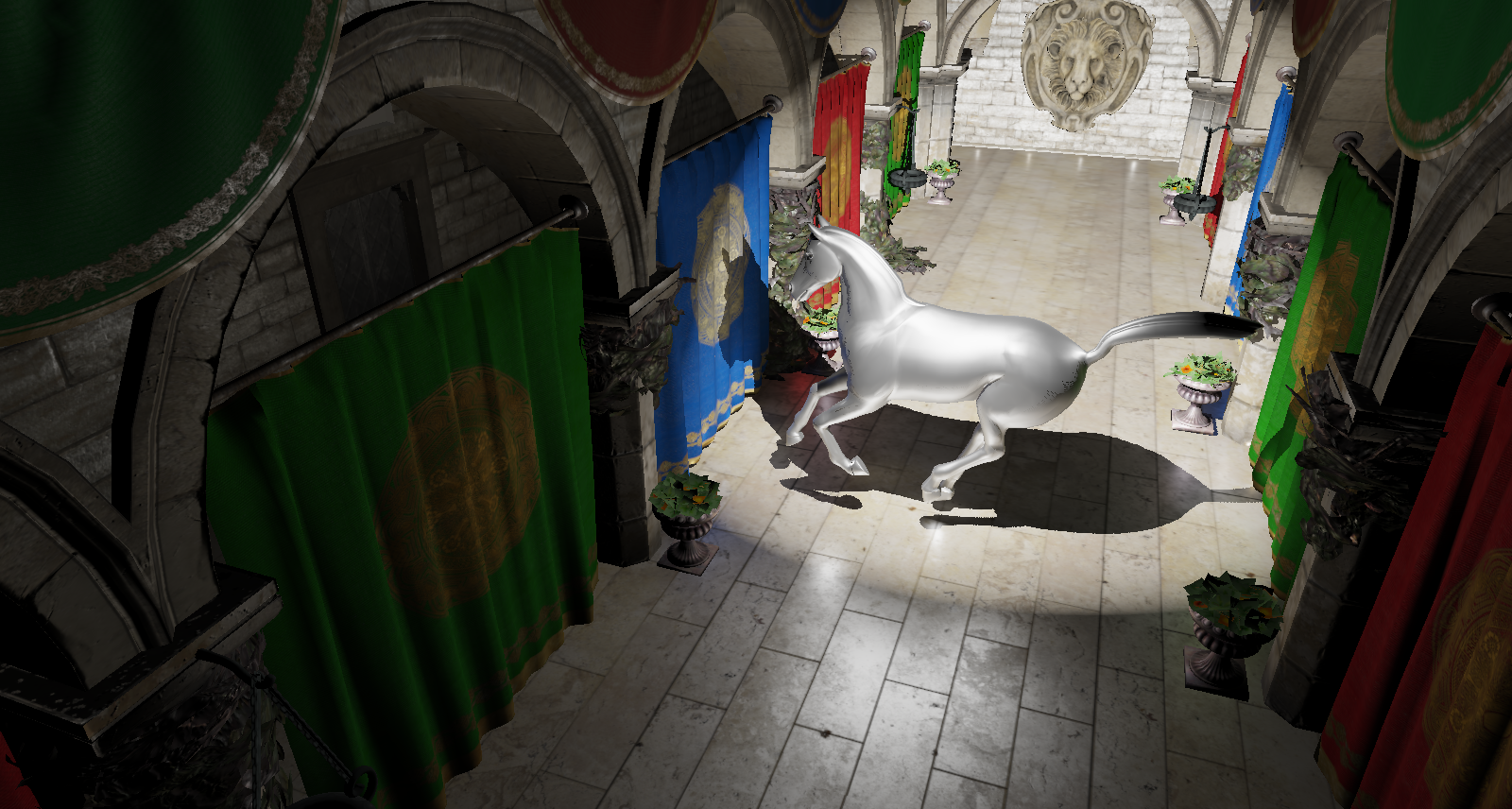}};
        \node [below right] at (img.152) [scale=1.5] {\hspace{0.2cm}\textcolor{white}{VPL}};
        \node [below left] at (img.28) [scale=1.5] {\textcolor{white}{39ms}};
        \spy [red] on (6.18,3.8) in node [left] at (3.4,2.5);
        \node [above left] at (img.-28) [scale=1.2] {\hspace{0.01cm}\textcolor{white}{\shortstack{rmse 0.2603\\ psnr 11.691\\ssim 0.8852}}};
  \end{scope}
        \end{tikzpicture}
    \end{subfigure}
    \hspace*{0.5cm}
    \\%
    \hspace*{0.5cm}
    \begin{subfigure}[b]{0.465\linewidth}%
        \begin{tikzpicture}
  \begin{scope}[spy using outlines=
      {rectangle, magnification=3, width=3.3cm, height=2.5cm, connect spies, rounded corners}]
        \node[anchor=south west,inner sep=0] (img) {\includegraphics[width=\textwidth]{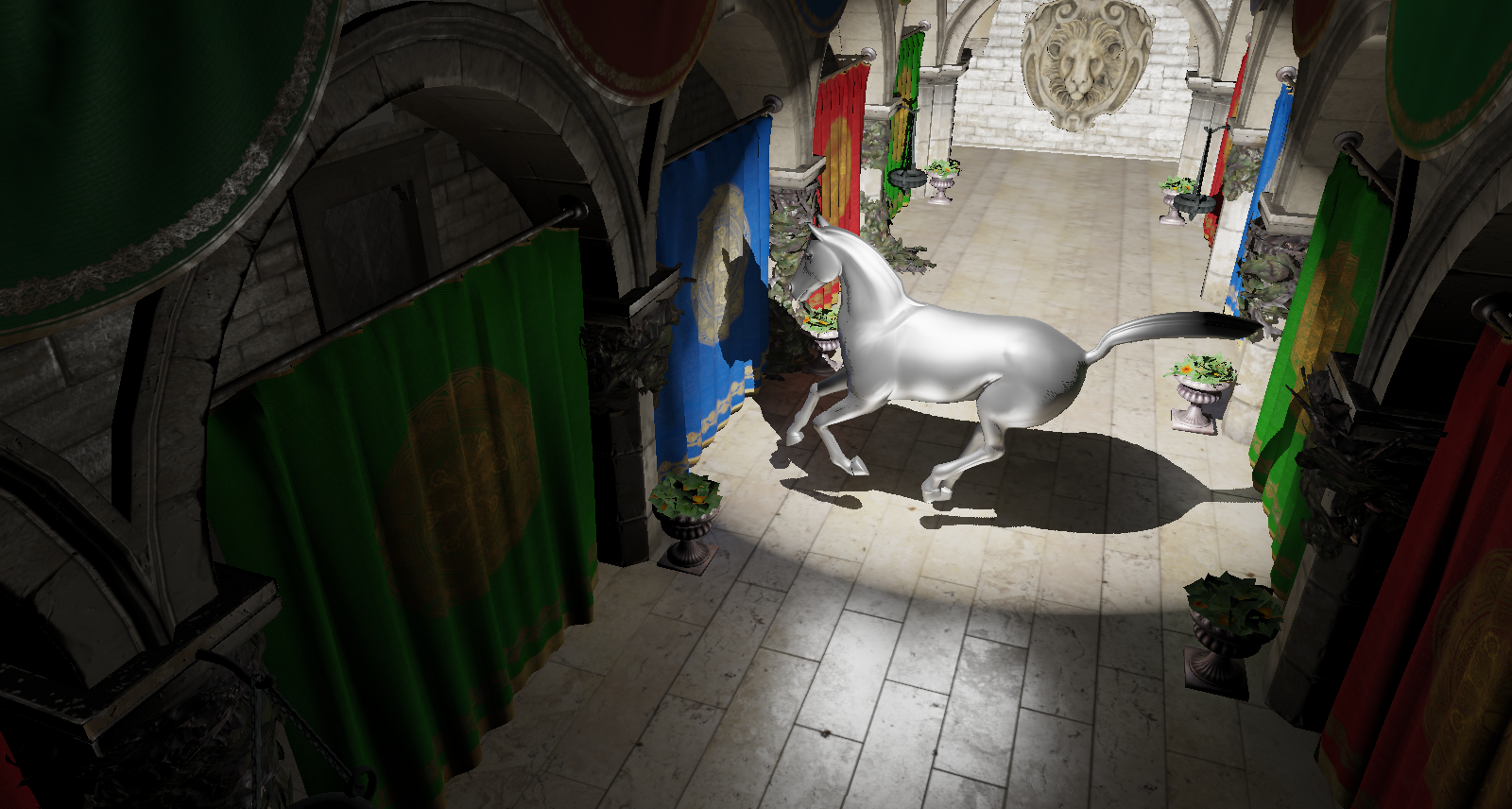}};
        \node [below right] at (img.152) [scale=1.5] {\hspace{0.2cm}\textcolor{white}{VSGL}};
        \node [below left] at (img.28) [scale=1.5] {\textcolor{white}{40ms}};
        \spy [red] on (6.18,3.8) in node [left] at (3.4,2.5);
        \node [above left] at (img.-28) [scale=1.2] {\hspace{0.01cm}\textcolor{white}{\shortstack{rmse 0.2539\\ psnr 11.907\\ssim 0.8740}}};
  \end{scope}
        \end{tikzpicture}
    \end{subfigure}%
    \hfill%
    \begin{subfigure}[b]{0.465\linewidth}%
        \begin{tikzpicture}
  \begin{scope}[spy using outlines=
      {rectangle, magnification=3, width=3.3cm, height=2.5cm, connect spies, rounded corners}]
        \node[anchor=south west,inner sep=0] (img) {\includegraphics[width=\textwidth]{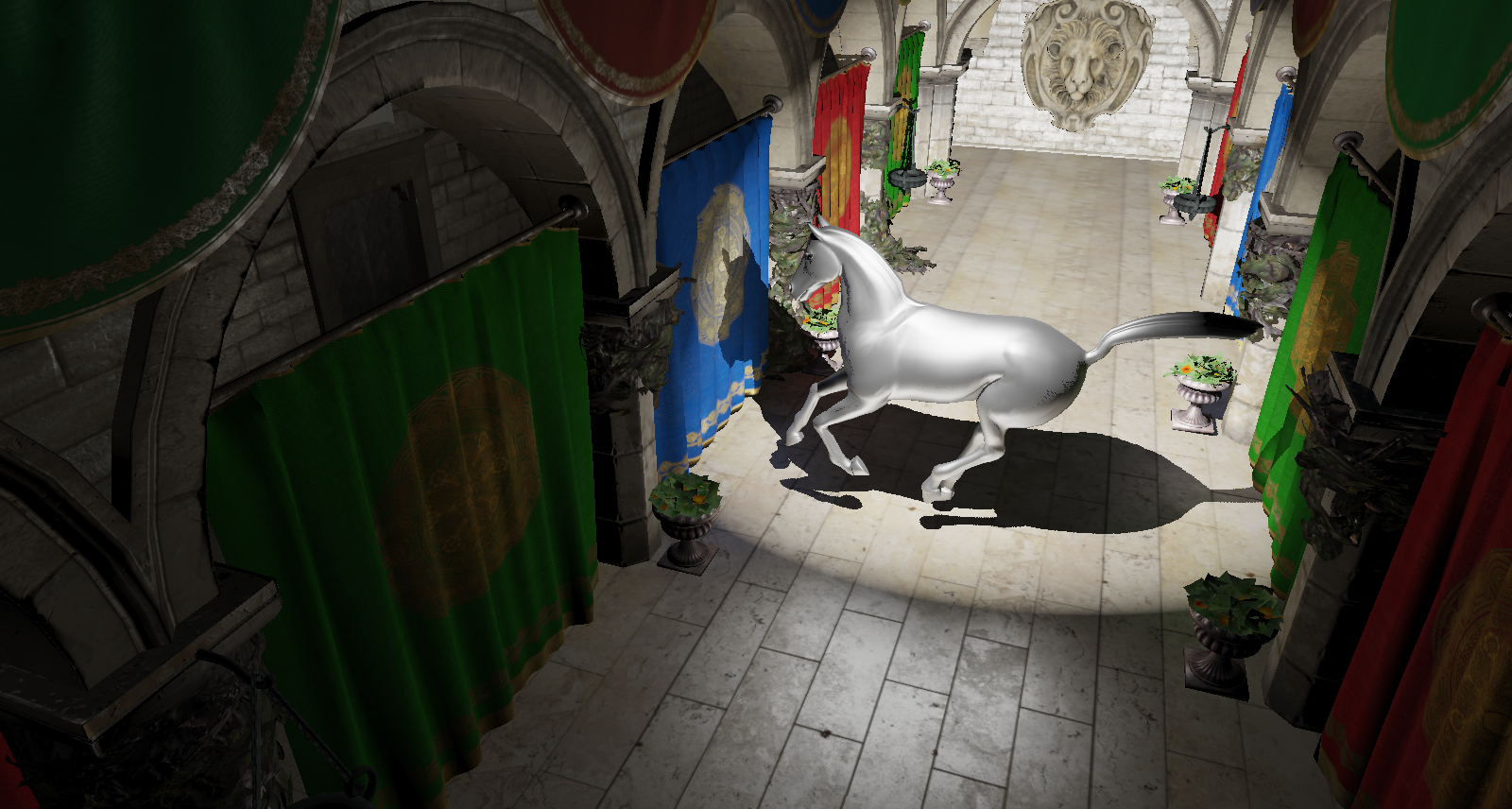}};
        \node [below right] at (img.152) [scale=1.5] {\hspace{0.2cm}\textcolor{white}{HVL}};
        \node [below left] at (img.28) [scale=1.5] {\textcolor{white}{40ms}};
        \spy [red] on (6.18,3.8) in node [left] at (3.4,2.5);
        \node [above left] at (img.-28) [scale=1.2] {\hspace{0.01cm}\textcolor{white}{\shortstack{rmse 0.2317\\ psnr 12.713\\ssim 0.8784}}};
  \end{scope}
        \end{tikzpicture}
    \end{subfigure}
    \hspace*{0.5cm}
    \\%
    \caption{Comparison with 900 VPL, 529 VSGL and 400 HVL, generated at equal time.
    Images are rendered at 1600$\times$900. The RSM resolution is $322^2$ pixels, and $14^2$ pixels for one VSGL.
    The contribution of HVLs is on 15 SH bands. In this rendering, HVL are limited to circularlt symmetric specular lobe to compute lighting only on ZH.  Insets show VPL spike artefacts and VSGL or HVL robustness to those artefacts. See additional material for more comparisons and implementation details. }%
    \label{fig:comparisonVSGL}%
\end{figure*}

Because there is no code available for the full VSGL method \cite{Yusuke2015}, we implemented a simpler version from the code available online from a paper on fast VSGL \cite{Yusuke2015f}. This allows us to produce a fair comparison between both raw methods, without importance or interleaved sampling.
We defined a scene similar to the one used in the VSGL paper \figp{\ref{fig:comparisonVSGL}} thanks to the data provided by the author. We accelerate lighting computation by using only ZH expansion in our code, thereby limiting convolution to materials with a symmetrical circular specular lobe which is fair compared to VSGL, being itself limited to BRDF with a single lobe :
\begin{itemize}
    \item We only use the ZH projection for the light using our proposal~\equp{\ref{eq:zonalHarmonicsLight}}.
    \item We modify the BRDF convolution to use the PRT framework proposed by Sloan \etal~\cite{Sloan2002} by adapting to our use case, \ie where the luminance and specular lobe of the BRDF are both circular symmetric functions :
    \begin{itemize}
        \item The diffuse contribution is computed using the work of Ramamoorthi and Hanrahan adapted when using only ZH~\cite{Ramamoorthi2001a}. (Appendix \ref{sec:BRDFconvolution})
        \item The specular contribution is computed using the convolution of two circular symmetric function on ZH~\cite{Dubouchet2019}. So, convolution complexity becomes $\mathcal{O}(n)$ instead of $\mathcal{O}(n^2)$, where $n$ is the number of SH bands. The weakening factor $\cos\theta_i$ is taken out of the integral \equp{\ref{eq:exactIndirect}} and calculated with the direction of the HVL center as input direction.
    \end{itemize}
    \item The HVLs directionnal emission is computed by evaluating a GGX BRDF.
\end{itemize}

In a high speed context, as VSGL approximate a set of nearby virtual light, it relies on RSM as it is the most efficient method to quickly access a VPL and its neighbors, in particular to take advantage of mipmap. Our HVL method extends VSL and only relies on RSM to sample the position of the virtual lights. As a result, HVL are less dependant on RSM making them more flexible than VSGL. 
As shown in \fig{\ref{fig:comparisonVSGL}}, HVL and VSGL exhibit similar results without any spikes. Here, both method are restricted to use only single-lobe BRDF as we use only one VSGL per pixel group, and HVL are restricted to it when using the acceleration method on ZH. Moreover, this acceleration on ZH makes the method much more competitive than the basic method (40ms for 15 SH bands). 
However, on this scene, VSGLs are less expensive because HVLs needs 15 bands to catch the strong specular reflection on the ground. On scenes with less specular effect, HVLs will require less bands, and therefore become faster than VSGLs.
The comparison between both method is pushed a little further into the supplemental, including rendering of the same scene with less SH bands showing that HVLs become less expensive than VSGLs.

\section{Limitations and Future Work}
\label{sec:limitations}

\paragraph*{Primary light}
As shown in  \sect{\ref{sec:HVLcontribution}}, our HVLs framework only considers simple primary light primitives whose emission is considered as a Dirac. In order to manage area lights, the process of computing HVLs luminance contribution should be reversed.
The SH coefficients of the incident luminance field due to the area light are to be stored on each HVL. Once the receiving fragment is known, the SH coefficients of the HVL emission are fetched according to the fragment direction and the convolution with the incident luminance field results in the incident luminance at the fragment. This will result in an increased computation time.

\paragraph*{HVLs visiblity}

The HVLs visibility is approximated by taking into account only its center and using shadow maps (or imperfect shadow map~\cite{Ritschel2008}), as VSL or VSGL already does~\cite{Hasan2009,Yusuke2015}.
However, this approximation is only valid when spherical lights are small.
For large HVLs, visibility management is more complex. 
It might be interesting to project the visibility on SH, as it's done by Ren~\etal~\cite{Ren2006} but this will require to evaluate a SH triple product ($\Visibility . \Radiance . \Reflectance$) instead of only a double product \equp{\ref{eq:LFdotproduct}}.
The extra cost of visibility projection and SH triple product would prevent to reach interactive rendering time. Indeed not like the SH double product, triple product does not reduce to a dot product of vectors.

\paragraph*{Efficient shading}

Our prototype implementation only use a simple gathering to compute HVLs contributions at each pixel, in which each pixel gets the contribution of all HVLs. This has a strong impact on the computation time when using numerous HVLs.
For increasing the efficiency of our framework, a fragment must ideally consider only HVLs that will have an influence on it.
As our proposal is orthogonal to any virtual lights gathering method, Lightcuts, proposed by Walter~\etal~\cite{Walter2005}, or ManyLoDs, proposed by Hollander~\etal~\cite{Hollander2011} and more interesting for interactive rendering, should be implemented over the HVLs set. Also, as done in VSGL paper \cite{Yusuke2015}, we believe that performance can be greatly improved by using interleaved sampling \cite{Keller2001}. HVLs exhibit band limit effect that can be useful for this technique. 

\paragraph*{Flickering}

Techniques dealing with efficient shading such as RSM clustering \cite{Prutkin2012} allow also solving the flickering problem.  As we do not apply normal filtering, temporal coherency management nor any filtering method, flickering becomes an issue with dynamic scenes. As HVLs inherit properties of VPLs, any method resolving flickering for VPLs should be adapted to HVLs, such as the temporally coherent sampling proposed by Bar{\'{a}}k \etal \cite{Barak2013}.


\paragraph*{Hemispherical bases}

Although spherical in shape, HVLs only emit light in an hemisphere. While the current implementation accounts for it with the geometric term, we thought about using the hemispherical basis proposed by Gautron~\etal~\cite{Gautron2007}. But this requires to correctly adapt the calculations to this basis, such as the integration of Legendre polynomials or the efficient evaluation of harmonics. The tricks used by Sloan~\cite{Sloan2013SH} to evaluate spherical harmonics efficiently do not apply directly on hemispherical harmonics. Alternatively, we could adapt the derivations to a more recent basis such as the $\mathcal{H}$-basis of Habel~\etal~\cite{Habel2010}.

\paragraph*{Soft angular kernels}
In our proposal, like for VSLs, we consider a constant emission on the solid angle of the spherical light. If VSLs rely on this hypothesis for efficient uniform sampling of VSLs for Monte-Carlo integration, we use this to derive an analytical solution for HVL \equp{\ref{eq:initialEquation}}. 
However, the solid angle boundary corresponds to an infinite frequency in the emission function that produce ringing artefacts in the SH approximation. 
To limit these ringing artefacts, one can add a soft angular kernel in \equ{\ref{eq:ZHcoeffsTheta}}.
However, finding an analytical, efficient to compute, solution with an arbitrary kernel is challenging but should improve the final result.
\paragraph*{Anisotropic materials}
As presented in \sect{\ref{sec:projRadAndMat}}, we use an equi-angular parameterization of the input direction to project materials on SH. If this memory efficient parameterization scales well for isotropic materials, the memory cost for anisotropic ones quickly become prohibitive \sectp{\ref{sec:distribution}}.
Investigating the parabolic parameterization proposed by Kautz~\etal~\cite{Kautz2002} should help to reduce this memory cost. Indeed, using their $128\times128$ parabolic map would reduce the cost of one anisotropic material on 10 SH bands from 388MB to 20MB. Doing so, it must be first demonstrated that the resolution will be sufficient to capture the full behaviour of the material according to the number of SH bands.

\section{Conclusion}

We have presented harmonics virtual lights, a novel model to capture and render indirect lighting in dynamic scenes with arbitrary materials.
Our HVLs are spherical, secondary light sources that we distribute on the surface of the scene after one bounce from the primary sources.
When viewed from a shaded point, their radially symmetric profile allows efficient projection of incoming luminance on the basis of zonal harmonics.
Using a fast rotation formula, we align this incoming luminance field with any reflectance function encoded on spherical harmonics, and perform shading with realistic materials as a simple dot product.
Thanks to our radius heuristics, the size of HVLs is determined adaptively to cover the scene as well as possible, thus avoiding VPLs artefacts.

Our results show that in comparison to virtual point lights, a much lower HVLs budget can be used to obtain images of similar quality.
Moreover, the output does not suffer artefacts such as unbounded contributions around the virtual sources, or undesirable reflections on glossy materials.
Our method is able to deliver such images at interactive framerates in fully dynamic scenes; the two hyperparameters, HVLs count and band limit, can be further tuned to achieve real-time rendering.

\section*{Acknowledgements}

Bunny and Dragon taken from the Stanford 3D Scanning Repository. Breakfast Room, Conference Room, Crytek Sponza, Dabrovic Sponza and Teapot taken from the McGuire Computer Graphics Archive. Suzanne taken from blender foundation.

\bibliographystyle{alpha}
\bibliography{references}       

\newcommand{\etalchar}[1]{$^{#1}$}
\begin{thebibliography}{HKWB09}

\bibitem[BBH13]{Barak2013}
Tom{\'{a}} Bar{\'{a}}k, Jir{\'{i}} Bittner, and Vlastimil
  Havran.
\newblock {Temporally coherent adaptive sampling for imperfect shadow maps}.
\newblock {\em Comput. Graph. Forum}, 32(4):87--96, 2013.

\bibitem[DJ18]{Dupuy2018}
Jonathan Dupuy and Wenzel Jakob.
\newblock {An adaptive parameterization for efficient material acquisition and
  rendering}.
\newblock In {\em SIGGRAPH Asia 2018 Tech. Pap. SIGGRAPH Asia 2018}, volume~37,
  pages 1--14. ACM New York, NY, USA, 2018.

\bibitem[DKH{\etalchar{+}}14]{Dachsbacher2014}
Carsten Dachsbacher, Jaroslav Křiv{\'{a}}nek, Milo{\v{s}} Ha{\v{s}}an, Adam
  Arbree, Bruce Walter, and Jan Nov{\'{a}}k.
\newblock {Scalable realistic rendering with many-light methods}.
\newblock {\em Comput. Graph. Forum}, 33(1):88--104, 2014.

\bibitem[DS05]{Dachsbacher2005}
Carsten Dachsbacher and Marc Stamminger.
\newblock {Reflective shadow maps}.
\newblock In {\em Proc. Symp. Interact. 3D Graph.}, pages 203--208, 2005.

\bibitem[DSJN19]{Dubouchet2019}
Adrien Dubouchet, Peter-Pike Sloan, Wojciech Jarosz, and Derek Nowrouzezahrai.
\newblock {Impulse Responses for Precomputing Light from Volumetric Media}.
\newblock 2019.

\bibitem[GKPB07]{Gautron2007}
Pascal Gautron, Jaroslav Krivanek, Sumanta Pattanaik, and Kadi Bouatouch.
\newblock {A novel hemispherical basis for accurate and efficient rendering}.
\newblock In {\em ACM SIGGRAPH 2007 Pap. - Int. Conf. Comput. Graph. Interact.
  Tech.}, volume 2004, pages 321--330, 2007.

\bibitem[HKWB09]{Hasan2009}
Milo{\v{s}} Ha{\v{s}}an, Jaroslav Křiv{\'{a}}nek, Bruce Walter, and Kavita
  Bala.
\newblock {Virtual Spherical Lights for Many-Light Rendering of Glossy Scenes}.
\newblock {\em ACM Trans. Graph.}, 28(5):1--6, 2009.

\bibitem[HREB11]{Hollander2011}
Matthias Holl{\"{a}}nder, Tobias Ritschel, Elmar Eisemann, and Tamy Boubekeur.
\newblock {ManyLoDs: Parallel many-view level-of-detail selection for real-time
  global illumination}.
\newblock {\em Comput. Graph. Forum}, 30(4):1233--1240, 2011.

\bibitem[HW10]{Habel2010}
Ralf Habel and Michael Wimmer.
\newblock {Efficient irradiance normal mapping}.
\newblock In {\em Proc. I3D 2010 ACM SIGGRAPH Symp. Interact. 3D Graph. Games},
  pages 189--195, 2010.

\bibitem[Kaj86]{Kajiya1986}
James~T Kajiya.
\newblock {The Rendering Equation}.
\newblock {\em Comput. Graph.}, 20(4):143--150, aug 1986.

\bibitem[Kel97]{Keller1997}
Alexander Keller.
\newblock {Instant radiosity}.
\newblock In {\em Proc. 24th Annu. Conf. Comput. Graph. Interact. Tech.
  SIGGRAPH 1997}, pages 49--56, 1997.

\bibitem[KH01]{Keller2001}
Alexander Keller and Wolfgang Heidrich.
\newblock {Interleaved Sampling}.
\newblock In {\em Eurographics Work. Render. Tech.}, pages 269--276. 2001.

\bibitem[KSS02]{Kautz2002}
Jan Kautz, Peter-Pike~Pike Sloan, and John Snyder.
\newblock {Fast, arbitrary BRDF shading for low-frequency lighting using
  spherical harmonics}.
\newblock In {\em Eurographics Work. Render.}, volume~2, pages 291--296, 2002.

\bibitem[LY19]{Lin2019}
Daqi Lin and Cem Yuksel.
\newblock {Real-Time Rendering with Lighting Grid Hierarchy}.
\newblock {\em Proc. ACM Comput. Graph. Interact. Tech.}, 2(1):1--17, 2019.

\bibitem[MPBM03]{Matusik2003}
Wojciech Matusik, Hanspeter Pfister, Matt Brand, and Leonard McMillan.
\newblock {A data-driven reflectance model}.
\newblock In {\em ACM Trans. Graph.}, volume~22, pages 759--769, 2003.

\bibitem[NSF12]{Nowrouzezahrai2012}
Derek Nowrouzezahrai, Patricio Simari, and Eugene Fiume.
\newblock {Sparse zonal harmonic factorization for efficient SH rotation}.
\newblock {\em ACM Trans. Graph.}, 31(3):1--9, 2012.

\bibitem[OS07]{Oat2007}
Christopher Oat and Pedro~V. Sander.
\newblock Ambient aperture lighting.
\newblock In {\em Proceedings of the 2007 Symposium on Interactive 3D Graphics
  and Games}, I3D '07, page 61–64, New York, NY, USA, 2007. Association for
  Computing Machinery.

\bibitem[PJH16]{pbrt-v3}
Matt Pharr, Wenzel Jakob, and Greg Humphreys.
\newblock {\em Physically Based Rendering: From Theory to Implementation (3rd
  ed.)}.
\newblock Morgan Kaufmann Publishers Inc., San Francisco, CA, USA, 3rd edition,
  November 2016.

\bibitem[PKD12]{Prutkin2012}
Roman Prutkin, Anton Kaplanyan, and Carsten Dachsbacher.
\newblock {Reflective Shadow Map Clustering for Real-Time Global Illumination.}
\newblock {\em Eurographics}, 2012.

\bibitem[RBMS17]{ribardiere2017}
Micka{\"{e}}l Ribardi{\`{e}}re, Benjamin Bringier, Daniel Meneveaux, and Lionel
  Simonot.
\newblock {STD: Student's t-Distribution of Slopes for Microfacet Based BSDFs}.
\newblock {\em Comput. Graph. Forum}, 36(2):421--429, 2017.

\bibitem[RDGK12]{Ritschel2012}
Tobias Ritschel, Carsten Dachsbacher, Thorsten Grosch, and Jan Kautz.
\newblock {The state of the art in interactive global illumination}.
\newblock {\em Comput. Graph. Forum}, 31(1):160--188, 2012.

\bibitem[RGK{\etalchar{+}}08]{Ritschel2008}
Tobias Ritschel, Thorsten Grosch, Min~H Kim, H.~P. Seidel, Carsten Dachsbacher,
  and Jan Kautz.
\newblock {Imperfect shadow maps for efficient computation of indirect
  illumination}.
\newblock In {\em ACM SIGGRAPH Asia 2008 Pap. SIGGRAPH Asia'08}, volume~27,
  pages 1--8. ACM New York, NY, USA, 2008.

\bibitem[RH01]{Ramamoorthi2001a}
Ravi Ramamoorthi and Pat Hanrahan.
\newblock {An efficient representation for irradiance environment maps}.
\newblock In {\em Proc. 28th Annu. Conf. Comput. Graph. Interact. Tech.
  SIGGRAPH 2001}, pages 497--500, 2001.

\bibitem[RWS{\etalchar{+}}06]{Ren2006}
Zhong Ren, Rui Wang, John Snyder, Kun Zhou, Xinguo Liu, Bo~Sun, Peter~Pike
  Sloan, Hujun Bao, Qunsheng Peng, and Baining Guo.
\newblock {Real-time soft shadows in dynamic scenes using spherical harmonic
  exponentiation}.
\newblock In {\em ACM SIGGRAPH 2006 Pap. SIGGRAPH '06}, pages 977--986. ACM,
  aug 2006.

\bibitem[SHD15]{Simon2015}
Florian Simon, Johannes Hanika, and Carsten Dachsbacher.
\newblock {Rich-VPLs for Improving the Versatility of Many-Light Methods}.
\newblock In {\em Comput. Graph. Forum}, volume~34, pages 575--584, 2015.

\bibitem[SKS02]{Sloan2002}
Peter~Pike Sloan, Jan Kautz, and John Snyder.
\newblock {Precomputed radiance transfer for real-time rendering in dynamic,
  low-frequency lighting environments}.
\newblock In {\em SIGGRAPH '02}, pages 527--536, 2002.

\bibitem[Slo08]{Sloan2008}
Peter-Pike Sloan.
\newblock {Stupid Spherical Harmonics ( SH ) Tricks}.
\newblock {\em Game Dev. Conf.}, 9:320--321, 2008.

\bibitem[Slo13]{Sloan2013SH}
Peter-Pike Sloan.
\newblock {Efficient Spherical Harmonic Evaluation}.
\newblock {\em J. Comput. Graph. Tech.}, 2(2):84--90, sep 2013.

\bibitem[SLS05]{Sloan2005}
Peter~Pike Sloan, Ben Luna, and John Snyder.
\newblock {Local, deformable precomputed radiance transfer}.
\newblock In {\em ACM Trans. Graph.}, volume~24, pages 1216--1224. ACM New
  York, NY, USA, 2005.

\bibitem[Tok15a]{Yusuke2015f}
Yusuke Tokuyoshi.
\newblock {Fast Indirect Illumination Using Two Virtual Spherical Gaussian
  Lights}.
\newblock In {\em SIGGRAPH Asia 2015 Posters}, SA '15, New York, NY, USA, 2015.
  Association for Computing Machinery.

\bibitem[Tok15b]{Yusuke2015}
Yusuke Tokuyoshi.
\newblock {Virtual Spherical Gaussian Lights for Real-time Glossy Indirect
  Illumination}.
\newblock In {\em Comput. Graph. Forum}, volume~34, pages 89--98, 2015.

\bibitem[WCZR20]{Lifan2020}
Lifan Wu, Guangyan Cai, Shuang Zhao, and Ravi Ramamoorthi.
\newblock {Analytic spherical harmonic gradients for real-time rendering with
  many polygonal area lights}.
\newblock {\em ACM Trans. Graph.}, 39(4), 2020.

\bibitem[WFA{\etalchar{+}}05]{Walter2005}
Bruce Walter, Sebastian Fernandez, Adam Arbree, Kavita Bala, Michael Donikian,
  and Donald~P Greenberg.
\newblock {Lightcuts: A scalable approach to illumination}.
\newblock In {\em ACM Trans. Graph.}, volume~24, pages 1098--1107, 2005.

\bibitem[WR18]{Wang2018}
Jingwen Wang and Ravi Ramamoorthi.
\newblock {Analytic spherical harmonic coefficients for polygonal area lights}.
\newblock {\em ACM Trans. Graph.}, 37(4), 2018.

\bibitem[XCM{\etalchar{+}}14]{Xu2014a}
Kun Xu, Yan~Pei Cao, Li~Qian Ma, Zhao Dong, Rui Wang, and Shi~Min Hu.
\newblock {A practical algorithm for rendering interreflections with
  all-frequency brdfs}.
\newblock {\em ACM Trans. Graph.}, 33(1):1--16, 2014.

\end{thebibliography}


\begin{thebibliography}{Tok15b}

\bibitem[Tok15a]{Yusuke2015f}
Yusuke Tokuyoshi.
\newblock {Fast Indirect Illumination Using Two Virtual Spherical Gaussian
  Lights}.
\newblock In {\em SIGGRAPH Asia 2015 Posters}, SA '15, New York, NY, USA, 2015.
  Association for Computing Machinery.

\bibitem[Tok15b]{Yusuke2015}
Yusuke Tokuyoshi.
\newblock {Virtual Spherical Gaussian Lights for Real-time Glossy Indirect
  Illumination}.
\newblock In {\em Comput. Graph. Forum}, volume~34, pages 89--98, 2015.

\end{thebibliography}
\appendix
\section{Legendre Polynomials Integration}
\label{sec:legendreIntegral}

We search for a closed form to the integral of Legendre polynomials over a spherical cap of angle $\angleProj$:
\begin{tightequation*}
    \int_{\theta=0}^\angleProj \legendre_\order(\cos(\theta))\sin(\theta) \diff{\theta}
\end{tightequation*}
Substituting the variable $u = \cos(\theta)$ yields
\begin{tightequation*}
    \int_{\theta = 0}^\angleProj \legendre_\order(\cos(\theta)) \sin(\theta) \diff{\theta} =
    \int_{u = \cos(\angleProj)}^1 \legendre_\order(u) \diff{u} 
\end{tightequation*}%
We get the primitive from the recursive formula
\begin{tightequation*}
    \int \legendre_\order(x) \diff{x} =
    \frac{\legendre_{\order + 1}(x) - \legendre_{\order - 1}(x)}{2\order + 1}
\end{tightequation*}%
and as $\forall\,l \geq 0,\,\legendre_\order(1) = 1$ we obtain
\begin{tightalign*}
    \int_{\cos(\angleProj)}^1 \legendre_\order(u) \diff{u} &=
    \Bigg[
        \frac{\legendre_{\order + 1}(u) - \legendre_{\order - 1}(u)}{2\order + 1}
    \Bigg]_{\cos(\angleProj)}^1 \\&=
    \frac{-\legendre_{\order + 1}(\cos(\angleProj))+\legendre_{\order-1}(\cos(\angleProj))}{2\order + 1}
\end{tightalign*}
In summary, we have established that
\begin{tightequation*}
\boxed{
    \int_{\theta = 0}^\angleProj \legendre_\order(\cos(\theta)) \sin(\theta) \diff{\theta} =
    \frac{-\legendre_{\order + 1}(\cos(\angleProj)) + \legendre_{\order - 1}(\cos(\angleProj))}{2\order + 1}
}
\end{tightequation*}

\section{Illuminance of a circular symmetric luminance}
\label{sec:BRDFconvolution}

Ramamoorthi and Hanrahan use a $4 \times 4$ matrix to compute illuminance~\cite{Ramamoorthi2001a}. It can be replace by a $2 \times 2$ matrix when using a circular symmetric luminance function as it can be only projected on ZH when orienting the projection according to the direction of luminance emission.
\begin{align*}
    M = \begin{pmatrix}
      c_3 \Radiance_{2}^0 & c_2 \Radiance_{1}^0 \\
      c_2 \Radiance_{1}^0 & c_4 \Radiance_{0}^0 - c_5 \Radiance_{2}^0 
    \end{pmatrix} \\
    c_2=0.511664 \hspace{0.3cm} c_3=0.743125 \\
    c_4=0.886227 \hspace{0.3cm} c_5=0.247708
\end{align*}
Initially, they evaluate illuminance only with the normal surface $\text{n}$. 
illuminance is evaluated accordingly to the direction of luminance emission $\omega_l$, thus if we note $z = \omega_l \cdot \text{n}$, illuminance is computed with 
\begin{align*}
    E(\text{o}) = \text{o}^t M \text{o} \quad \mbox{with} \quad
    \text{o}^t = \begin{pmatrix}
        z & 1
    \end{pmatrix}
\end{align*}
Hence, we can write too
\begin{equation*}
    E(z) = c_3 \Radiance_{2}^0z^2 + 2c_2\Radiance_{1}^0z  + c_4 \Radiance_{0}^0 - c_5 \Radiance_{2}^0
\end{equation*}

\section{Materials used from MERL/RGL database}
\label{sec:materialsList}

{\small
\paragraph*{MERL}\cite{Matusik2003} (1)~alum-bronze, (2)~alumina-oxide, (3)~aluminium, (4)~beige-fabric, (5)~blue-acrylic, (6)~blue-fabric, (7)~blue-rubber, (8)~gold-metallic-paint, (9)~gold-paint, (10)~green-plastic, (11)~red-fabric2, (12)~red-phenolic, (13)~red-specular-plastic, (14)~silver-metallic-paint, (15)~silver-metallic-paint2, (16)~teflon, (17)~two-layer-gold, (18)~white-fabric, (19)~white-marble, (20)~white-paint, (21)~yellow-phenolic.

\paragraph*{RGL}\cite{Dupuy2018} (22)~acrylic-felt-green, (23)~acrylic-felt-orange, (24)~acrylic-felt-pink, (25)~acrylic-felt-purple, (26)~acrylic-felt-white, (27)~cardboard, (28)~cc-iris-purple-gem, (29)~ilm-l3-37-metallic, (30)~ilm-solo-m-68, (31)~ilm-solo-illenium-falcon, (32)~irid-flake-paint2, (33)~paper-blue, (34)~paper-red, (35)~paper-yellow, (36)~sating-gold, (37)~vch-frozen-amethyst.}

{
\begin{table}[h]%
    \setlength{\tabcolsep}{8.7pt}%
    \centering%
    \small
    \begin{tabular}{|c |c | c |}%
        \hline%
        Fig. & Object & materials\\\hline\hline%
        
        \multirow{2}{*}{\Fig{\ref{fig:Teaser}} } & Dragon & 12, 18, 32, 33, 36 \\\cline{2-3}
        & Atrium &  1, 2, 6, 11, 18\\\hline
        
        \multirow{2}{*}{ \Fig{\ref{fig:GIexamples}}} & \multirow{2}{*}{Sponza} & 10, 11, 18, 19, 20, 22, 23\\
        & &  24, 25, 26, 29, 34, 35, 37 \\\hline
        
        \Fig{\ref{fig:radiusHVL}} & - & 1, 4, 11 \\\hline
         
        \multirow{3}{*}{\Fig{\ref{fig:differentMaterial}}} & Table & 7, 16, 34 \\\cline{2-3}
        & Chairs & 16, 23, 30 \\\cline{2-3}
        & Other & 3, 8, 15, 17, 18, 21, 22, 27, 31   \\\hline
        
        \Fig{\ref{fig:comparisonHVLVPLVSL}}, \ref{fig:comparisonHVLVPLVSL2} & - & 5, 9, 13, 14 \\\hline

    \end{tabular}%
    \label{table:materialsData}%
\end{table}}

\end{document}


\title{Harmonics Virtual Lights : fast projection of luminance field on spherical harmonics for efficient rendering \\ (Supplemental Material)}
\date{}

\newcommand{\IRIT}{IRIT, Universit\'{e} de Toulouse, CNRS, UT3, Toulouse, France}

\author{ 
    \href{https://orcid.org/0000-0003-4981-4649}{\includegraphics[scale=0.06]{orcid.pdf}\hspace{1mm}}Pierre Mézières\\
\IRIT \\
\texttt{pierre.mezieres@irit.fr} \\
\And
\href{https://orcid.org/0000-0002-6246-7146}{\includegraphics[scale=0.06]{orcid.pdf}\hspace{1mm}} François Desrichard \\
\IRIT \\
\texttt{françois.desrichard@irit.fr}
\And
\href{https://orcid.org/0000-0003-0506-6036}{\includegraphics[scale=0.06]{orcid.pdf}\hspace{1mm}} David Vanderhaeghe \\
\IRIT \\
\texttt{david.vanderhaeghe@irit.fr}
\And
\href{https://orcid.org/0000-0001-5606-9654}{\includegraphics[scale=0.06]{orcid.pdf}\hspace{1mm}}Mathias Paulin \\
\IRIT \\
\texttt{mathias.paulin@irit.fr} \\
}

\renewcommand{\undertitle}{}
\renewcommand{\shorttitle}{Harmonics Virtual Lights (Supplemental Material)}
\renewcommand{\headeright}{}
\maketitle

\section{Implementation details of VSGL}

We implemented a simpler version of VSGL from the code available online of a paper on fast VSGL \cite{Yusuke2015}, because there is no code available for the full VSGL method \cite{Yusuke2015f}.
We precise that one VSGL use two spherical gaussians, one for diffuse and the other one for specular information. 
What changes compared to the full version is that it is only one VSGL without extras, no importance sampling and no interleaved sampling. 

For a fair comparison between our method and VSGL, we slightly modified the code published by\cite{Yusuke2015} to generate several VSGLs, while the author generates only one VSGL on the whole RSM. 
We distribute VSGLs and HVLs uniformly on the RSM. We typically use $14^2$ pixels for each VSGL, and an RSM of $280^2$ pixels for 400 VSGLs and $322^2$ pixels for 529 VSGLs. 
We place HVLs or VPLs at the center of the pixel group used to define a VSGL when comparing with the same number of virtual lights.

\begin{figure*}
    \centering%
    \begin{subfigure}[b]{0.499\linewidth}%
        \begin{tikzpicture}
          \begin{scope}[spy using outlines=
              {rectangle, magnification=3, width=3.5cm, height=2cm, connect spies, rounded corners}]
                \node[anchor=south west,inner sep=0] (img) {\includegraphics[width=\textwidth]{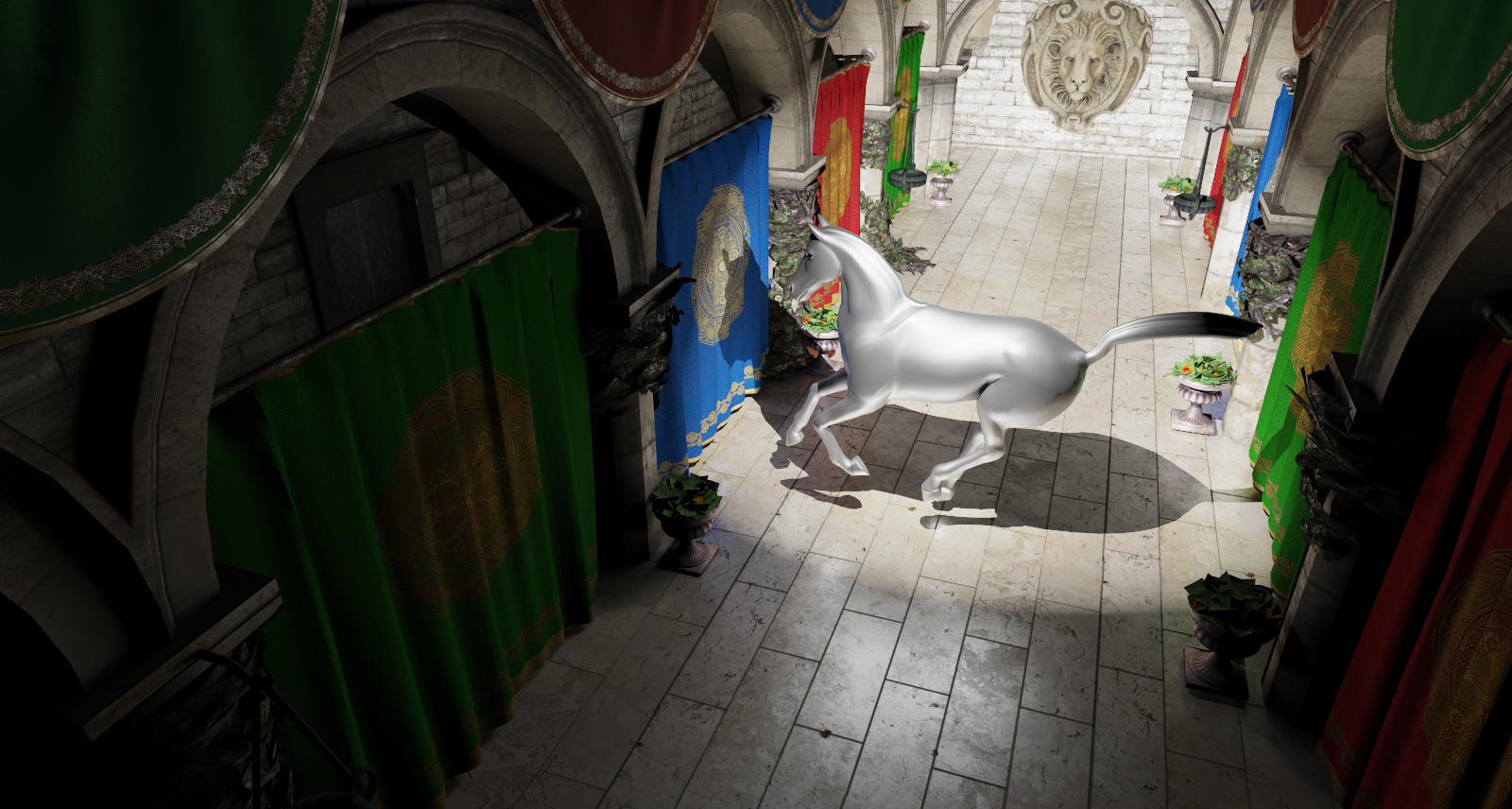}};
                \node [below right] at (img.152) [scale=1.2] {\hspace{0.01cm}\textcolor{white}{Path-tracing}}; 
                \node [above right] at (img.208) [scale=1.2] {\hspace{0.01cm}\textcolor{white}{Reference}};
                
          \end{scope}
        \end{tikzpicture}
    \end{subfigure}%
    \hfill%
    \begin{subfigure}[b]{0.499\linewidth}%
        \begin{tikzpicture}
          \begin{scope}[spy using outlines=
              {rectangle, magnification=3, width=3.1cm, height=1.7cm, connect spies, rounded corners}]
                \node[anchor=south west,inner sep=0] (img) {\includegraphics[width=\textwidth]{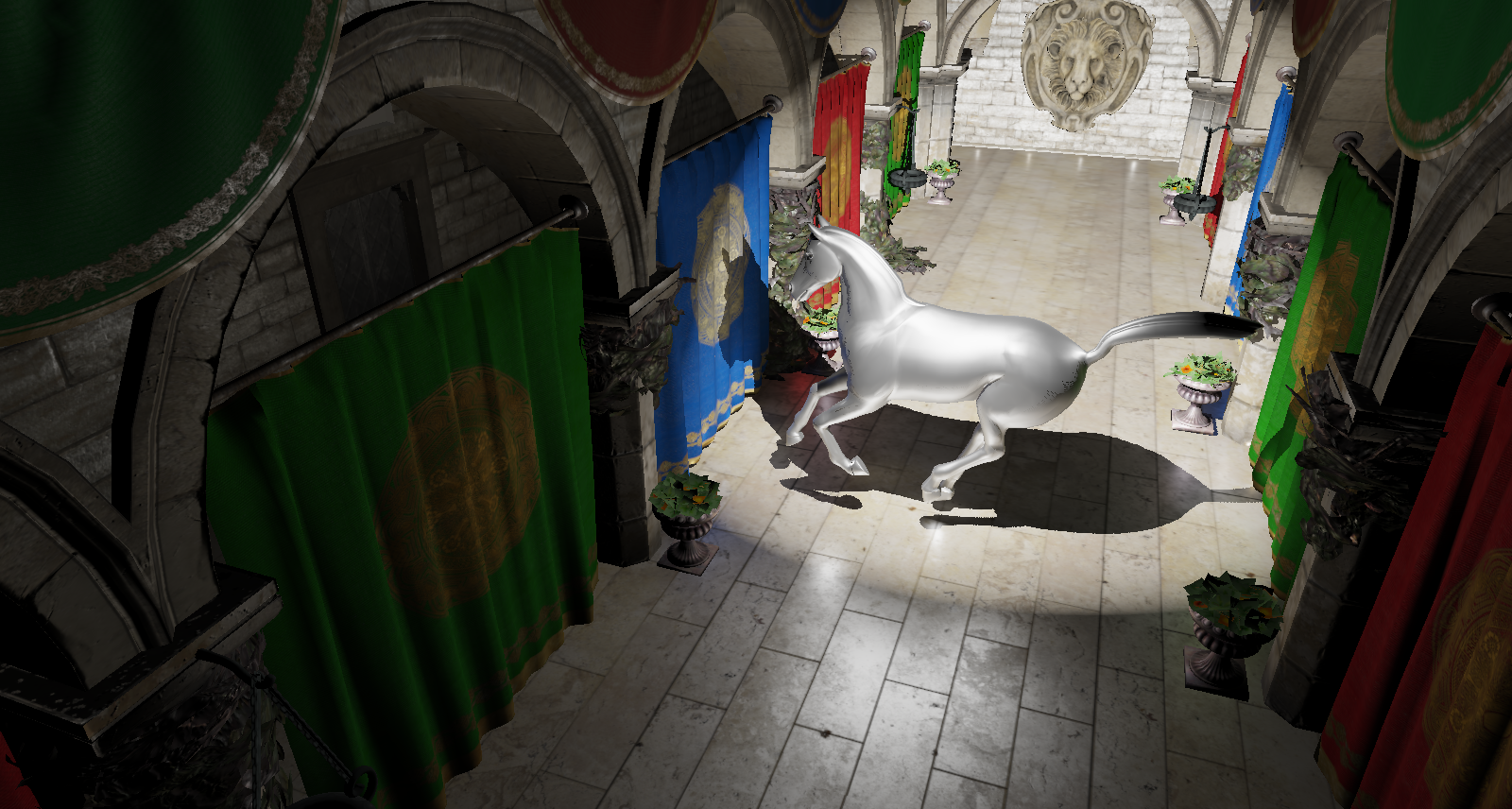}};
                \node [below right] at (img.152) [scale=1.2] {\hspace{0.01cm}\textcolor{white}{900 VPL}};
                \node [below left] at (img.28) [scale=1.2] {\textcolor{white}{39ms}}; 
                \node [above right] at (img.208) [scale=1.2] {\hspace{0.01cm}\textcolor{white}{\shortstack{rmse 0.2603\\ psnr 11.691\\ ssim 0.8852}}};
                \coordinate (DG) at (img.west);
                \coordinate (DD) at (img.east);
          \end{scope}
        \end{tikzpicture}
    \end{subfigure}\\%
    \begin{subfigure}[b]{0.499\linewidth}%
        \begin{tikzpicture}
          \begin{scope}[spy using outlines=
              {rectangle, magnification=3, width=3.5cm, height=2cm, connect spies, rounded corners}]
                \node[anchor=south west,inner sep=0] (img) {\includegraphics[width=\textwidth]{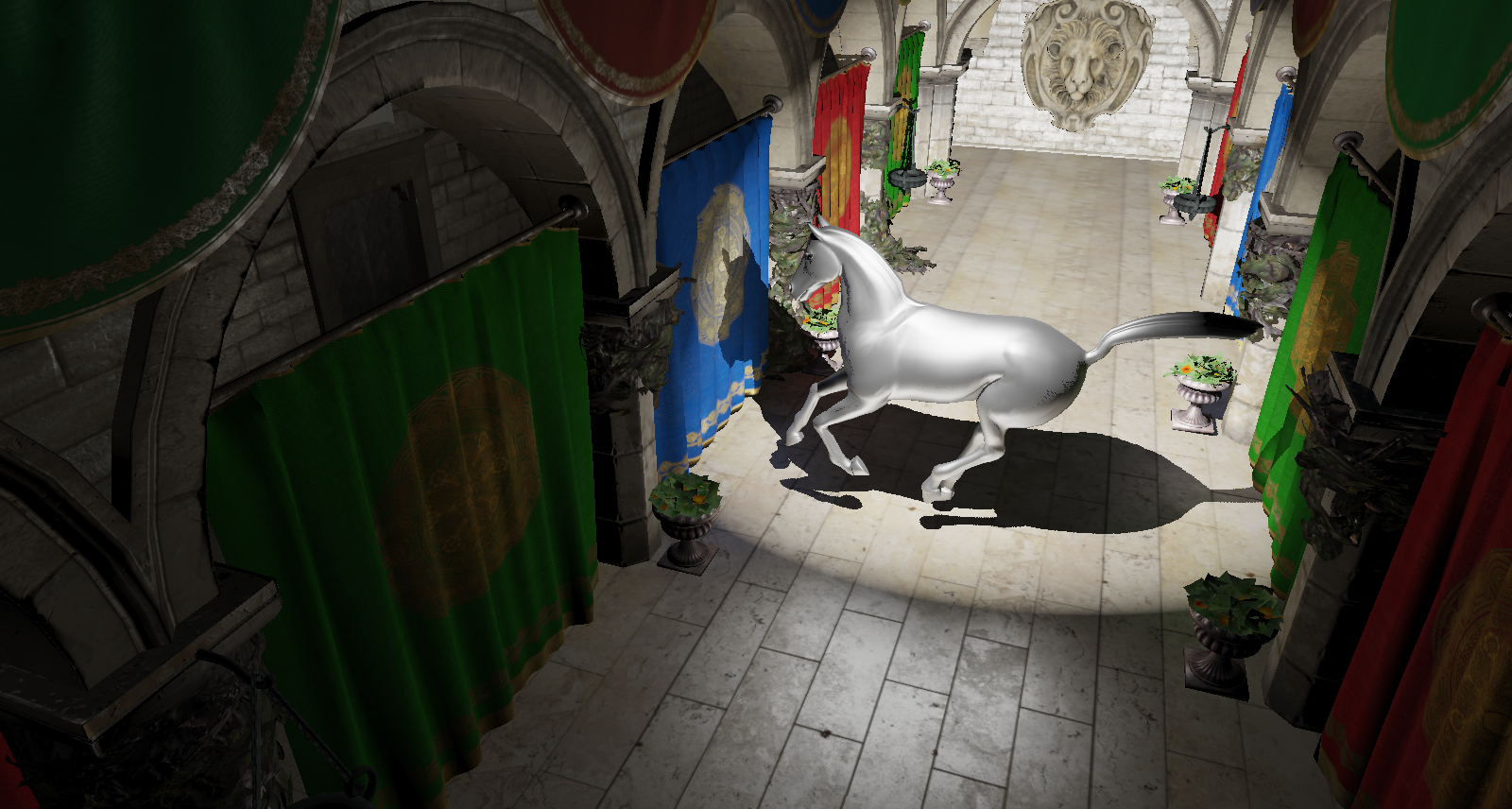}};
                \node [below right] at (img.152) [scale=1.2] {\hspace{0.01cm}\textcolor{white}{400 HVL - 15 ZH}};
                \node [below left] at (img.28) [scale=1.2] {\textcolor{white}{40ms}};
                \node [above right] at (img.208) [scale=1.2] {\hspace{0.01cm}\textcolor{white}{\shortstack{rmse 0.2317\\ psnr 12.703\\ssim 0.8785}}};
          \end{scope}
        \end{tikzpicture}
    \end{subfigure}%
    \hfill%
    \begin{subfigure}[b]{0.499\linewidth}%
        \begin{tikzpicture}
          \begin{scope}[spy using outlines=
              {rectangle, magnification=3, width=3.5cm, height=2cm, connect spies, rounded corners}]
                \node[anchor=south west,inner sep=0] (img) {\includegraphics[width=\textwidth]{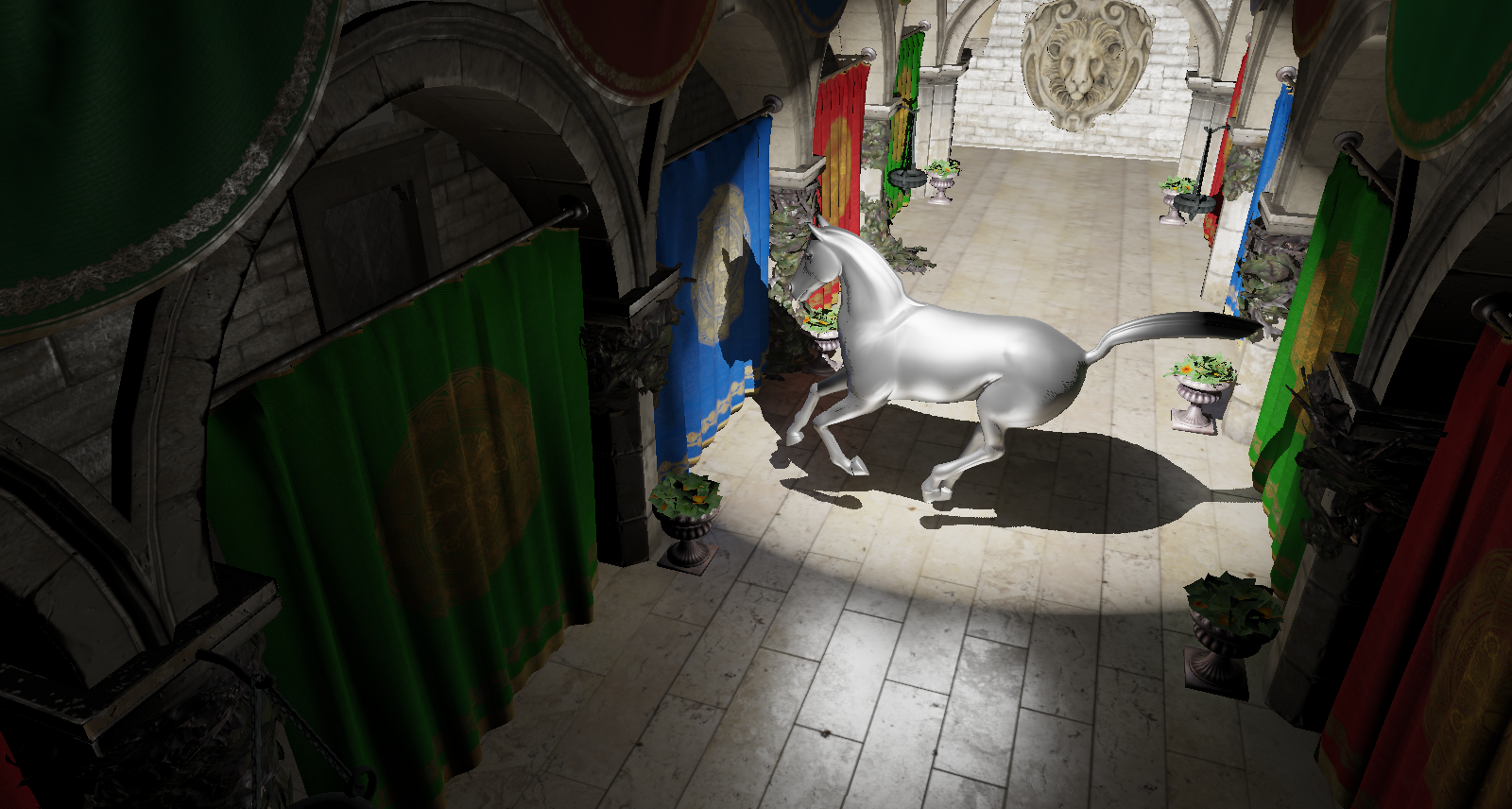}};
                \node [below right] at (img.152) [scale=1.2] {\hspace{0.01cm}\textcolor{white}{529 VSGL}};
                \node [below left] at (img.28) [scale=1.2] {\textcolor{white}{40ms}};
                \node [above right] at (img.208) [scale=1.2] {\hspace{0.01cm}\textcolor{white}{\shortstack{rmse 0.2539\\ psnr 11.907\\ssim 0.8740}}};
          \end{scope}
        \end{tikzpicture}
    \end{subfigure} %
    \caption{HVL and VSGL are able to generate similar quality in same time and without producing traditionnal VPL artefacts. }
    \label{fig:1}
\end{figure*}

\vspace{-0.25em}
\section{Results and comparisons}

Results are generated with a RTX 2080 on 1600$\times$900 resolution. The strong specular reflection on the ground of Sponza is the most visible effect which varies depending on the rendering technique. 
We consider the result generated with path-tracing as a reference to compute the error metrics \figp{\ref{fig:1}}.
We compiled in Table \ref{table1} the timings to generate these results and the resulting scores on RMSE, PSNR and SSIM.

\subsection{Effect of the number of Zonal Harmonics}

\begin{figure}[]%
    \centering%
    \hspace{0.5cm}
    \begin{subfigure}[b]{0.465\linewidth}%
        \begin{tikzpicture}
          \begin{scope}[spy using outlines=
              {rectangle, magnification=3, width=3.1cm, height=1.7cm, connect spies, rounded corners}]
                \node[anchor=south west,inner sep=0] (img) {\includegraphics[width=\textwidth]{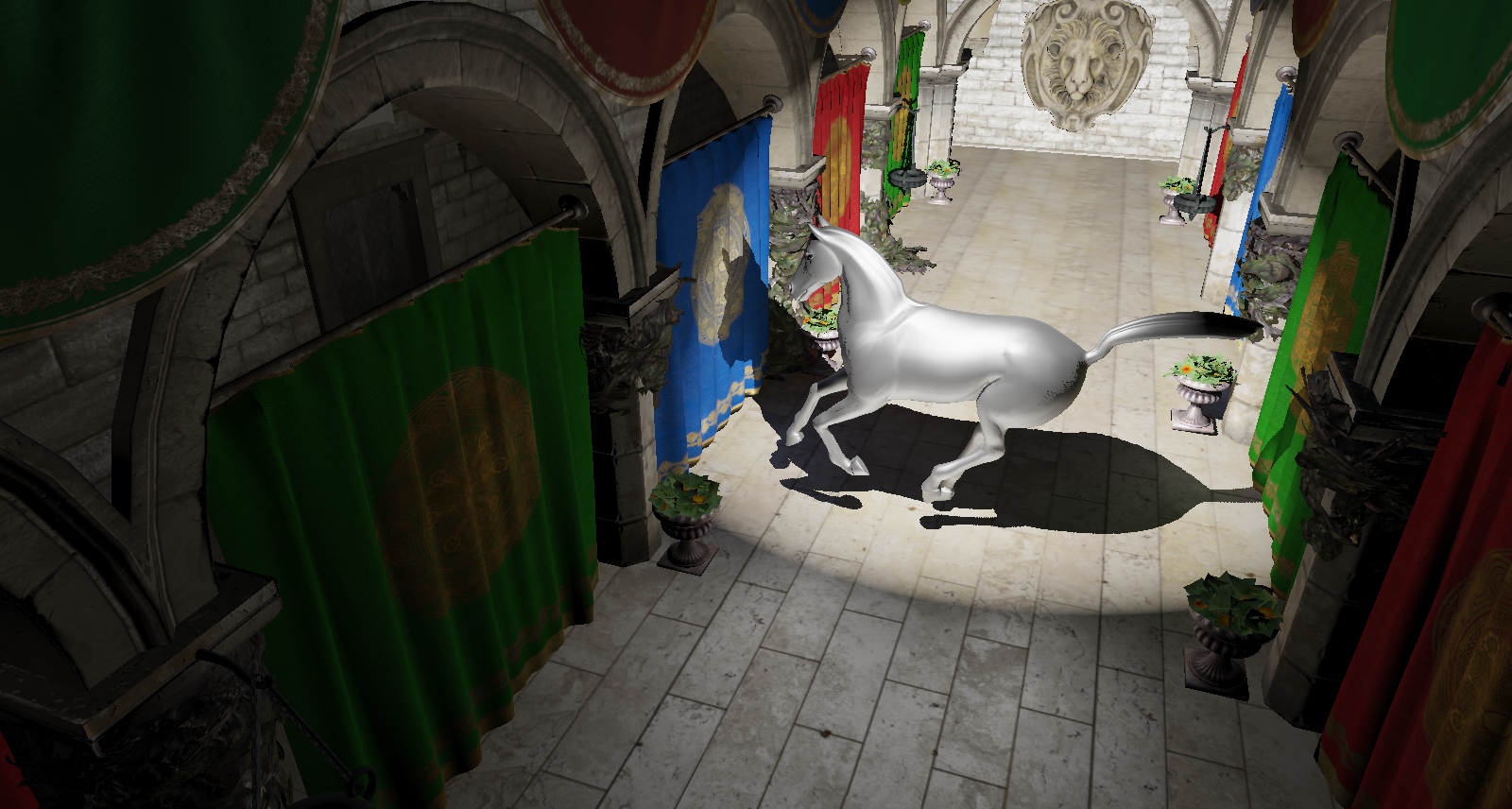}};
                \node [below right] at (img.152) [scale=1.2] {\hspace{0.01cm}\textcolor{white}{\large\textcolor{red}{\textbf{5 ZH}}}};
                \node [below left] at (img.28) [scale=1.2] {\textcolor{white}{23ms}};
                \node [above right] at (img.208) [scale=1.2] {\hspace{0.01cm}\textcolor{white}{\shortstack{rmse 0.2417\\ psnr 12.336 \\ssim 0.8592}}};
                \draw[red,ultra thick,rounded corners] (4,0.05) rectangle (6,1.3);
          \end{scope}
        \end{tikzpicture}
    \end{subfigure}
    \begin{subfigure}[b]{0.465\linewidth}%
        \begin{tikzpicture}
          \begin{scope}[spy using outlines=
              {rectangle, magnification=3, width=3.1cm, height=1.7cm, connect spies, rounded corners}]
                \node[anchor=south west,inner sep=0] (img) {\includegraphics[width=\textwidth]{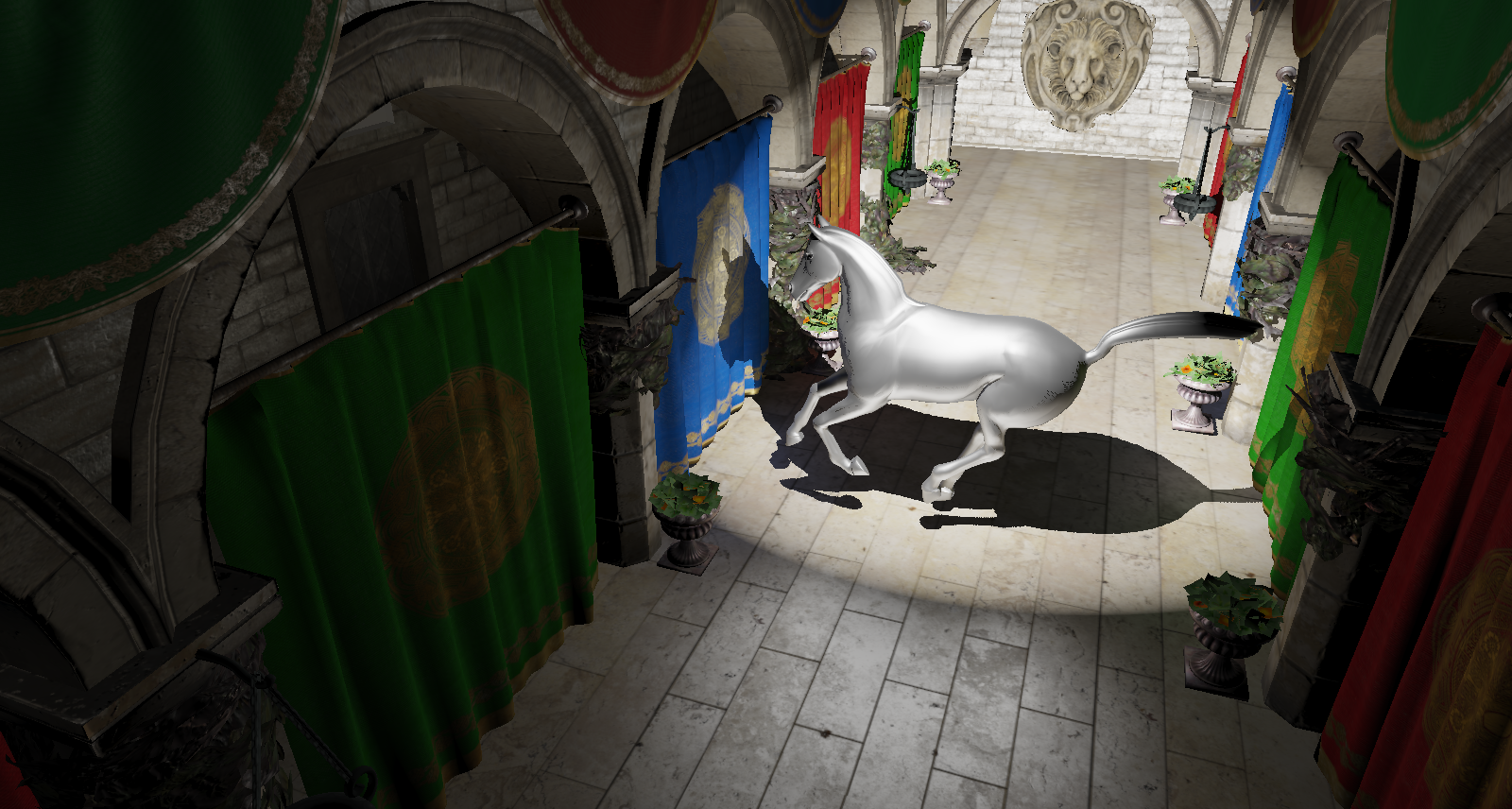}};
                \node [below right] at (img.152) [scale=1.2] {\hspace{0.01cm}\textcolor{white}{\large\textcolor{red}{\textbf{10 ZH}}}};
                \node [below left] at (img.28) [scale=1.2] {\textcolor{white}{32ms}};
                \node [above right] at (img.208) [scale=1.2] {\hspace{0.01cm}\textcolor{white}{\shortstack{rmse 0.2336\\ psnr 12.629\\ssim 0.8733}}};
                \draw[red,ultra thick,rounded corners] (4,0.05) rectangle (6,1.3);
          \end{scope}
        \end{tikzpicture}
    \end{subfigure} 
    \hspace*{0.5cm}
    \\%
    \hspace*{0.5cm}
    \begin{subfigure}[b]{0.465\linewidth}%
        \begin{tikzpicture}
          \begin{scope}[spy using outlines=
              {rectangle, magnification=3, width=5cm, height=4cm, connect spies, rounded corners}]
                \node[anchor=south west,inner sep=0] (img) {\includegraphics[width=\textwidth]{graphics/VSGLcomparisonCGF/sponza-hvl400-15.png}};
                \node [below right] at (img.152) [scale=1.2] {\hspace{0.01cm}\textcolor{white}{\large\textcolor{red}{\textbf{15 ZH}}}};
                \node [below left] at (img.28) [scale=1.2] {\textcolor{white}{40ms}};
                \node [above right] at (img.208) [scale=1.2] {\hspace{0.01cm}\textcolor{white}{\shortstack{rmse 0.2317\\ psnr 12.703\\ssim 0.8785}}};
                \draw[red,ultra thick,rounded corners] (4,0.05) rectangle (6,1.3);
          \end{scope}
        \end{tikzpicture}
    \end{subfigure} 
    \begin{subfigure}[b]{0.465\linewidth}%
        \begin{tikzpicture}
          \begin{scope}[spy using outlines=
              {rectangle, magnification=3, width=3.5cm, height=2cm, connect spies, rounded corners}]
                \node[anchor=south west,inner sep=0] (img) {\includegraphics[width=\textwidth]{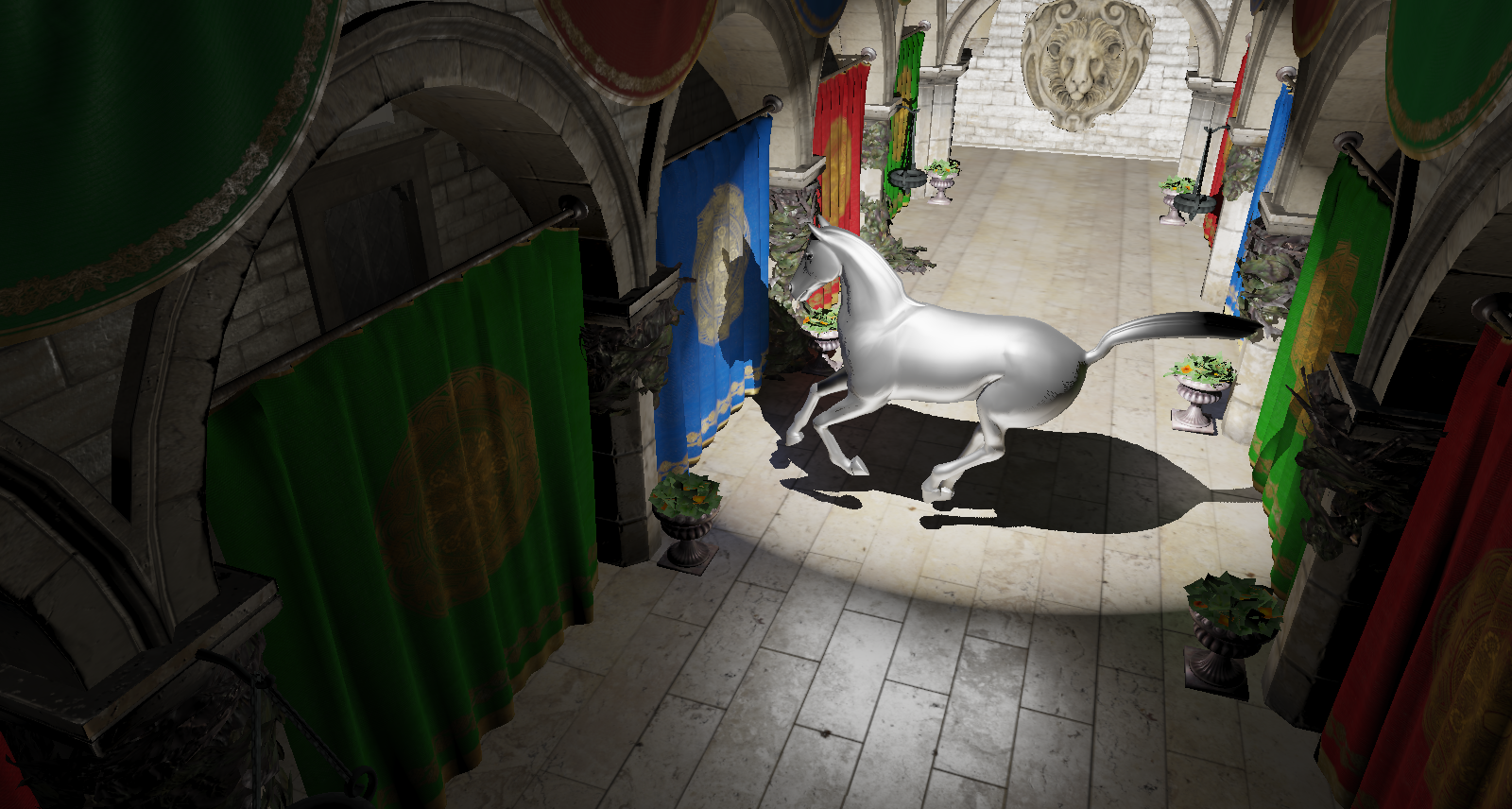}};
                \node [below right] at (img.152) [scale=1.2] {\hspace{0.01cm}\textcolor{white}{\large\textcolor{red}{\textbf{20 ZH}}}};
                \node [below left] at (img.28) [scale=1.2] {\textcolor{white}{47ms}};
                \node [above right] at (img.208) [scale=1.2] {\hspace{0.01cm}\textcolor{white}{\shortstack{rmse 0.2314\\ psnr 12.713\\ssim 0.8804}}};
                \draw[red,ultra thick,rounded corners] (4,0.05) rectangle (6,1.3);
          \end{scope}
        \end{tikzpicture}
    \end{subfigure}
    \hspace*{0.5cm}
    \\%
    \caption{Observation of the effect of the number of zonal harmonics using 400 HVLs. We clearly see the specular reflection on the ground becoming higher with higher frequency. However, we can also see that the rendering does not need to go beyond 15 ZH. Indeed, between 15 and 20 ZH, we do not observe any visible difference. The error values show it as well.}%
    \label{fig:HVL}
\end{figure}

The scene of \fig{\ref{fig:HVL}} shows the effect of the number of SH bands on to the strong specular effect on the ground. We do not observe any difference between 15 and 20 ZH, the error values show it as well. This means that for this scene we do not need to go beyond 15 bands.

\subsection{Comparison with VSGL and VPL}

\begin{figure*}[]%
    \centering%
    \hspace{0.5cm}
    \begin{subfigure}[b]{0.465\linewidth}%
        \begin{tikzpicture}
          \begin{scope}[spy using outlines=
              {rectangle, magnification=3, width=3.1cm, height=1.7cm, connect spies, rounded corners}]
                \node[anchor=south west,inner sep=0] (img) {\includegraphics[width=\textwidth]{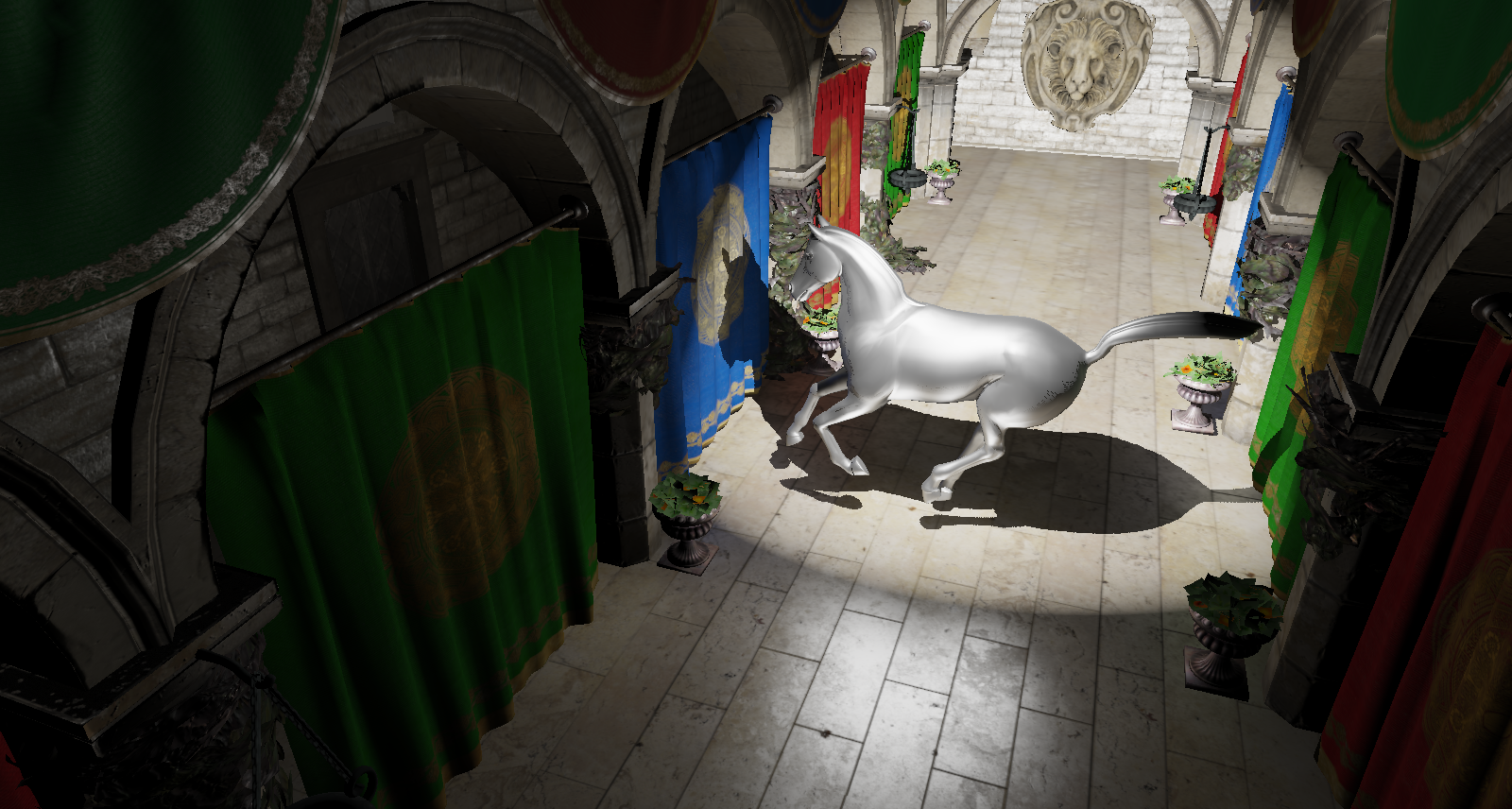}};
                \node [below right] at (img.152) [scale=1.2] {\hspace{0.01cm}\textcolor{white}{400 VSGL}};
                \node [below left] at (img.28) [scale=1.2] {\textcolor{white}{32ms}};
                \node [above right] at (img.208) [scale=1.2] {\hspace{0.01cm}\textcolor{white}{\shortstack{rmse 0.2617\\ psnr 11.645\\ssim 0.8756}}};
                \draw[red,ultra thick,rounded corners] (4,0.05) rectangle (6.5,1.5);
          \end{scope}
        \end{tikzpicture}
    \end{subfigure}%
    \hfill%
    \begin{subfigure}[b]{0.465\linewidth}%
        \begin{tikzpicture}
          \begin{scope}[spy using outlines=
              {rectangle, magnification=3, width=3.1cm, height=1.7cm, connect spies, rounded corners}]
                \node[anchor=south west,inner sep=0] (img) {\includegraphics[width=\textwidth]{graphics/VSGLcomparisonCGF/sponza-hvl400-5.png}};
                \node [below right] at (img.152) [scale=1.2] {\hspace{0.01cm}\textcolor{white}{400 HVL - 5 ZH}};
                \node [below left] at (img.28) [scale=1.2] {\textcolor{white}{23ms}};
                \node [above right] at (img.208) [scale=1.2] {\hspace{0.01cm}\textcolor{white}{\shortstack{rmse 0.2417\\ psnr 12.336\\ssim 0.8592}}};
                \draw[red,ultra thick,rounded corners] (4,0.05) rectangle (6.5,1.5);
          \end{scope}
        \end{tikzpicture}
    \end{subfigure} 
    \hspace*{0.5cm}
    \\%
    \hspace*{0.5cm}
    \begin{subfigure}[b]{0.465\linewidth}%
        \begin{tikzpicture}
          \begin{scope}[spy using outlines=
              {rectangle, magnification=3, width=3.1cm, height=1.7cm, connect spies, rounded corners}]
                \node[anchor=south west,inner sep=0] (img) {\includegraphics[width=\textwidth]{graphics/VSGLcomparisonCGF/sponza-hvl400-10.png}};
                \node [below right] at (img.152) [scale=1.2] {\hspace{0.01cm}\textcolor{white}{400 HVL - 10 ZH}};
                \node [below left] at (img.28) [scale=1.2] {\textcolor{white}{32ms}};
                \node [above right] at (img.208) [scale=1.2] {\hspace{0.01cm}\textcolor{white}{\shortstack{rmse 0.2336\\ psnr 12.629\\ssim 0.8733}}};
                \draw[red,ultra thick,rounded corners] (4,0.05) rectangle (6.5,1.5);
          \end{scope}
        \end{tikzpicture}
    \end{subfigure}%
    \hfill%
    \begin{subfigure}[b]{0.465\linewidth}%
        \begin{tikzpicture}
          \begin{scope}[spy using outlines=
              {rectangle, magnification=3, width=3.5cm, height=2cm, connect spies, rounded corners}]
                \node[anchor=south west,inner sep=0] (img) {\includegraphics[width=\textwidth]{graphics/VSGLcomparisonCGF/sponza-hvl400-15.png}};
                \node [below right] at (img.152) [scale=1.2] {\hspace{0.01cm}\textcolor{white}{400 HVL - 15 ZH}};
                \node [below left] at (img.28) [scale=1.2] {\textcolor{white}{40ms}};
                \node [above right] at (img.208) [scale=1.2] {\hspace{0.01cm}\textcolor{white}{\shortstack{rmse 0.2317\\ psnr 12.713\\ssim 0.8785}}};
                \draw[red,ultra thick,rounded corners] (4,0.05) rectangle (6.5,1.5);
          \end{scope}
        \end{tikzpicture}
    \end{subfigure}
    \hspace*{0.5cm}
    \\%
    \caption{Comparison between VSGL and HVL. HVL need use enough ZH bands to match the rendering generated with VSGL.}%
    \label{fig:HVLVSGL}
\end{figure*}

\Fig{\ref{fig:1}} shows that at equal time, HVL and VSGL are able to render images with similar quality.
With the same amount of virtual light and considering the same quality on this scene, VSGL are faster \figp{\ref{fig:HVLVSGL}}. 
However, on mainly diffuse scenes, we could lower the number of ZH bands while keeping a similar quality. And so, HVL would become faster than VSGL. Hence, the number of ZH bands can be used as a slider to manage the quality according to the scene used, being a trade-off between quality and rendering time.

\begin{figure*}[]%
    \hspace{0.5cm}
    \begin{subfigure}[b]{0.465\linewidth}%
        \begin{tikzpicture}
          \begin{scope}[spy using outlines=
              {rectangle, magnification=6, width=1.5cm, height=1.5cm, connect spies, rounded corners}]
                \node[anchor=south west,inner sep=0] (img) {\includegraphics[width=\textwidth]{graphics/VSGLcomparisonCGF/sponza-path.png}};
                \node [below right] at (img.152) [scale=1.2] {\hspace{0.01cm}\textcolor{white}{Path-tracing}};
                \node [above right] at (img.208) [scale=1.2] {\hspace{0.01cm}\textcolor{white}{Reference}};
                \spy [red] on (5.93,3.7) in node (a) at (8,3.3);
                \spy [blue] on (4.65,3.73) in node (b) at (1,3.42);
                \spy [green] on (6.98,2.87) in node (c) at (8,1.5);
                \spy [orange] on (4.93,3.2) in node (d) at (1,1.8);
                \coordinate (DG) at (img.west);
                \coordinate (DD) at (img.east);
          \end{scope}
          \node[circle,draw=red,ultra thick,inner sep=0,minimum size=5pt,
      scale=7,xshift=-0cm,yshift=-0.001cm] (circle) at (a.center) {};
          \node[circle,draw=blue,ultra thick,inner sep=0,minimum size=5pt,
      scale=4,xshift=0.005cm,yshift=0.014cm] (circle) at (b.center) {};
          \node[circle,draw=green,ultra thick,inner sep=0,minimum size=5pt,
      scale=5.5,xshift=-0cm,yshift=-0.015cm] (circle) at (c.center) {};
          \node[circle,draw=orange,ultra thick,inner sep=0,minimum size=5pt,
      scale=4,xshift=-0cm,yshift=-0cm] (circle) at (d.center) {};
        \end{tikzpicture}
    \end{subfigure}%
    \hfill%
    \begin{subfigure}[b]{0.465\linewidth}%
        \begin{tikzpicture}
          \begin{scope}[spy using outlines=
              {rectangle, magnification=6, width=1.5cm, height=1.5cm, connect spies, rounded corners}]
                \node[anchor=south west,inner sep=0] (img) {\includegraphics[width=\textwidth]{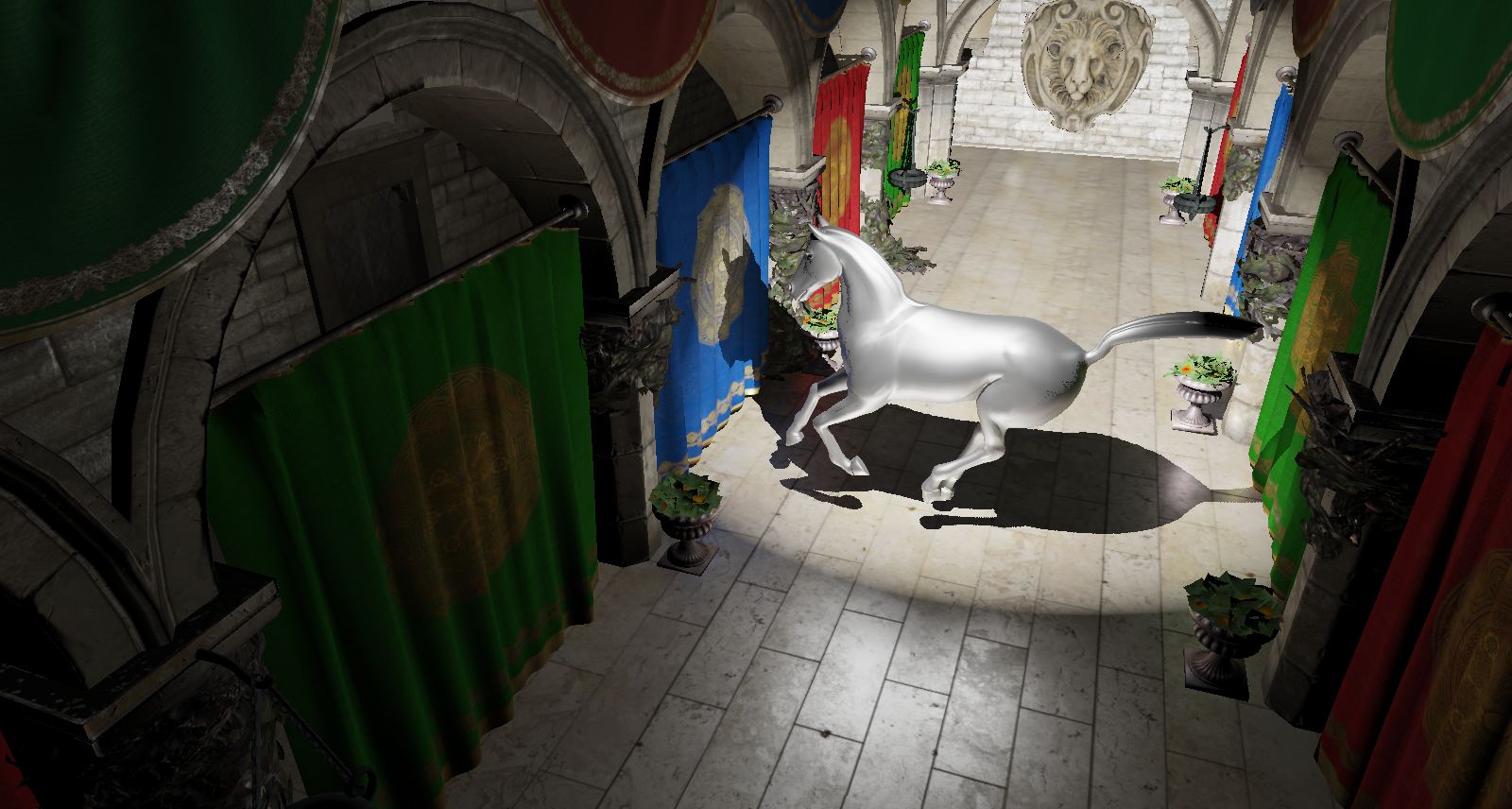}};
                \node [below right] at (img.152) [scale=1.2] {\hspace{0.01cm}\textcolor{white}{400 VPL}};
                \node [above right] at (img.208) [scale=1.2] {\hspace{0.01cm}\textcolor{white}{\shortstack{psnr 12.296\\ssim 0.8765}}};
                \node [below left] at (img.28) [scale=1.2] {\textcolor{white}{17ms}};
                \spy [red] on (5.93,3.7) in node (a) at (8,3.3);
                \spy [blue] on (4.65,3.73) in node (b) at (1,3.42);
                \spy [green] on (6.98,2.87) in node (c) at (8,1.5);
                \spy [orange] on (4.93,3.2) in node (d) at (1,1.8);
                \coordinate (FG) at (img.135);
          \end{scope}
          \node[circle,draw=red,ultra thick,inner sep=0,minimum size=5pt,
      scale=7,xshift=-0cm,yshift=-0.001cm] (circle) at (a.center) {};
          \node[circle,draw=blue,ultra thick,inner sep=0,minimum size=5pt,
      scale=4,xshift=0.005cm,yshift=0.014cm] (circle) at (b.center) {};
          \node[circle,draw=green,ultra thick,inner sep=0,minimum size=5pt,
      scale=5.5,xshift=-0cm,yshift=-0.015cm] (circle) at (c.center) {};
          \node[circle,draw=orange,ultra thick,inner sep=0,minimum size=5pt,
      scale=4,xshift=-0cm,yshift=-0cm] (circle) at (d.center) {};
        \end{tikzpicture}
    \end{subfigure} 
    \hspace*{0.5cm}
    \\%
    \hspace*{0.5cm}
    \begin{subfigure}[b]{0.465\linewidth}%
        \begin{tikzpicture}
          \begin{scope}[spy using outlines=
              {rectangle, magnification=6, width=1.5cm, height=1.5cm, connect spies, rounded corners}]
                \node[anchor=south west,inner sep=0] (img) {\includegraphics[width=\textwidth]{graphics/VSGLcomparisonCGF/sponza-hvl400-15.png}};
                \node [below right] at (img.152) [scale=1.2] {\hspace{0.01cm}\textcolor{white}{400 HVL - 15 ZH}};
                \node [above right] at (img.208) [scale=1.2] {\hspace{0.01cm}\textcolor{white}{\shortstack{psnr 12.703\\ssim 0.8785}}};
                \node [below left] at (img.28) [scale=1.2] {\textcolor{white}{40ms}};
                \spy [red] on (5.93,3.7) in node (a) at (8,3.3);
                \spy [blue] on (4.65,3.73) in node (b) at (1,3.42);
                \spy [green] on (6.98,2.87) in node (c) at (8,1.5);
                \spy [orange] on (4.93,3.2) in node (d) at (1,1.8);
          \end{scope}
          \node[circle,draw=red,ultra thick,inner sep=0,minimum size=5pt,
      scale=7,xshift=-0cm,yshift=-0.001cm] (circle) at (a.center) {};
          \node[circle,draw=blue,ultra thick,inner sep=0,minimum size=5pt,
      scale=4,xshift=0.005cm,yshift=0.014cm] (circle) at (b.center) {};
          \node[circle,draw=green,ultra thick,inner sep=0,minimum size=5pt,
      scale=5.5,xshift=-0cm,yshift=-0.015cm] (circle) at (c.center) {};
          \node[circle,draw=orange,ultra thick,inner sep=0,minimum size=5pt,
      scale=4,xshift=-0cm,yshift=-0cm] (circle) at (d.center) {};
        \end{tikzpicture}
    \end{subfigure}%
    \hfill%
    \begin{subfigure}[b]{0.465\linewidth}%
        \begin{tikzpicture}
          \begin{scope}[spy using outlines=
              {rectangle, magnification=6, width=1.5cm, height=1.5cm, connect spies, rounded corners}]
                \node[anchor=south west,inner sep=0] (img) {\includegraphics[width=\textwidth]{graphics/VSGLcomparisonCGF/sponza-vsgl400.png}};
                \node [below right] at (img.152) [scale=1.2] {\hspace{0.01cm}\textcolor{white}{400 VSGL}};
                \node [above right] at (img.208) [scale=1.2] {\hspace{0.01cm}\textcolor{white}{\shortstack{psnr 11.645\\ ssim 0.8756}}};
                \node [below left] at (img.28) [scale=1.2] {\textcolor{white}{32ms}};
                \spy [red] on (5.93,3.7) in node (a) at (8,3.3);
                \spy [blue] on (4.65,3.73) in node (b) at (1,3.42);
                \spy [green] on (6.98,2.87) in node (c) at (8,1.5);
                \spy [orange] on (4.93,3.2) in node (d) at (1,1.8);
          \end{scope}
          \node[circle,draw=red,ultra thick,inner sep=0,minimum size=5pt,
      scale=7,xshift=-0cm,yshift=-0.001cm] (circle) at (a.center) {};
          \node[circle,draw=blue,ultra thick,inner sep=0,minimum size=5pt,
      scale=4,xshift=0.005cm,yshift=0.014cm] (circle) at (b.center) {};
          \node[circle,draw=green,ultra thick,inner sep=0,minimum size=5pt,
      scale=5.5,xshift=-0cm,yshift=-0.015cm] (circle) at (c.center) {};
          \node[circle,draw=orange,ultra thick,inner sep=0,minimum size=5pt,
      scale=4,xshift=-0cm,yshift=-0cm] (circle) at (d.center) {};
        \end{tikzpicture}
    \end{subfigure}
    \hspace*{0.5cm} 
    \\%
    \caption{Comparison between VPL, HVL and VSGL. Compared to VPL, HVL and VSGL do not produce artifacts with a low number of virtual sources, HVLs reduce artefacts through frequency smoothing and the spherical nature of the virtual sources.}%
    \label{fig:VPLHVLVSGL}
\end{figure*}

Finally we add a comparison with the standard VPL method \figp{\ref{fig:VPLHVLVSGL}}. 
This method is obviously faster than HVL or VSGL but generates some artefacts.

\begin{table}[]%
    \setlength{\tabcolsep}{4pt}%
    \centering%
    \begin{tabular}{|c|c | c| c | c | c | c |}%
        \hline%
        \multicolumn{1}{|c|}{\multirow{2}{*}{Method}} & \multicolumn{2}{c|}{Parameters} & \multirow{2}{*}{Timing}  & \multirow{2}{*}{rmse} & \multirow{2}{*}{psnr} & \multirow{2}{*}{ssim}\\\cline{2-3}
         & number & other* &  & & & \\\hline\hline%
        
        \multirow{3}{*}{VPL}  & 400 & - & 17 & 0.2428 & 12.296 & 0.8765 \\\cline{2-7}
         & 900 & - & 39 & 0.2603 & 11.691 & 0.8852 \\\cline{2-7}
         & 1156 & - & 52 & 0.2425 & 12.308 & 0.8831 \\\hline
         
        \multirow{8}{*}{HVL}  & \multirow{4}{*}{400} & 5 & 23 & 0.2417 & 12.336 & 0.8592 \\\cline{3-7}
         &  & 10 & 32 & 0.2336 & 12.629 & 0.8733 \\\cline{3-7}
         &  & 15 & 40 & 0.2317 & 12.703 & 0.8785 \\\cline{3-7}
         &  & 20 & 47 & 0.2314 & 12.713 & 0.8804 \\\cline{2-7}
         & \multirow{4}{*}{1156}  & 5 & 68 & 0.2406 & 12.374 & 0.8643\\\cline{3-7}
         &  & 10 & 93 & 0.2321 & 12.687 & 0.8806 \\\cline{3-7}
         &  & 15 & 117 & 0.2297 & 12.776 & 0.8851\\\cline{3-7}
         &  & 20 & 140 & 0.2292 & 12.796 & 0.8859 \\\hline
         
        \multirow{3}{*}{VSGL}  & 400 & - & 32 & 0.2617 & 11.645 & 0.8756\\\cline{2-7} 
        & 529 & - & 40 & 0.2539 & 11.907 & 0.8740 \\\cline{2-7}
         & 1156 & - & 87 & 0.2336 & 12.632 & 0.8762\\\hline
        
        \multicolumn{7}{c}{\small* HVL = number of bands}
         
    \end{tabular}%
    \caption{Timings (ms) and metrics for the figures of this supplemental. Images that are not shown here are available in the images archive joined to the paper.}%
    \label{table1}
\end{table}

\bibliographystyle{alpha} 
\bibliography{references}